\documentclass{aa}  

\usepackage{graphicx}
\usepackage{txfonts}
\usepackage{xcolor}
\usepackage{textcomp}
\usepackage{amsmath}	% Advanced maths commands
\usepackage{amssymb}	% Extra maths symbols
\usepackage{upgreek}
\usepackage{placeins}
\usepackage{hyperref}

\newcommand{\orcid}[1]{\href{https://orcid.org/#1}{\includegraphics[width=10pt]{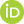}}}

\begin{document} 

   \title{The enigmatic double-peaked stripped-envelope SN~2023aew}

   %\subtitle{}

   \author{T. Kangas
          \inst{1,2}\orcid{0000-0002-5477-0217}
          \and
          H. Kuncarayakti\inst{2,1}\orcid{0000-0002-1132-1366}
          \and
          T. Nagao\inst{2,3,4}\orcid{0000-0002-3933-7861}
          \and
          R. Kotak\inst{2}\orcid{0000-0001-5455-3653}
          \and
          E. Kankare\inst{2}\orcid{0000-0001-8257-3512}
          \and
          M. Fraser\inst{5}\orcid{0000-0003-2191-1674} 
          \and
          H. Stevance\inst{6}\orcid{0000-0002-0504-4323}
          \and
          S. Mattila\inst{2,7}\orcid{0000-0001-7497-2994}
          \and
          K. Maeda\inst{8}\orcid{0000-0003-2611-7269}
          \and
          M. Stritzinger\inst{9}\orcid{0000-0002-5571-1833} 
          \and
          P. Lundqvist\inst{10}\orcid{0000-0002-3664-8082}
          \and
          N. Elias-Rosa\inst{11,12}\orcid{0000-0002-1381-9125}
          \and
          L. Ferrari\inst{13,14}\orcid{0009-0000-6303-4169}
          \and
          G. Folatelli\inst{13,14,15}\orcid{0000-0001-5247-1486}
          \and
          C. Frohmaier\inst{16}\orcid{0000-0001-9553-4723}
          \and
          L. Galbany\inst{12,17}\orcid{0000-0002-1296-6887}
          \and
          M. Kawabata\inst{18}\orcid{0000-0002-4540-4928}
          \and
          E. Koutsiona\inst{19}
          \and
          T. E. M\"uller-Bravo\inst{12,17}\orcid{0000-0003-3939-7167}
          \and
          L. Piscarreta\inst{12,17}\orcid{0009-0006-4637-4085}
          \and
          M. Pursiainen\inst{20}\orcid{0000-0003-4663-4300}
          \and
          A. Singh\inst{21}\orcid{0000-0003-2091-622X}
          \and
          K. Taguchi\inst{8}\orcid{0000-0002-8482-8993}
          \and
          R. S. Teja\inst{22,23}\orcid{0000-0002-0525-0872}
          \and
          G. Valerin\inst{11}\orcid{0000-0002-3334-4585}
          \and
          A. Pastorello\inst{11}\orcid{0000-0002-7259-4624}
          \and
          S. Benetti\inst{11}
          \orcid{0000-0002-3256-0016}
          \and
          Y.-Z. Cai \inst{24,25,26}\orcid{0000-0002-7714-493X}
          \and
          P. Charalampopoulos\inst{2}\orcid{0000-0002-0326-6715}
          \and
          C. P. Guti\'errez\inst{17,12}\orcid{0000-0003-2375-2064}
          \and
          T. Kravtsov \inst{27,2}\orcid{0000-0003-0955-9102}
          \and
          A. Reguitti\inst{28,11}\orcid{0000-0003-4254-2724}
          %ORDER NOT FINAL
          }

   \institute{Finnish Centre for Astronomy with ESO (FINCA), FI-20014 University of Turku, Finland
            \email{tjakangas@gmail.com}
            \and
            Department of Physics and Astronomy, FI-20014 University of Turku, Finland
            \and
            Aalto University Metsähovi Radio Observatory, Metsähovintie 114, 02540 Kylmälä, Finland
            \and
            Aalto University Department of Electronics and Nanoengineering, PO Box 15500, 00076 Aalto, Finland
            \and
            School of Physics, University College Dublin, L.M.I. Main Building, Beech Hill Road, Dublin 4, D04 P7W1, Ireland
            \and
            Department of Physics, University of Oxford, Denys Wilkinson Building, Keble Road, Oxford OX1 3RH, UK
            \and
            School of Sciences, European University Cyprus, Diogenes Street, Engomi, 1516 Nicosia, Cyprus
            \and
            Department of Astronomy, Kyoto University, Kitashirakawa-Oiwake-cho, Sakyo-ku, Kyoto 606-8502, Japan
            \and
            Department of Physics and Astronomy, Aarhus University, Ny Munkegade 120, 8000 Aarhus C, Denmark
            \and
            Oskar Klein Centre, Department of Astronomy, Stockholm University, Albanova University Centre, SE-106 91 Stockholm, Sweden
            \and
            INAF - Osservatorio Astronomico di Padova, Vicolo dell’Osservatorio 5, I-35122 Padova, Italy
            \and
            Institute of Space Sciences (ICE, CSIC), Campus UAB, Carrer de Can Magrans s/n, 08193 Barcelona, Spain
            \and
            Facultad de Ciencias Astron\'{o}micas y Geof\'{i}sicas, Universidad Nacional de La Plata, Paseo del Bosque s/n, B1900FWA La Plata, Argentina
            \and
            Instituto de Astrof\'{i}sica de La Plata (IALP), CCT-CONICET-UNLP, Paseo del Bosque S/N, B1900FWA, La Plata, Argentina
            \and
            Kavli Institute for the Physics and Mathematics of the Universe (WPI), The University of Tokyo, 5-1-5 Kashiwanoha, Kashiwa, Chiba, 277-8583, Japan
            \and
            School of Physics and Astronomy, University of Southampton, Southampton, SO17 1BJ, UK
            \and
            Institut d’Estudis Espacials de Catalunya (IEEC), Gran Capit\'{a}, 2-4, Edifici Nexus, Desp. 201, E-08034 Barcelona, Spain
            \and
            Nishi-Harima Astronomical Observatory, Center for Astronomy, University of Hyogo, 407-2 Nishigaichi, Sayo-cho, Sayo, Hyogo 679-5313, Japan
            \and
            Department of Physics, University of Crete, Vasilika Bouton, 70013 Heraklion, Greece
            \and
            Department of Physics, University of Warwick, Gibbet Hill Road, Coventry, CV4 7AL, UK
            \and
            Hiroshima Astrophysical Science Center, Hiroshima University, Higashi-Hiroshima, Hiroshima 739-8526, Japan
            \and
            Indian Institute of Astrophysics, II Block, Koramangala, Bengaluru-560034, Karnataka, India
            \and
            Pondicherry University, Chinna Kalapet, Kalapet, Puducherry 605014, India
            \and
            Yunnan Observatories, Chinese Academy of Sciences, Kunming 650216, P.R. China
            \and
            International Centre of Supernovae, Yunnan Key Laboratory, Kunming 650216, P.R. China
            \and
            Key Laboratory for the Structure and Evolution of Celestial Objects, Chinese Academy of Sciences, Kunming 650216, P.R. China
            \and
            European Southern Observatory, Alonso de C\'ordova 3107, Casilla 19, Santiago, Chile
            \and
            INAF – Osservatorio Astronomico di Brera, Via E. Bianchi 46, I-23807, Merate (LC), Italy
            }

   \date{Received ,; accepted ,}

% \abstract{}{}{}{}{} 
% 5 {} token are mandatory
 
  \abstract
  % context heading (optional)
  % {} leave it empty if necessary  
   {
   We present optical and near-infrared photometry and spectroscopy of SN 2023aew and our findings on its remarkable properties. This event, initially resembling a Type IIb supernova (SN), rebrightens dramatically $\sim$90~d after the first peak, at which time its spectrum transforms into that of a SN~Ic. The slowly evolving spectrum specifically resembles a post-peak SN~Ic with relatively low line velocities even during the second rise. The second peak, reached 119~d after the first peak, is both more luminous ($M_r = -18.75\pm0.04$~mag) and much broader than those of typical SNe Ic. Blackbody fits to SN~2023aew indicate that the photosphere shrinks almost throughout its observed evolution, and the second peak is caused by an increasing temperature. Bumps in the light curve after the second peak suggest interaction with circumstellar matter (CSM) or possibly accretion. We consider several scenarios for producing the unprecedented behavior of SN~2023aew. Two separate SNe, either unrelated or from the same binary system, require either an incredible coincidence or extreme fine-tuning. A pre-SN eruption followed by a SN requires an extremely powerful, SN-like eruption (consistent with $\sim$10$^{51}$~erg) and is also disfavored. We therefore consider only the first peak a true stellar explosion. The observed evolution is difficult to reproduce if the second peak is dominated by interaction with a distant CSM shell. A delayed internal heating mechanism is more likely, but emerging embedded interaction with a CSM disk should be accompanied by CSM lines in the spectrum, which are not observed, and is difficult to hide long enough. A magnetar central engine requires a delayed onset to explain the long time between the peaks. Delayed fallback accretion onto a black hole may present the most promising scenario, but we cannot definitively establish the power source.
   }

   \maketitle
%
%-------------------------------------------------------------------
\section{Introduction}

Supernovae (SNe), the luminous explosive deaths of stars, are an important ingredient in the chemical evolution of galaxies and an indicator of the final stages of the evolution of the exploding star. Core-collapse SNe (CCSNe), in particular, are the deaths of massive stars with zero-age main-sequence (ZAMS) masses of $\gtrsim8$~M$_\odot$. CCSNe are divided into types Ib (H-poor), Ic (H- and He-poor), IIb (a small amount of H), and II (H-rich) based on the pre-SN structures and compositions of their progenitor stars \citep{filippenko97}, which in turn is based on different kinds of mass loss during the lives of these stars. Mass loss is an especially murky aspect of stellar evolution \citep[e.g.,][]{smith14} that has a profound effect on the properties of the SN. SNe can thus be used to study the poorly understood late phases of the evolution of the massive progenitor stars.

Interaction with a binary companion can strip the hydrogen and/or helium envelope of a star, resulting in a stripped-envelope SN (SE-SN) of type IIb, Ib, or Ic depending on how much of the envelope is removed \citep{pods92, yoon15, drout23}. In addition to binary interaction, winds \citep[at least those of very massive single stars; e.g.,][]{yoon15} and pre-SN eruptions \citep[e.g.,][]{pastorello07} can also strip the envelope and create circumstellar matter (CSM) which the ejecta of the SN can then interact with, resulting in extra luminosity and various signatures of interaction in light curves and spectra, such as narrow emission lines \citep[e.g., in SNe Ibn and Icn, where ``n'' stands for narrow;][]{pastorello08,galyam22}, but which mass-loss processes affect which progenitor stars is still somewhat unclear \citep{smith14}.

\citet{das23} recently examined a sample of double-peaked SNe Ibc detected by the Zwicky Transient Facility \citep[ZTF;][]{bellm19,graham19}. The first peak in these SNe is interpreted as the shock-cooling phase after a shock breakout from an extended envelope of material previously ejected in an eruption, with the second peak corresponding to the typical $^{56}$Ni-powered peak of a SN Ibc. In superluminous SNe (SLSNe), early bumps in the light curve \citep[e.g.,][]{chen23b} have been suggested to result from such shock cooling \citep{piro15}, the shock possibly driven by a magnetar \citep{kasen16}. Another double-peaked SN Ic was studied by \citet{kuncarayakti23}, who attribute the first peak to $^{56}$Ni decay and the second to circumstellar interaction (CSI) instead. Early peaks and bumps in SN light curves can also result from pre-SN eruptions, at least in H-rich interacting SNe IIn \citep[e.g., in SN~2009ip and similar events;][]{mauerhan13,pastorello18}. In either case, double-peaked and other SNe with CSM-related features in their light curves can be used to probe the recent mass-loss history of the SN progenitor.

It is also possible that the later peaks in double-peaked events are not powered by CSI, but by a central engine. SN 2005bf \citep{anupama05,tominaga05,folatelli06} is a double-peaked SN Ib that was suggested, among other things, to be powered by a double-peaked $^{56}$Ni distribution \citep{tominaga05} or a magnetar central engine \citep{maeda07}. \citet{taddia18} also suggest these scenarios, the latter possibly created by jets, to be responsible for the second peak in the SN~Ic PTF11mnb (but the $^{56}$Ni required an initial progenitor mass of 85 $\mathrm{M}_\odot$, while the magnetar explanation could involve a less massive progenitor). SN 2019cad \citep{gutierrez21} is another double-peaked SN Ic in which a magnetar is considered a possibility for powering the second peak. In SN~2022jli, also of Type Ic, the favored power source is super-Eddington accretion from a companion star onto the newborn compact object instead \citep{chen23}. These examples demonstrate the diversity of double-peaked SE-SNe and their proposed mechanisms.

Here we study SN~2023aew, a remarkable seemingly type-changing double-peaked SE-SN with several peculiar properties. The discovery of SN 2023aew was reported by ZTF on 2023 Jan 23 \citep[MJD\,=\,59968;][]{munoz23}, and it was assigned the internal name ZTF23aaawbsc. However, even before this, on 2023 Jan 21 (MJD\,=\,59965), it was also detected by the \emph{Gaia} mission \citep{gaia16} and assigned the \emph{Gaia} Alerts\footnote{\url{http://gsaweb.ast.cam.ac.uk/alerts/alertsindex}} \citep{hodgkin21} name Gaia23ate. The early evolution of the SN, including its initial peak, was missed by ZTF because of solar obscuration and because of technical problems and bad weather at Palomar Observatory (J. Sollerman, priv. comm.); the last \emph{Gaia} non-detection is from 2022 Dec 18 (MJD\,=\,59932, i.e. over a month prior to discovery). However, after the original submission and arXiv release of this paper, \citet{sharma24} released another study of SN~2023aew, with some additional data including a serendipitous Transiting Exoplanet Survey Satellite (TESS) light curve that covers the first peak until shortly before the discovery.

The SN was initially classified as a SN IIb \citep{wise23}. However, after $\sim$90~d of slow decline, SN~2023aew exhibits a considerable rebrightening to a second peak brighter than the discovery magnitude, during which it resembles a SN~Ic \citep{frohmaier23,hoogendam23}. The second peak is in itself both spectroscopically and photometrically peculiar for a SE-SN: the light-curve peak is broader than those of normal SNe~Ic, even of ``luminous'' SNe (LSNe) of similar peak luminosity \citep{gomez22}, and the spectral evolution quite slow. The delay between the two peaks is, furthermore, very long compared to other double-peaked SE-SNe \citep[e.g.,][]{das23}.

In this paper, we present our photometric and spectroscopic follow-up observations of this remarkable transient in the optical and infrared. We analyze the light curve and spectral evolution, comparing both to other SNe Ic (especially double-peaked SNe), in order to assess numerous different physical scenarios. We present the observations we use, including proprietary and public ones, in Sect. \ref{sec:data}, our analysis in Sect. \ref{sec:analysis}, models of the photometry in Sect. \ref{sec:models}, and a discussion of our results in Sect. \ref{sec:disco}. We present our conclusions in Sect. \ref{sec:concl}. Throughout this paper, magnitudes are in the AB system \citep{okegunn83} unless otherwise noted, and we assume a $\Lambda$CDM cosmology with parameters $H_0 = 69.6$~km~s$^{-1}$~Mpc$^{-1}$, $\Omega_M = 0.286$ and $\Omega_\Lambda = 0.714$ \citep{bennett14}. Reported uncertainties correspond to 68\% ($1\sigma$) and upper limits to $3\sigma$. 

%--------------------------------------------------------------------
\section{Observations and data reduction}
\label{sec:data}

We made use of the public $gr$-band ZTF light curve of SN~2023aew -- through the Automatic Learning for the Rapid Classification of Events (ALeRCE) broker \citep{förster21}\footnote{\url{https://alerce.science/}} -- and the public \emph{Gaia} Alerts $G$-band light curve, and also retrieved the public classification spectrum of SN~2023aew \citep{wise23} from the Transient Name Server (TNS\footnote{\url{https://www.wis-tns.org/}}). %We also checked the ZTF forced photometry service\footnote{\url{https://ztfweb.ipac.caltech.edu/cgi-bin/requestForcedPhotometry.cgi}} \citep{masci19} for additional light-curve constraints, and found no data points between the last \emph{Gaia} non-detection and the discovery. 
We also extracted the public TESS light curve from Sector 60 and converted it to AB magnitudes (roughly equivalent to the $i$ band) using \texttt{TESSreduce}\footnote{\url{https://github.com/CheerfulUser/TESSreduce}} \citep{ridden21}{, which performs difference imaging, aperture photometry, and flux calibration based on the PANSTARRS1\footnote{\url{https://catalogs.mast.stsci.edu/panstarrs/}} \citep{ps1_cat} catalog}; the 200-second individual exposures were combined into 12-hour bins. Finally, we obtained public near-ultraviolet (NUV) images from the Ultra-violet/Optical Telescope (UVOT) on the Neil Gehrels \emph{Swift} Observatory \citep{gehrels04} between MJD\,=\,60076.5 and MJD\,=\,60085.4 and a late-time template image from the NASA High Energy Astrophysics Science Archive Research Center (HEASARC) Data Archive.\footnote{\url{https://heasarc.gsfc.nasa.gov/cgi-bin/W3Browse/swift.pl}} The TESS and \emph{Swift} data were first presented by \citet{sharma24}. In addition to public data, we have obtained an extensive set of imaging and spectroscopic observations in the optical and infrared.

\subsection{Imaging observations and photometry}

The full logs of our photometric measurements, including the UVOT and TESS magnitudes, are shown in Tables \ref{tab:growth} to \ref{tab:nics}. Optical and near-infrared (NIR) imaging of SN~2023aew was done using the following instruments:
\begin{itemize}
    \item the Alhambra Faint Object Spectrograph and Camera (ALFOSC) on the 2.56-m Nordic Optical Telescope (NOT) located at the Roque de los Muchachos Observatory, La Palma, as part of the NOT Unbiased Transient Survey 2 (NUTS2) collaboration\footnote{\url{https://nuts.sn.ie/}} (proposal 66-506, PIs Kankare, Stritzinger, Lundqvist): $ugriz$
    \item the Las Cumbres Observatory Global Telescope (LCOGT) network (proposal OPTICON 22A/012, PI Stritzinger): $uBgri$
    \item the 0.7-m GROWTH-India Telescope \citep[GIT,][]{GIT2022} situated at Indian Astronomical Observatory (IAO), Hanle, India: $griz$. The data were obtained under regular proposals covering three cycles (C): GIT-2023-C1/C2/C3-P03 (PI: RS Teja)
    \item the IO:O instrument on the 2.0-m robotic Liverpool Telescope (LT) on La Palma (proposal PL23A18, PI Frohmaier): $gri$
    \item the Near Infrared Camera Spectrometer \citep[NICS;][]{baffa01} on the 3.6-m Telescopio Nazionale Galileo (TNG) on La Palma (proposal A47TAC\_37, PI Valerin): one epoch of $JHKs$
    \item the Nordic Optical Telescope near-infrared Camera and spectrograph (NOTCam) on the NOT as part of NUTS2: $JHKs$
    \item the Nishiharima Infrared Camera (NIC) on the 2.0-m Nayuta telescope at the Nishi-Harima Astronomical Observatory: $JHKs$.
\end{itemize}

The ALFOSC images were reduced using the custom pipeline \texttt{Foscgui},\footnote{\texttt{Foscgui} is a graphic user interface aimed at extracting SN spectroscopy and photometry obtained with FOSC-like instruments. It was developed by E. Cappellaro. A package description can be found at {\url http://sngroup.oapd.inaf.it/foscgui.html}.} which performs bias subtraction and flat-field correction. The preprocessing and photometry of the GIT images were performed using a Python-based custom pipeline described in \citet{GIT2022}. LCOGT data were reduced by the automatic pipeline \texttt{BANZAI},\footnote{\url{https://github.com/LCOGT/banzai}} which performs similar steps, also including astrometry calibration and bad-pixel masking. The LT images were similarly automatically reduced by the IO:O pipeline.\footnote{\url{https://telescope.livjm.ac.uk/TelInst/Pipelines/}} The \texttt{Image Reduction and Analysis Facility} (\texttt{IRAF}\footnote{\url{https://iraf-community.github.io/}}) package \texttt{notcam}\footnote{\url{https://www.not.iac.es/instruments/notcam/quicklook.README}} was used to perform sky subtraction, flat-field correction, distortion correction, bad-pixel masking, and co-addition of the NOTCam NIR images. TNG/NICS NIR images were reduced in a similar manner using \texttt{IRAF} tasks and the photometry performed using the \texttt{SNOoPY}\footnote{\url{http://sngroup.oapd.inaf.it/snoopy.html}.} pipeline. Similar reduction and point-spread-function photometry was performed for Nayuta/NIC data as well. 

Aperture photometry was performed on the LCOGT, NOT, and LT images using Starlink\footnote{\url{http://starlink.eao.hawaii.edu/starlink}} \texttt{Gaia}\footnote{\url{http://star-www.dur.ac.uk/~pdraper/gaia/gaia.html}} \citep{currie14}. The zero points of each image were calibrated using field stars from the Sloan Digital Sky Survey (SDSS) Data Release 17 \citep{sdss-dr17} in the $ugriz$ filters, from the Two Micron All Sky Survey \citep[2MASS;][]{skrutskie06} in the $JHKs$ bands, and from the AAVSO Photometric All-Sky Survey\footnote{\url{https://www.aavso.org/apass}} (APASS) data release 9 \citep{henden16} in the $B$ band. Color terms were taken into account for the LCOGT\footnote{\url{https://lco.global/observatory/photometric-coefficients/}} and LT\footnote{\url{https://telescope.livjm.ac.uk/Pubs/LTTechNote1_TelescopeThroughput.pdf}} photometry; zero color terms were assumed for NOT and ZTF photometry. Host template subtraction was not performed in the optical, as the location of the SN has very little contribution from the host. No source is detected at the location of the SN in pre-explosion PANSTARRS1 $r$-band images, down to a limiting magnitude of $23.2$~mag. 

For NUV photometry, we used the tasks \texttt{uvotimsum} and \texttt{uvotsource} in \texttt{HEASOFT}\footnote{\url{https://heasarc.gsfc.nasa.gov/}} v.6.31.1 to co-add individual exposures and measure the source brightness, respectively. A 5-arcsecond circular aperture was used to extract the source flux, with background flux subtracted using a 10-arcsecond circular region; the same aperture was used in the NUV templates from 2023 Dec 14 in order to subtract any host contamination from the earlier measured fluxes, similarly to \citet{sharma24}.

\begin{figure}
\centering
\includegraphics[width=0.95\linewidth]{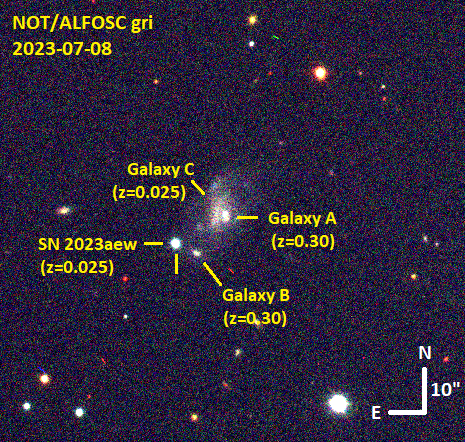}
\caption{Three-color (blue: $g$, green: $r$, red: $i$) cutout of the field of SN~2023aew, observed using NOT/ALFOSC on MJD=60134.0. Nuclei of host galaxy candidates, Galaxy A and Galaxy B, are labeled. These galaxies are located at $z=0.30$ behind the true host galaxy, Galaxy C (see Sect. \ref{sec:host}).}
\label{fig:fieldRGB}
\end{figure}

In Fig. \ref{fig:fieldRGB}, we show a three-color ($gri$) image of the field around SN~2023aew, created using our NOT/ALFOSC images on MJD\,=\,60134.0, when the seeing full width half-maximum (FWHM) was $\sim$0\farcs4. This figure shows the location of the SN, seemingly in the outskirts of a spiral galaxy (SDSS J174049.60+661229.1) and close to a smaller galaxy (SDSS J174050.55+661220.7). These host galaxy candidates are examined in Sect. \ref{sec:host}. We refer to the apparent nucleus of the spiral galaxy as Galaxy A, the smaller galaxy as Galaxy B, and the spiral galaxy itself as Galaxy C.

\subsection{Spectroscopic observations}

We obtained a total of 26 optical spectra and one NIR spectrum of SN~2023aew during and after its second peak. The full log of spectroscopic observations is shown in Table \ref{tab:specs}. These include 
\begin{itemize}
    \item eight spectra obtained using NOT/ALFOSC between MJD\,=\,60061 and MJD\,=\,60169 via NUTS2
    \item six spectra between MJD\,=\,60061 and MJD\,=\,60113 using the fiber-fed KOOLS-IFU \citep{kools} integral field unit on the 3.8-m Seimei telescope \citep{kurita20} at the Okayama Astrophysical Obsevatory in Japan (proposals 23A-K-0001, 23A-N-CT10, PI Maeda, and 23A-K-0017, PI Kawabata)
    \item four spectra between MJD\,=\,60055 and MJD\,=\,60075 with the SPectrograph for the Rapid Acquisition of Transients \citep[SPRAT;][]{sprat} on the LT (proposal PL23A18, PI Frohmaier)
    \item four spectra between MJD\,=\,60095 and MJD\,=\,60126 with the Intermediate Dispersion Spectrograph (IDS) on the 2.5-m Isaac Newton Telescope (INT) on La Palma (proposal C54, PI M\"{u}ller-Bravo)
    \item three spectra with the Optical System for Imaging and low-Intermediate-Resolution Integrated Spectroscopy (OSIRIS) spectrograph on the 10.4-m Gran Telescopio Canarias (GTC), also on La Palma, between MJD\,=\,60184 and MJD\,=\,60227 (program GTCMULTIPLE2A-23A, PI Elias-Rosa)
    \item one spectrum on MJD\,=\,60147 with the Gemini Multi-Object Spectrograph (GMOS) on the 8.1-m Gemini-North telescope at the Mauna Kea Observatories, Hawaii (program GN-2023A-Q-113, PI Ferrari)
    \item one NIR spectrum (11700--24700~\AA) using TNG/NICS on MJD\,=\,60064.
\end{itemize}

%\textcolor{red}{INT reduction missing}

NOT/ALFOSC and GTC/OSIRIS spectra were reduced -- bias-subtracted, flat-fielded, wavelength- and flux-calibrated, and corrected for telluric absorption -- using \texttt{Foscgui}. LT/SPRAT spectra were automatically reduced by the LT pipeline \citep{spratpipe}. Performing the same steps, the TNG/NICS NIR spectrum was reduced through standard \texttt{IRAF} tasks. KOOLS-IFU spectra were bias-subtracted using \texttt{IRAF}, then reduced, sky-subtracted, and stacked using the \texttt{Hydra} package \citep{hydra1,hydra2}. INT/IDS spectra were reduced using the package \texttt{idsred}.\footnote{\url{https://github.com/temuller/idsred}} The GMOS spectrum was bias-subtracted, flat-fielded, and wavelength- and flux-calibrated using the \texttt{gemini-gmos}\footnote{\url{https://www.gemini.edu/observing/phase-iii/reducing-data/gemini-iraf-data-reduction-software}} package in \texttt{IRAF}. All spectra of SN~2023aew we have obtained are available on the Weizmann Interactive Supernova Data Repository\footnote{\url{https://www.wiserep.org/}} \citep[WISeREP;][]{yaron12}.

%--------------------------------------------------------------------
\section{Analysis}
\label{sec:analysis}

\begin{figure*}
\centering
\includegraphics[width=0.9\linewidth]{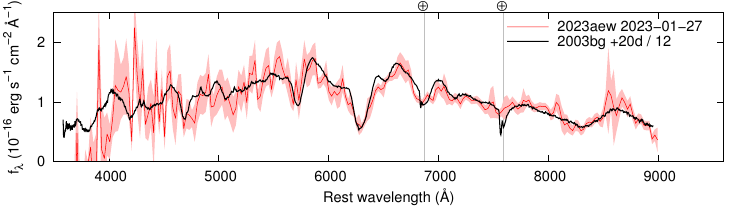}
\caption{Comparison of the SN~2023aew classification spectrum \citep[red;][]{wise23} and SN~2003bg \citep[black;][]{hamuy09}, a SN~IIb, at +20~d. The resemblance is fairly strong, but the SN~2023aew spectrum is quite noisy (uncertainties are shown as shaded regions). Telluric absorption lines ($\bigoplus$) are marked with vertical gray lines. }
\label{fig:03bg}
\end{figure*}

\begin{figure*}
\centering
\includegraphics[width=0.95\linewidth]{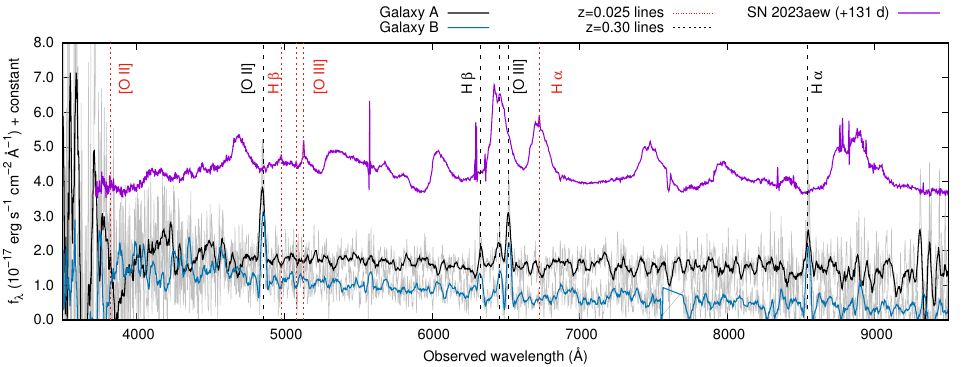}
\caption{NOT/ALFOSC spectra of Galaxies A (black) and B (blue) and the GTC/OSIRIS spectrum of SN~2023aew at +131~d showing narrow host lines (purple). Savitzky-Golay smoothing has been applied to the original ALFOSC spectra (gray). Identified lines of [O{\sc ii}], [O{\sc iii}], and H at $z=0.30$ are plotted as black dashed vertical lines. The same lines are also plotted at the redshift of SN~2023aew, $z=0.025$, as red dotted vertical lines. No lines are detected at this redshift in Galaxies A and B.}
\label{fig:hostspec}
\end{figure*}

\subsection{Host galaxy and redshift}
\label{sec:host}

We used the \texttt{SuperNova IDentification} \citep[\texttt{SNID};][]{snid} software to fit our spectra. The first spectrum was obtained with the low-resolution ($R\sim100$) Spectral Energy Distribution Machine \citep[SEDM;][]{sedm} on MJD\,=\,59971.5, 18 rest-frame days after the first peak and 101 rest-frame days before the second (main) peak (see Sect. \ref{sec:LC}). It was reported by \citet{wise23} to resemble that of a SN~IIb at $z=0.025$. We find that this fit is only found by loosening the fit quality requirement parameter to \texttt{rlapmin}=2.0 and forcing \texttt{SNID} to use type IIb, in which case the returned redshift is $z=0.026$. However, the best matches from the GEneric cLAssification TOol (GELATO)\footnote{\url{https://gelato.tng.iac.es/}} service using this redshift are also of type IIb, strengthening this association. By eye, the first spectrum of SN~2023aew 18~d after the first peak provides a reasonable match to SN~2003bg \citep{hamuy09}, a SN~IIb, 20~d post-peak, as shown in Fig. \ref{fig:03bg}. This SN is returned as the best fit by \texttt{SNID}. However, the SEDM spectrum has both a low resolution and a low signal-to-noise (S/N) ratio, making it not particularly constraining. The best \texttt{SNID} fits to our spectra taken during the second peak, on the other hand, are consistently those of post-peak SNe~Ic\footnote{Specifically, until +58~d, SNe~1995F and 1997ei at +11 to +57~d -- but we note that for these SNe, the peak epoch is not known and this is a lower limit for the phase. At late times, matches to SNe~IIb also reappear, presumably driven by an emission feature at $\sim6550$~\AA.} between $z=0.026$ and $z=0.036$. An exception is the first LT/SPRAT spectrum on MJD\,=\,60055, which can fit either a post-peak SN~Ib or a Ic in the same redshift range. 

The originally reported redshift of $z = 0.025$ indeed turns out to be correct. This conclusion is based on our GTC/OSIRIS spectrum from MJD\,=\,60208.9, in which we find faint narrow (unresolved) emission lines that we identify as host contamination: H$\alpha$ ($z=0.0252$), H$\beta$ ($z=0.0248$), and [O~{\sc iii}]~$\lambda\lambda 4959,5007$ ($z=0.0248$ and 0.0247 respectively). H$\alpha$ and [O~{\sc iii}]~$\lambda\lambda 4959,5007$ are also seen in the MJD\,=\,60226.9 spectrum, at $z=0.0254$ and $z=0.0251$ respectively. We also see lines at this redshift in our GTC/OSIRIS 2D spectra at the position of the large spiral galaxy close to the SN. We conclude that the spiral galaxy (Galaxy C in Fig. \ref{fig:fieldRGB}) is the host galaxy of SN~2023aew. In the following, we thus assume a redshift of $z=0.0250\pm0.0003$ for SN~2023aew, and all epochs, wavelengths, and absolute magnitudes are calculated with this assumption in mind. This redshift results in a luminosity distance of $109.8\pm1.4$~Mpc and a distance modulus of $\mu = 35.20\pm0.03$~mag.

Because of a lack of narrow emission lines from the host galaxy in our SN spectra before MJD\,=\,60200 (see Sect. \ref{sec:specs}), we initially attempted to constrain the redshift of the SN by setting the position angle of the slit to intersect Galaxies A and B, which appeared to be the nuclei of the obvious host candidates. We positioned the slit on Galaxy A on MJD\,=\,60068, 60077, and 60084 and on Galaxy B on MJD\,=\,60117 and 60133. We then stacked the host galaxy spectra for total exposure times of 2700~s for Galaxy A and 4200~s for Galaxy B. The resulting combined spectra are shown in Fig. \ref{fig:hostspec}. It is evident that neither object has clear emission lines at the \texttt{SNID}-suggested redshift of SN~2023aew, $z\approx0.03$ (red dotted lines in Fig. \ref{fig:hostspec}). Instead, in both galaxies we do see clear emission lines of [O~{\sc ii}]~$\lambda$3727, [O~{\sc iii}]~$\lambda\lambda$4959,5007, H$\beta$, and H$\alpha$ -- typical to star-forming SN host galaxies -- but lines in both objects are at a redshift of $z=0.302\pm0.001$. 

The GTC/OSIRIS 2D spectrum from MJD\,=\,60184.0 includes $z=0.30$ emission lines from Galaxy A in addition to the lines from Galaxy C (while the spectrum at MJD\,=\,60208.9 only includes $z=0.025$ lines from Galaxy C, owing to a different slit position angle), and of the two sets of lines, the ones at $z=0.30$ are considerably brighter. The fainter low-redshift lines from the true host galaxy are not seen in the earlier spectra (even stacked), most likely because of the lesser depth of the NOT/ALFOSC spectra. We thus infer that Galaxy A, which appears to be the nucleus of the host, Galaxy C, is, in fact, a separate galaxy shining through the disk of Galaxy C, most likely part of the same cluster as Galaxy B at $z=0.30$.

\subsection{Light curve}
\label{sec:LC}

We performed numerical interpolation of the light curves using a Gaussian process (GP) regression algorithm \citep{rasmussen06} to obtain the peak magnitudes and epochs. We used the Python-based {\tt george} package \citep{hodlr}, which implements various different kernel functions. The widely used Mat{\'e}rn kernel with $\nu$ parameter of 3/2 was applied. The uncertainty in the peak epoch was estimated as the time range when the GP light curve is brighter than the $1\sigma$ lower bound on the peak brightness. The first peak is only covered by the TESS observations. We obtain a first-peak epoch of MJD$_{TESS,\mathrm{peak}} = 59953.4^{+8.2}_{-2.4}$. In the well-observed $r$ band, we obtain a second-peak epoch of MJD$_{r,\mathrm{peak}} = 60074.9\pm3.1$. We use the latter as the reference date for all phases in this paper. We show the measured second-peak epochs and magnitudes in all bands in Table \ref{tab:peaks}.

\begin{figure*}
\centering
\includegraphics[width=0.9\linewidth]{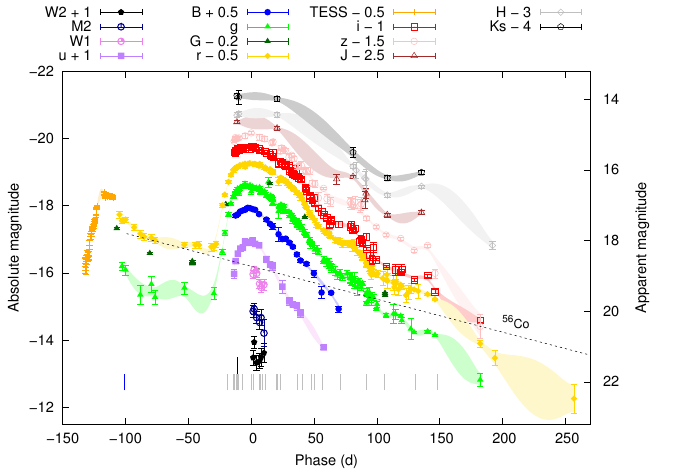}
\caption{NUV, optical, and NIR light curve of SN~2023aew. $G$-band errors are not reported at \emph{Gaia} Alerts, and only the distance uncertainty is included. The shaded regions correspond to the 68\% uncertainties of our GP fits. The decay rate of $^{56}$Co is indicated with a dashed black line. Our optical spectral coverage is indicated with gray vertical lines and our NIR spectrum with a black vertical line. The epoch of the classification spectrum \citep{wise23} is marked with a blue vertical line. All magnitudes have been corrected for Galactic extinction. Phases refer to the second $r$-band peak (MJD\,=\,60074.9).}
\label{fig:absLC}
\end{figure*}

The ZTF discovery magnitude is $r = 18.05 \pm 0.10$~mag, and the first $g$-band magnitude ($\sim2$~d later) is $g = 19.13 \pm 0.12$~mag, indicating a very red color soon after the first peak. The difference between the last TESS magnitudes, close to the $i$ band in effective wavelength, and the first $r$ points ($r - \mathrm{TESS}\sim0.5$~mag, including a possible decline over 5~d) after the first peak, is also consistent with a very red color. We applied a Milky Way extinction correction of $A_{V,\mathrm{MW}} = 0.106$~mag \citep{schlafly11}, using the \citet{cardelli89} extinction law. The location of the SN at the outskirts of the host galaxy (see Fig. \ref{fig:fieldRGB}) and a lack of observed narrow Na~{\sc i}~D absorption argue against a high extinction. Therefore we assumed a negligible host galaxy extinction, despite the red color after the first peak, which may be intrinsic to the event. This results in peak absolute magnitudes of $M_{TESS,\mathrm{peak}} = -17.88 \pm 0.12$~mag for the first peak and $M_{r,\mathrm{peak}} = -18.75 \pm 0.04$~mag for the more luminous second peak (the $i$ band, more comparable to TESS data, peaks at a similar $M_{i,\mathrm{peak}} = -18.74 \pm 0.04$~mag). We note that \citet{sharma24} obtain different TESS magnitudes: their values are roughly 0.5~mag fainter than the ones returned by \texttt{TESSreduce}. Based on the $g-r$ color of $\sim1$~mag, one might expect the TESS magnitudes to be brighter than in the $r$ band, which is not the case in their Fig. 1, but such a difference can be seen in our values. The absolute-magnitude light curve, along with our GP fits, is shown in Fig. \ref{fig:absLC}.

\begin{table}
\centering
\caption{Epochs and absolute magnitudes at the second peak in the optical bands.}
\label{tab:peaks}
\begin{tabular}{ccc}
\hline
Filter & MJD$_\mathrm{peak}$ & $M_{\mathrm{peak}}$ \\
 &  & (mag) \\
\hline
\hline
$u$ & $60074.8^{+3.6}_{-3.7}$ & $-17.93 \pm 0.05$ \\
$B$ & $60073.3\pm3.4$ & $-18.41 \pm 0.04$ \\
$g$ & $60070.8^{+3.6}_{-3.0}$ & $-18.60\pm0.04$ \\
$r$ & $60074.9\pm3.1$ & $-18.75 \pm 0.04$ \\
$i$ & $60073.9^{+4.9}_{-6.1}$ & $-18.74 \pm 0.04$ \\
$z$ & $60074.7\pm5.3$ & $-18.63 \pm 0.05$ \\
\hline
\end{tabular}
\tablefoot{The $r$-band peak is used as the reference epoch throughout the paper.}
\end{table}

The $\sim$90~d after the first peak show first a decline similar to the end of the $^{56}$Ni bump and the beginning of the tail phase in a SE-SN, then a flattening inconsistent with $^{56}$Co decay. The apparent undulations in the GP fit between $-$75 and $-$25~d are only a result of the slight over-fitting of the $g$-band light curve in this range (see Fig. \ref{fig:absLC}): this is expected behavior, as there are few data points to anchor the fit and the evolution of the light curve in this range is slower than around and after peak \citep{stevance23}. The rebrightening begins between $-$28 and $-$23~d, the last epoch of the plateau\footnote{This point does show a slight upturn in brightness, but is still 1$\sigma$ consistent with the previous points.} and the first detection of the rebrightening, respectively. This is thus the constraint we obtain for the rise time to the second peak. As can be seen in Fig. \ref{fig:absLC} and in Table \ref{tab:peaks}, in bands where we have enough data to reliably fit for the peak epoch -- in the NIR and the NUV, there are not enough data points for this -- the peak epochs are all consistent with the $r$ band within 1$\sigma$. This is somewhat unusual if the second peak is a SN: in most SNe the bluer wavelengths tend to reach their peak first because of rapidly declining temperatures, but in SN~2023aew it is reached roughly simultaneously in all optical bands indicating an unusual temperature evolution.

After the second peak, the $u$ and $B$ bands decline linearly at $0.058\pm0.002$ and $0.043\pm0.002$~mag~d$^{-1}$ (rest-frame) respectively. The NUV bands similarly decline quickly, and are faint and red (W1$-u \approx 2$~mag and W2$-\mathrm{W1} \approx 1.5$~mag) even near the peak. The $g$ and especially the $r$ and $i$ bands initially decline more slowly, then show a steepening in the light curve around +40~d and a slight bump around +80~d. The bump is replicated in the $J$ band as well. Such features could indicate the presence of CSM interaction or, since the $gr$-band bump corresponds roughly to the same brightness as the flattening after the first peak, possibly the continuing contribution of an earlier emission process. The post-peak decline is too swift for $^{56}$Co decay, and a clearly $^{56}$Co-powered tail phase is not reached during our observations. Our last photometry point at +257~d, consistent with the PANSTARRS1 detection limit, may be severely contaminated by host-galaxy emission (lines from the galaxy are detectable in our spectra $\sim$100~d before this; Sect. \ref{sec:host}), and the decline rate at this point is likely underestimated. A faster decline, signifying dust formation or non-negligible gamma-ray leakage, is seen in, for example, SN~2020wnt \citep{gutierrez22,tinyanont23}, but in SN~2020wnt the fully trapped decay is observed before this. Sometime between $+$110 and $+$135~d, the $JHKs$ light curve starts rising again. Roughly simultaneously (between 130 and 140~d), the optical light curves briefly flatten. Even the NIR rebrightening has subsided by $\sim$190~d, at which point the decline has become faster in all bands. The gap in the NIR coverage prevents determining the duration of the rebrightening.

\begin{figure}
\centering
\includegraphics[width=0.99\linewidth]{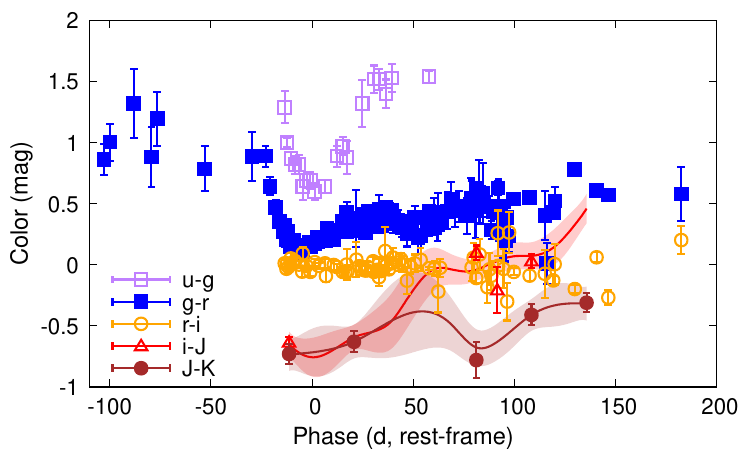}
\caption{Measured colors of SN~2023aew in various bands when roughly simultaneous ($\Delta t <0.1$~d) epochs are available (points) and subtractions between our GP fits of individual light curves when simultaneous photometry points at the different bands are scarce (lines and shaded regions).}
\label{fig:colors}
\end{figure}

The color evolution of SN~2023aew, based on measured colors -- when roughly simultaneous multi-band observations are available -- and on our GP fits, is shown in Fig.~\ref{fig:colors}. The $g-r$ color, the only color available before the second peak, stays fairly constant at $\sim1$~mag from discovery until the sudden rise begins, at which point the SN becomes considerably bluer, reaching $g-r \approx 0.2$~mag at peak. This rapid evolution to bluer colors and higher temperatures is reminiscent of the second peak of the double-peaked SN~2022xxf \citep{kuncarayakti23}. During the decline after the peak, the $g-r$ color slowly reddens, reaching $g-r \sim 0.6$~mag at $+$150~d. Meanwhile, the $r-i$ color evolves even slower and becomes slightly bluer: at maximum light, $r-i \approx 0$, but at $+$150~d, $r-i \sim -0.1$~mag. This evolution is not typical for SE-SNe, which tend to quickly redden after the peak in all the optical colors \citep{stritzinger18}. The $u-g$ color is more typical: it becomes bluer during the rise to the second peak, $u-g \sim 0.65$~mag at its bluest, and subsequently reddens much faster than $g-r$, reaching $u-g \sim 1.6$~mag around $+$30~d. The $i-J$ and $J-K$ colors are similar around the peak ($\sim -0.7$~mag), but diverge around $+$40~d, with $i-J$ reddening to $\sim0.4$~mag around $+$130~d and $J-K$ only reaching $\sim-0.3$~mag.

\begin{figure*}
\centering
\includegraphics[width=0.99\linewidth]{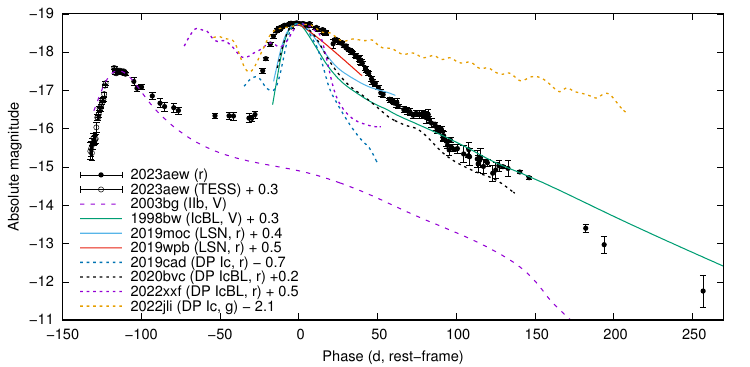}
\caption{Comparison of the light curve of SN~2023aew (points) to other SE-SNe (lines). The first peak at first resembles SN~2003bg \citep[dashed purple line;][]{hamuy09}, until the flattening around $-$100~d. The second peak has a luminosity similar to the Type Ic-BL SN~1998bw \citep{galama98,mckenzie99} and LSNe \citep{gomez22} (solid lines), but a much slower evolution. Of the double-peaked SE-SNe (DP; dotted lines) pictured here -- SNe~2019cad \citep{gutierrez21}, 2020bvc \citep{ho20,izzo20}, 2022xxf \citep{kuncarayakti23}, and 2022jli \citep{chen23} -- only SN~2020bvc resembles the late-time light curve of SN~2023aew, and the separation between peaks is much shorter in all of them than in SN~2023aew.}
\label{fig:lccomp}
\end{figure*}

We compare the light-curve behavior of SN~2023aew to other SE-SNe in Fig. \ref{fig:lccomp}. The first peak is similar in luminosity to SN~2003bg \citep{hamuy09}, but declines considerably slower from the beginning, both in and before the tail phase. The phase compared to SN2003bg also roughly matches that obtained from the \texttt{SNID} fit to the classification spectrum (see Fig. \ref{fig:03bg}). The second peak, on the other hand, is compared to several SNe Ic. While the maximum luminosity of the second peak is close to that of SN~1998bw \citep{galama98,mckenzie99}, a broad-lined SN Ic (SN Ic-BL), and the LSNe\footnote{We obtained the public ZTF light curves of two LSNe considered by \citet{gomez22} as slow and Ic-like; SN~2019moc and SN~2019wpb. Other SNe in this subgroup do not have useful public ZTF light curves.} studied by \citet{gomez22}, clear differences are apparent as well. During our observations, SN~2023aew does not reach the $^{56}$Co-powered tail phase of SN~1998bw. The second peak of SN~2023aew evolves much slower than SNe~Ic that attain a similar peak luminosity. 

We also show a comparison to four other double-peaked SNe~Ic: SNe~2019cad \citep{gutierrez21}, 2020bvc \citep{ho20,izzo20}, 2022jli \citep{chen23}, and 2022xxf \citep{kuncarayakti23}. All show a much shorter delay and/or a smaller contrast between the peaks than SN~2023aew. SNe~2019cad and 2022xxf evolve much faster than the second peak of SN~2023aew, while SN~2022jli shows an even slower, linear decline and periodic undulations that are not seen in SN~2023aew (an $\sim$80-day period is technically possible, but with only two bumps we cannot say this for certain). All three of these double-peaked SNe~Ic have a different light-curve evolution and different proposed physical scenarios, but none of them resemble SN~2023aew. SN~2020bvc does show a similar peak luminosity and bumpy light curve as SN~2023aew, but its light-curve peak is again narrower and its spectrum that of a SN~Ic-BL, with hints of an accompanying long gamma-ray burst that remained undetected \citep{ho20,izzo20}.

The double-peaked SE-SNe studied by \citet{das23} mostly do not closely resemble the light-curve evolution of SN~2023aew either; a similar delay between peaks is only seen in one event in the ZTF sample, and in this case (SN~2021uvy) the rise to both peaks is much slower and the ``dip'' between the peaks much less pronounced. However, the objects in \citet{das23} do exhibit a wide range of decline rates.

\subsection{Spectroscopy}
\label{sec:specs}

\begin{figure*}
\centering
\includegraphics[width=0.87\linewidth]{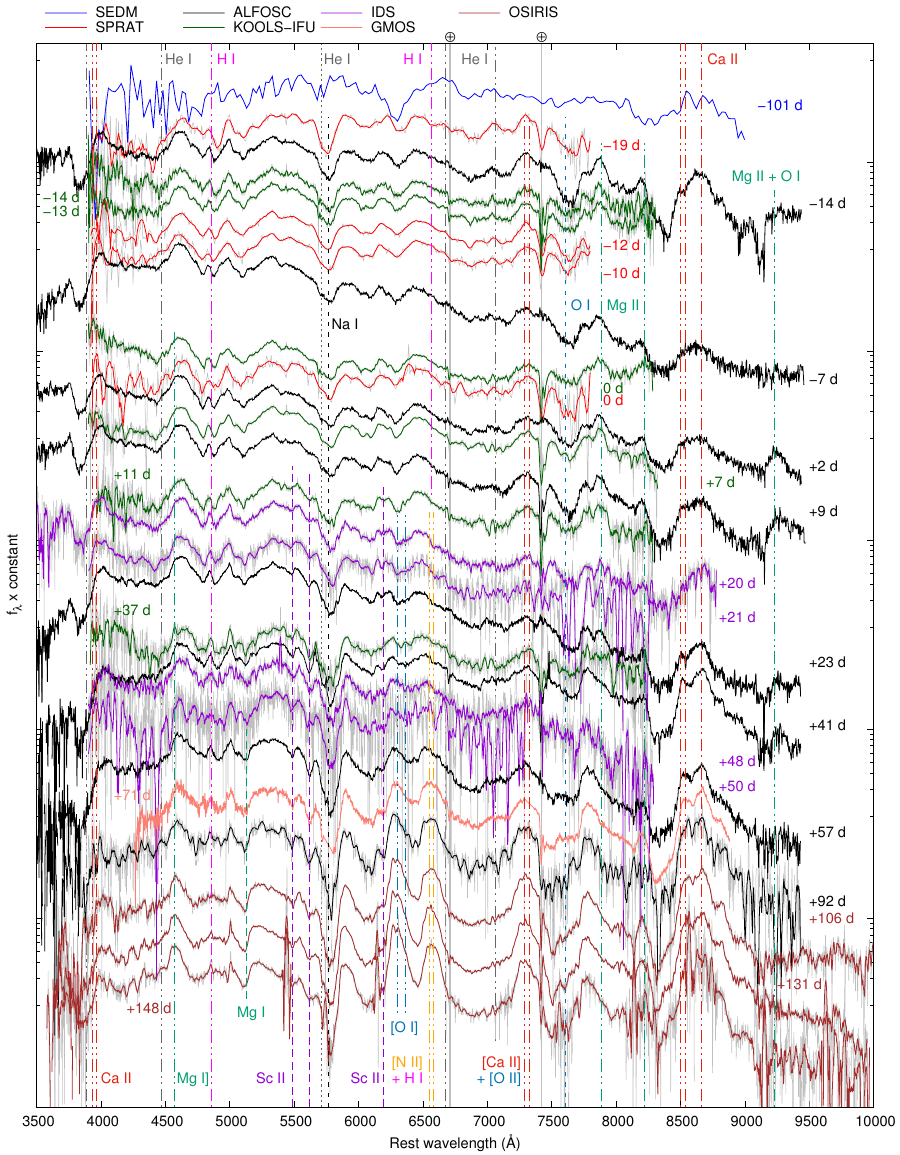}
\caption{Spectral evolution of the first \citep[blue;][]{wise23} and second peak of SN 2023aew. Different instruments are color-coded, and epochs refer to the second peak. Telluric absorption lines ($\bigoplus$) are marked with solid light gray lines. Dash-dotted lines are at zero velocity (of the stronger line in case of a close doublet); long dashes at $-$2500 km~s$^{-1}$; short dashes at $-$6500 km~s$^{-1}$; and dotted lines at $-$8500 km~s$^{-1}$. When Savitzky-Golay smoothing has been applied, the original spectrum is plotted in gray.}
\label{fig:specseq}
\end{figure*}

We present our optical spectra of SN~2023aew and the \citet{wise23} SEDM spectrum of the first peak in Fig. \ref{fig:specseq}. Here we describe the main features of the spectrum and their evolution.

A clear change is apparent between spectra taken during the first and second peak. The classification spectrum at $-101$~d relative to the second peak is characterized by lines of He~{\sc i}, Ca~{\sc ii} and H$\alpha$ (possibly contaminated by Si~{\sc ii}~$\lambda\lambda$6347,6371). 
The He~{\sc i}~$\lambda$5876 line is seen in absorption around $-9000$~km~s$^{-1}$, and the Ca~{\sc ii} NIR triplet absorption exhibits a similar velocity, while the 6678 and 7065~\AA~lines are dominated by emission. These features are similar to the Type~IIb SN~2003bg around 10--20~d past maximum light \citep{hamuy09,mazzali09}, which is also suggested by the \texttt{SNID} fit at this epoch. The absorption minimum of the H$\alpha$ line is at $-$12\,500~km~s$^{-1}$, assuming no contribution from Si~{\sc ii}~$\lambda\lambda$6347,6371 -- a line dominated by Si~{\sc ii} is unlikely, as its velocity would be only $\sim-$3000~km~s$^{-1}$. A comparison of the pseudo-equivalent width of the H$\alpha$+Si~{\sc ii}~$\lambda\lambda$6347,6371 feature ($\sim160$~\AA) to the ones presented in \citet{holmbo23} indicates a type IIb and an H$\alpha$-dominated feature as well, while the strength of the He~{\sc i}~$\lambda$5876+Na~{\sc i}~$\lambda\lambda$5890,5986 feature is more typical of SNe Ic. We note, though, that the spectrum of SN~2003bg (see Fig. \ref{fig:03bg}) differs from prototypical SNe IIb  \citep[e.g.,][]{filippenko94,ergon14} in that the H$\alpha$ line is still stronger than He lines at this phase. Some  C~{\sc ii}~$\lambda$6580 contribution cannot be excluded, though.

In the $-$19~d spectrum, taken only a few days after the start of the rebrightening, and other spectra taken before and at the second peak, we see a different set of lines, as indicated by the reasonable \texttt{SNID} matches to SNe~Ic (Sect.~\ref{sec:host}).\footnote{Even forcing \texttt{SNID} to use Type~Ic does not result in reasonable fits to the $-$101~d spectrum despite its low S/N.} An absorption feature close to H$\alpha$ and an emission feature close to He~{\sc i}~$\lambda$7065 are still present, but the He~{\sc i}~$\lambda$6678 line (unfortunately disturbed by a telluric line) seems weak or nonexistent and He~{\sc i}~$\lambda$4471 is not visible. Lines of O~{\sc i} and Mg~{\sc ii}, and the Ca~{\sc ii} H\&K lines, which are not (clearly) present in the $-$101~d spectrum, also appear, as does [Ca~{\sc ii}]~$\lambda\lambda$7291,7323 possibly blended with [O~{\sc ii}]~$\lambda\lambda$7320,7330. The blue part of the spectrum is now dominated by a plethora of iron lines \citep{dessart21,dessart23}. Some of the features visible at $-$14~d could be lost in the low S/N of the $-$101~d spectrum, but the disappearance of He~{\sc i}~$\lambda$6678 and the appearance of O~{\sc i} lines do not seem to be attributable to this.

The He~{\sc i}~$\lambda$5876 absorption is now blended with the Na~{\sc i}~$\lambda\lambda$5890,5986 doublet, and the absorption trough likely has a contribution from both. The velocity of the absorption minimum is $\sim-6800$~km~s$^{-1}$ if dominated by Na~{\sc i}, or $\sim-6000$~km~s$^{-1}$ if dominated by He~{\sc i}. This velocity is comparable to the absorption minimum of the Ca~{\sc ii} NIR triplet at $-$14~d, $\sim-6000$~km~s$^{-1}$, and that of the strong O~{\sc i}~$\lambda$7774 line, $\sim-5500$~km~s$^{-1}$. If one assumes the absorption line close to H$\alpha$ is still H$\alpha$, its velocity is $\sim-11\,000$~km~s$^{-1}$ (a slightly higher velocity of $\sim-11\,500$~km~s$^{-1}$ is obtained assuming C~{\sc ii}~$\lambda$6580). This absorption and the accompanying emission peak might be a combination of H$\alpha$ and the Si~{\sc ii}~$\lambda\lambda$6347,6371 feature, but H$\alpha$ emission would then be much more blueshifted than other emission lines, suggesting a different origin. Spectral models by \citet{dessart19} and \citet{dessart23} show blends of Fe~{\sc ii} emission lines at 6148, 6149, 6238, 6247, 6417, 6433, 6456, and 6516~\AA. These could produce the observed features, which peak at 6190 and 6480~\AA.

\begin{figure*}
\centering
\includegraphics[width=0.36\linewidth]{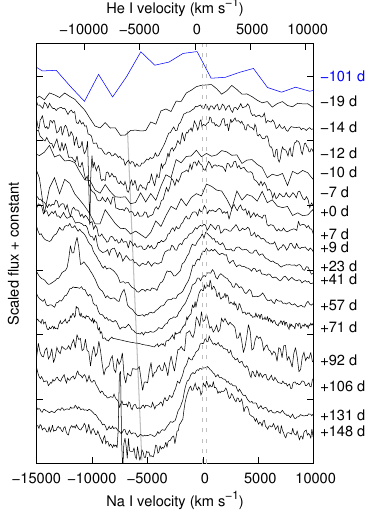}
\includegraphics[width=0.325\linewidth]{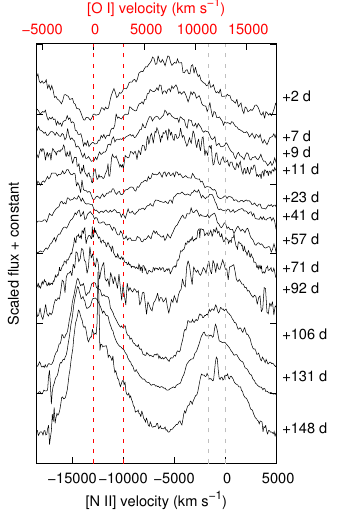}
\includegraphics[width=0.29\linewidth]{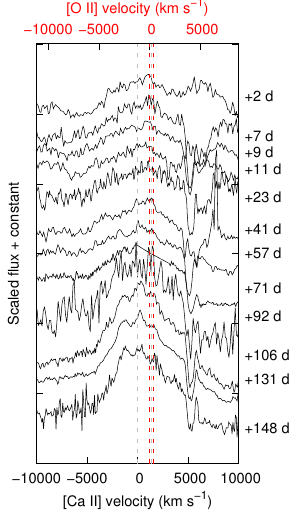}
\caption{Left: evolution of the Na~{\sc i}~$\lambda\lambda$5890,5986 line (dashed gray lines), possibly blended with He~{\sc i}~$\lambda$5876. The approximate evolution of the absorption minimum from a Gaussian fit after $-$19~d is marked with a solid gray line. The velocity evolution is slow during the second peak and the subsequent decline. Middle: evolution of the putative [O~{\sc i}]~$\lambda\lambda$6300,6364 (dashed red lines) and [N~{\sc ii}]~$\lambda\lambda$6548,6584 (dashed gray lines) doublets after the second peak. Both start to appear between +41 and +71~d and strengthen over time. Right: evolution of the [Ca~{\sc ii}]~$\lambda\lambda$7291,7323 (dashed gray lines) + [O~{\sc ii}]~$\lambda\lambda$7320,7330 (dashed red lines) feature. Zero velocity refers to the stronger line of each doublet. All forbidden features shown here are double-peaked at late times.}
\label{fig:naid}
\end{figure*}

\begin{figure*}
\centering
\includegraphics[width=0.8\linewidth]{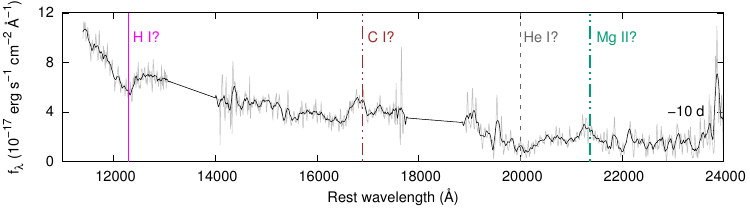}
\caption{TNG/NICS NIR spectrum of SN~2023aew at $-$10~d. Possible features of H, He, Mg and C are seen, but the spectrum is noisy and atypical for SE-SNe. Line velocities are as in Fig. \ref{fig:specseq}, with the addition of $-$12\,500~km~s$^{-1}$ as a solid line. Savitzky-Golay smoothing has been applied.}
\label{fig:specNIR}
\end{figure*}

Spectroscopic evolution is, afterwards, quite slow, and the velocities of the Na/He and O~{\sc i} lines decline only by about 1500~km~s$^{-1}$ from $-$19~d to +148~d. The slow evolution is emphasized by the fact that the spectra of the second peak consistently resemble SNe~Ic at post-peak epochs, even before this peak. The main changes between these epochs include the appearance of low-velocity ($-2500$~km~s$^{-1}$) Sc~{\sc ii} absorption lines; the gradual appearance of strong emission features at Na~{\sc i}~$\lambda\lambda$5890,5986, [O~{\sc i}]~$\lambda\lambda$6300,6364, and at $\sim6550$~\AA; and the fading of Mg~{\sc ii}~$\lambda$7887 while Mg~{\sc ii}~$\lambda$8224 persists. The initially iron-dominated peak at $\sim4600$~\AA~gradually moves slightly bluewards as it starts being dominated by Mg~{\sc i}]~$\lambda$4571. Mg~{\sc i} lines around 5170~\AA~also appear in a P~Cygni feature around the same time, with an absorption minimum at $\sim-3000$~km~s$^{-1}$. 

The $\sim6550$~\AA~feature appears roughly simultaneously with [O~{\sc i}]~$\lambda\lambda$6300,6364 and Mg~{\sc i}]~$\lambda$4571 while the Fe~{\sc ii} features in these regions fade. An obvious possibility for this line would be H$\alpha$; by definition this line is not seen in a normal SN~Ic, but could be a result of interaction. At such a late stage of evolution, C~{\sc ii}~$\lambda$6580 is unlikely. Alternatively, if the first peak is a true SN~IIb (see Sect. \ref{sec:disco}), the [O~{\sc i}], Mg~{\sc i}], and $\sim$6550~\AA~features appear $\lesssim200$~d after that explosion, when forbidden/nebular lines from the first SN could be seen. The timing also coincides with the bump phase where the SN reaches the same brightness as the flattening between the two peaks (Sect. \ref{sec:LC}) and where nebular features from the first SN might start contributing noticeably if the flat underlying light curve has continued. 

Such a strong late-time H$\alpha$ line is not typical for a SN~IIb, and [N~{\sc ii}] is in any case considered the more likely origin by \citet[][]{jerkstrand15} and \citet{fang18}. H$\alpha$ from CSI was seen in late-time spectra of the prototypical SN~1993J, but it was also much weaker than the line seen in SN~2023aew, and the shape of its profile was boxy as expected from a CSM shell \citep[e.g.,][]{patat95}. [N~{\sc ii}] is typically very weak in late-time SN~Ic spectra compared to [O~{\sc i}], but can be comparable to [O~{\sc i}] in SNe~IIb with low progenitor masses. A [N~{\sc ii}]/[O~{\sc i}] ratio of $\sim2/3$ is seen in SN~2023aew at +148~d; a 12-M$_\odot$ progenitor star model can produce such a ratio \citep{jerkstrand15}, and \citet{fang19} report some SNe~IIb with similar ratios, though such SNe also tend to exhibit [O~{\sc i}]/[Ca~{\sc ii}] ratios below 1, whereas in SN~2023aew this ratio is $\sim1.7$ at +148~d. [N~{\sc ii}] is produced in the helium layer of the ejecta, where CNO burning results in a high nitrogen abundance \citep{jerkstrand15}, which is compatible with the helium contribution in the relatively early spectra. Since the line is quite strong for [N~{\sc ii}] given the high [O~{\sc i}]/[Ca~{\sc ii}] ratio, a combination of [N~{\sc ii}]~$\lambda\lambda$6548,6584 and H$\alpha$ is the most likely option; similar cases of H$\alpha$ ``contamination'' are noted by \citet{fang19}. 

Fig. \ref{fig:naid} shows the evolution of the Na/He absorption, the putative [N~{\sc ii}] and [O~{\sc i}] lines, and the [Ca~{\sc ii}]~$\lambda\lambda$7291,7323 + [O~{\sc ii}]~$\lambda\lambda$7320,7330 feature. At late times ($>+130$~d), all of these features evolve to show a slightly double-peaked profile (i.e., two overlapping doublets at different velocities). The widths of the three features are similar ($\sim5000$~km~s$^{-1}$ after +130~d), and they seem to exhibit additional peaks blueshifted by $\sim1500$~km~s$^{-1}$ relative to each line. The locations of these peaks compared to the other lines are consistent with an [N~{\sc ii}] origin for the $\sim$6550~\AA~feature. Furthermore, if Fe~{\sc ii} lines were still responsible for the putative [O~{\sc i}] and [N~{\sc ii}], this would result in a redshift of several thousand km~s$^{-1}$ for the peaks of these blends, which is not seen in other spectral lines. We thus consider it more likely that these emission features are correctly identified as [O~{\sc i}]~$\lambda\lambda$6300,6364 and [N~{\sc ii}]~$\lambda\lambda$6548,6584 (likely with H$\alpha$ contamination). We do note, though, that the latter is difficult to verify as the [N~{\sc ii}]~$\lambda5754$ feature falls on the absorption trough of the Na/He absorption and is not clearly seen. 

\begin{figure}
\centering
\includegraphics[width=0.99\linewidth]{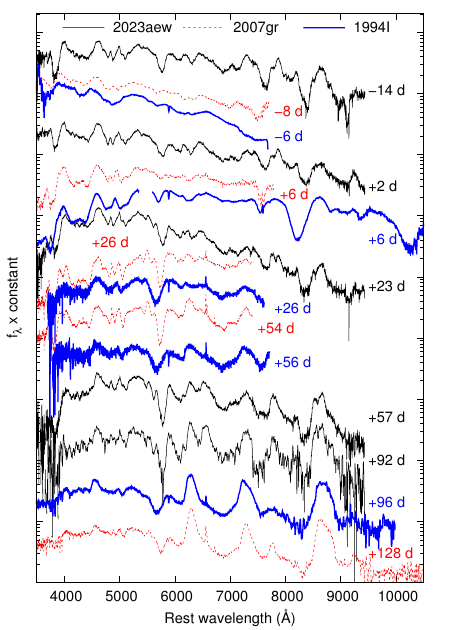}
\caption{Spectra of SN~2023aew compared to well-observed ordinary SNe Ic. While neither SN~1994I {(thick blue line)} nor SN~2007gr {(dashed red line)} matches the early-time features of SN~2023aew, the latter does resemble SN~2007gr at later epochs. The feature at $\sim6500$~\AA~is not seen in late-time spectra of either SN~1994I or SN~2007gr. Overall the evolution of SN~2023aew is relatively slow.}
\label{fig:speccomp}
\end{figure}

\begin{figure}
\centering
\includegraphics[width=0.95\linewidth]{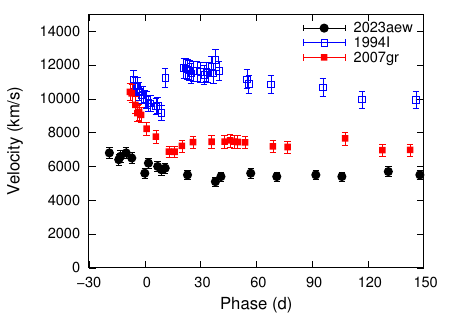}
\caption{Evolution of the Na~{\sc i}~$\lambda\lambda$5890,5896 absorption minimum velocity in SN~2023aew (black) after the rebrightening, compared to SNe~1994I (blue) and 2007gr (red). Uncertainties are estimated at 5\%. For SN~2023aew, the phase refers to the second peak.}
\label{fig:nai_vel}
\end{figure}

We present our NIR spectrum of SN~2023aew at $-$10~d in Fig.~\ref{fig:specNIR}. This spectrum is rather noisy, but we see some features of the same elements as in the optical. If the absorption feature at $\sim$12\,200~\AA~is identified with Pa$\beta$, its velocity is $\sim-12\,500$~km~s$^{-1}$, close to the putative H$\alpha$ velocity and suggesting a hydrogen origin for that line, but other NIR lines of hydrogen are not clearly seen -- Pa$\alpha$ is located in a gap in the spectrum, but Br$\gamma$ may be lost in the noise. Absorption consistent with He~{\sc i}~$\lambda$20\,581 at $\sim-8500$~km~s$^{-1}$ is seen in the spectrum as well as the emission of Mg~{\sc ii}~$\lambda$21\,369. We also see a possible C~{\sc i}~$\lambda$16\,890 emission line, perhaps accompanied by weak P~Cygni absorption. Both putative emission lines are blueshifted by $\sim1500$~km~s$^{-1}$. These lines are not seen in the NIR spectra of typical SNe~Ic shown by \citet{shahbandeh22}, which instead show Mg~{\sc i}~$\lambda$14\,878 and C~{\sc i}~$\lambda$14\,543. SNe~IIb do show Pa$\beta$ and He~{\sc i}~$\lambda$20\,581, but the latter mostly in absorption. As the spectrum is quite noisy, these identifications are uncertain.

For comparison between the second peak of SN 2023aew and normal SNe~Ic, we obtained publicly available spectra of SNe~1994I \citep{filippenko95,modjaz14} and 2007gr \citep{valenti08,modjaz14,shivvers19} from WISeREP. We show this comparison at different epochs in Fig. \ref{fig:speccomp}. It is clear from this figure that SN~2023aew spectroscopically evolves slower than either comparison SN. Even the first spectrum of SN~2023aew, 19~d before the second peak, resembles that of SN~2007gr at much later times, with much stronger features. After the second peak, and especially after a few weeks post-peak, SN~2023aew begins to spectroscopically resemble SN~2007gr, with similar strengths and widths in the absorption lines; however, the broad emission peak at $\sim$6500~\AA~is not seen in late-time ($>$60~d) spectra of either SN~1994I or SN~2007gr. Instead, only a narrow H$\alpha$ line from the host galaxy is seen in the latter two. The resemblance to SN~2007gr is generally stronger than SN~1994I in terms of velocity and line strength. SN~2007gr still tends toward higher velocities than SN~2023aew, while SN~1994I exhibits even higher velocities with fewer and, until late times, generally weaker lines. The velocity comparison is illustrated in Fig. \ref{fig:nai_vel}, where we show the evolution of the Na~{\sc i}~$\lambda\lambda$5890,5896 absorption minimum.

This is somewhat similar to what was seen in the unusual SLSN~I, SN~2020wnt \citep{gutierrez22}, which also showed a similarity to SN~2007gr, but with slower evolution and a $\sim$6500~\AA~feature. This object also shared some light-curve features in common with the second peak of SN~2023aew (Sect. \ref{sec:LC}). The $\sim$6500~\AA~feature was attributed to [Fe~{\sc ii}]~$\lambda$6456 and $\lambda$6518 in SN~2020wnt, but this is unlikely in SN~2023aew, as the line profile would have to be redshifted by $\sim$4000 km~s$^{-1}$.

\begin{figure}
\centering
\includegraphics[width=\linewidth]{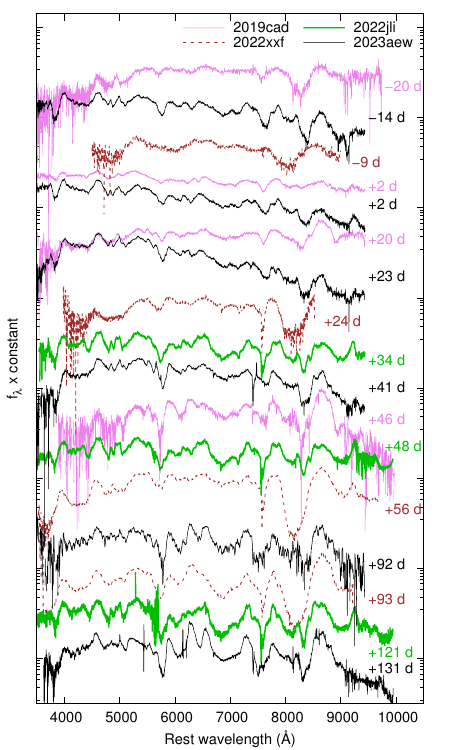}
\caption{Spectra of SN~2023aew (black) compared to double-peaked SE-SNe SN~2019cad (purple), SN~2022jli ({thick green line}) and SN~2022xxf ({dashed brown line}). All epochs refer to the second peak. SN~2023aew evolves slower than SNe~2019cad or 2022xxf and exhibits lower velocities, but an excellent match is seen with the latest spectrum of SN~2019cad. The 6500~\AA~feature, in particular, is replicated. The match with SN~2022xxf similarly improves at late times, although this feature does not appear. SN~2022jli provides a good match at $\sim$30--40 days and evolves slower than other comparison objects, but lacks a strong [O~{\sc i}] doublet at late times.}
\label{fig:speccomp2}
\end{figure}

We have also compared SN~2023aew to other double-peaked SE-SNe: SN~2019cad \citep{gutierrez21}, SN~2022jli \citep{chen23,moore23}, and SN~2022xxf \citep{kuncarayakti23}. We show this comparison in Fig. \ref{fig:speccomp2}, where all epochs refer to the second peak, which is the brighter one in all objects. These comparison objects, as stated in Sect. \ref{sec:LC}, are fainter than SN~2023aew, except SN~2022xxf, and exhibit a shorter delay between the two peaks. In general, once again, we see a faster evolution in the spectra as well. However, some similarities are also seen: the last spectrum of SN 2019cad, especially, where the 6500~\AA~feature also appears (albeit weaker and unidentified), matches the spectrum of SN~2023aew well. 

\section{Modeling}
\label{sec:models}

\subsection{Blackbody fits}
\label{sec:BB}

\begin{figure}
\centering
\includegraphics[width=0.99\linewidth]{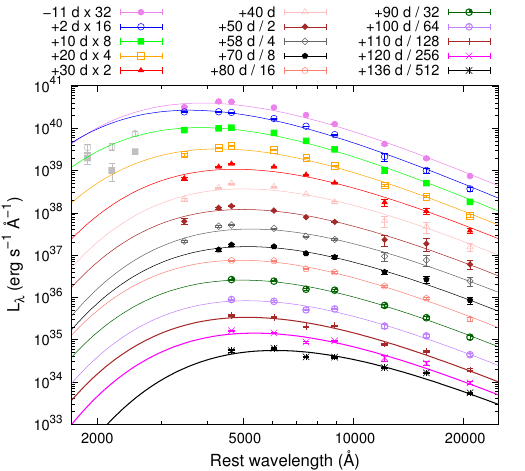}
\caption{Blackbody fits (lines) to the observed SEDs (points) at roughly 10-day intervals. NUV points (light gray) at +2 and +10~d clearly deviate from a blackbody because of line blanketing, and were ignored in the fits.}
\label{fig:bbfits}
\end{figure}

We used the GP-interpolated light curves (see Sect. \ref{sec:LC}) to construct spectral energy distributions (SEDs) at roughly 10-day intervals between the first $JHKs$-band epoch ($-$11.4~d) and the last $JHKs$ epoch at +135.6~d. We note, however, that we do not have NIR data at the optical peak, and close to that epoch, the NIR fluxes are most likely underestimated by the GP fit, leading to overestimated temperatures. Conversely, in the epochs after +70~d we lack the $uB$ bands, and temperatures can be affected from the other direction. Additionally, we used the preceding epochs where only ZTF $gr$-band data are available and where the parameters are thus highly uncertain. We show the blackbody function at each epoch compared to the interpolated SED in Fig. \ref{fig:bbfits}. Line blanketing in the NUV results in clear deviation from a blackbody, and the NUV bands were ignored in the fits.

\begin{figure}
\centering
\includegraphics[width=0.99\linewidth]{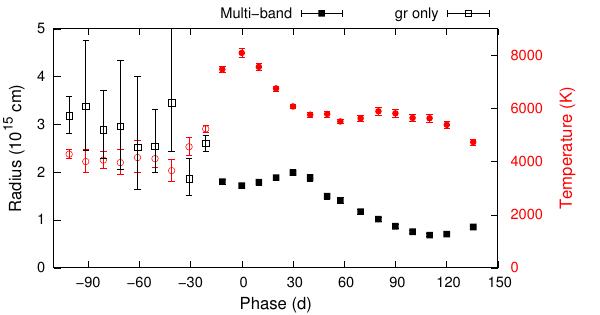}
\caption{Evolution of the blackbody parameters over time. The temperature from the fit rises during the rebrightening from a preceding plateau and is highest at the peak -- albeit likely slightly overestimated -- then declines to $\sim$5500~K and stays constant until $\gtrsim$120~d. The radius slowly declines almost throughout the observed evolution.}
\label{fig:bbparams}
\end{figure}

\subsubsection{Evolution of the blackbody}
\label{sec:BBparams}

The evolution of the blackbody parameters, the temperature and radius, is shown in Fig. \ref{fig:bbparams}. Initially, in the epochs where we can only use $gr$ photometry, the color stays constant implying a temperature around 4000~K, while the radius decreases from $3\times10^{15}$~cm to about $2\times10^{15}$~cm close to the rebrightening. In the well-measured epochs, the radius stays fairly constant around $2\times10^{15}$~cm, until a decline around +40~d, which coincides with the steepening of the $ri$-band light curve (Fig. \ref{fig:absLC}). The radius then appears to slowly decline from $2\times10^{15}$~cm at +40~d to $\sim6\times10^{14}$~cm at +120~d. Meanwhile, the temperature seems to reach a peak of $\sim$8100~K at the optical light curve peak (but note that this is where the NIR interpolation is likely increasing the temperature) before declining slowly to $\sim$5500~K at +40~d, after which it stays roughly constant. The last epoch at +136~d seems to exhibit a lower temperature, around 4700~K, and an expanding radius, but at late times the SED deviates somewhat from a blackbody, and the resulting parameters may be unreliable.

The mostly continuous decrease of the radius seems to indicate that the photosphere moves inward through our observations. The rebrightening is mostly caused by an increase in temperature, with possibly a pause in the shrinking of the photosphere as well; hence, the blue bands did not peak before the red ones (Sect. \ref{sec:LC}). Even if one ignores the fits using only $gr$ data, a radius of $\sim$2$\times10^{15}$~cm is already reached at $-$11~d. The rebrightening starts between $-$28~d and $-$23~d; if this corresponds to the explosion date of a SN, the ejecta velocity required to reach the blackbody radius ranges from $\sim$14\,000 to $\sim$19\,000~km~s$^{-1}$. Such velocities are not supported by our spectra, where the absorption minimum of the Na~{\sc i}~$\lambda\lambda$5890,5896 doublet is at 6800~km~s$^{-1}$ at $-$19~d (or, if contaminated by He~{\sc i}~$\lambda$5876, even lower). If the feature at $\sim6500$~\AA~in the $-$101~d spectrum is H$\alpha$ as the IIb classification suggests, the velocity of the absorption minimum is $-$12\,500~km~s$^{-1}$. As photospheric velocities tend to be smaller than the H$\alpha$ velocity, reaching a radius of $\sim3\times10^{15}$~cm would take $\gtrsim28$~d, which is compatible with the epoch of the spectrum $\sim$34~d after the explosion leading to the first peak.

\subsubsection{Bolometric light curve}
\label{sec:bolom}

\begin{figure}
\centering
\includegraphics[width=0.99\linewidth]{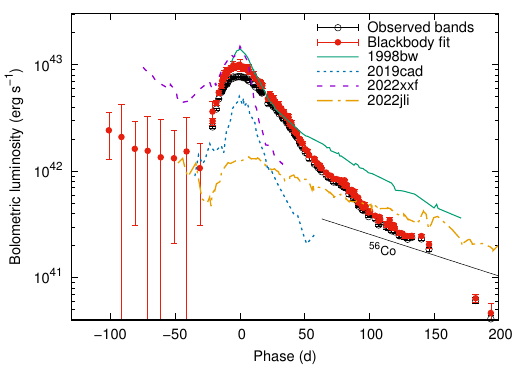}
\caption{\texttt{SuperBol} \citep{nicholl18} pseudo-bolometric light curve of the second peak integrated over observed bands (black open circles), and a bolometric light curve based on blackbody fits (red filled circles), highly uncertain before the rebrightening. The difference is notable only around maximum light. The late-time decline is inconsistent with $^{56}$Co decay (solid black line). Pseudo-bolometric light curves of SNe~1998bw \citep{patat2001}, 2019cad \citep{gutierrez21}, 2022jli \citep{moore23}, and 2022xxf \citep{kuncarayakti23} are included for comparison.}
\label{fig:bolLC}
\end{figure}

We used \texttt{SuperBol} \citep{nicholl18} to construct the bolometric light curve of the second peak of SN~2023aew, where multi-band photometry is available. The $r$ band was used as the baseline, and the optical and NIR light curves in other filters were interpolated to the $r$-band epochs using a polynomial fit within \texttt{SuperBol}. Extrapolation of the very early and/or late light curve was performed by assuming a constant color, as polynomial fits often diverged wildly from the optical light curves. NUV data were not included due to the short time baseline that would necessitate extrapolation over most of the light curve, but \texttt{SuperBol} blackbody fits include simulated line blanketing below 3000~\AA\, to account for the observed deviation of the NUV fluxes from a blackbody. We constructed both a pseudo-bolometric light curve using only the observed bands and a bolometric light curve corrected for missed emission using \texttt{SuperBol}'s blackbody fit. Additionally, we used our $gr$-band blackbody fits to estimate the bolometric luminosity before the rebrightening. These are displayed in Fig. \ref{fig:bolLC}, along with comparison events. Out of these, only SN~1998bw is more energetic than the second peak because of its slower decline.

The blackbody correction is small except near maximum light, when the temperature also peaked (see Sect. \ref{sec:BBparams}). The evolution of the bolometric light curve is very similar to that of the $r$-band light curve, with bumps around 80 and 140 days -- this is expected, as the bumps are replicated in the NIR as well. A tail phase with a decline consistent with the decay of $^{56}$Co at full gamma-ray trapping is not reached during our observations. We also integrated over both light curves to determine the total radiated energy during the second peak. In the observed filters, we obtain a total of $(3.67\pm0.01)\times10^{49}$~erg, while the blackbody-corrected total radiated energy is $(4.34\pm0.04)\times10^{49}$~erg. 

\subsection{Light-curve fitting}
\label{sec:mosfit}

We used the publicly available code Modular Open Source Fitter for Transients \citep[\texttt{MOSFiT}\footnote{\url{https://mosfit.readthedocs.io/en/latest/index.html}};][]{guillochon18} to fit the second peak of the light curve, assuming it was a H-poor SN. The code includes the $^{56}$Ni decay model \citep{arnett82,nadyozhin94}, labeled \texttt{default}; a CSI model \citep{chatz13,villar17,jiang20}, labeled \texttt{csm}; a fallback accretion model \citep{moriya18b}, labeled \texttt{fallback}; and a magnetar central engine model with line blanketing correction \citep{nicholl17}, labeled \texttt{slsn}. These models are fitted using the dynamic nested sampling package \texttt{dynesty}.\footnote{\url{https://dynesty.readthedocs.io/en/latest/}} We ran each fitting process until convergence, which typically took between 25\,000 and 30\,000 iterations.

We set simple uninformative uniform or log-uniform priors for each free parameter, summarized in Table \ref{tab:priors} for each model. We set the lower limit on the characteristic ejecta velocity to be close to the Na~{\sc i} absorption minimum in our earliest spectrum, and the upper limit at twice the lower limit. An upper limit for the column density of neutral hydrogen in the host galaxy, $n_{H,\mathrm{host}}$, was set at $2\times10^{21}$~cm$^{-2}$ \citep[corresponding to $A_V \lesssim 1$~mag according to][]{guver09}, as the host extinction is likely to be low. Other parameters common to all models include the explosion time before observations $t_\mathrm{expl}$, the minimum temperature $T_\mathrm{min}$, and the ejecta mass $M_\mathrm{ej}$. The nickel model additionally includes the nickel fraction in the ejecta $f_{\mathrm{Ni}}$, the opacity $\kappa$, and the opacity to $\gamma$ rays $\kappa_\gamma$. The magnetar model, on the other hand, includes $\kappa$, the spin period $P_\mathrm{spin}$, the magnetic field perpendicular to the spin axis $B_\perp$, the neutron star mass $M_\mathrm{NS}$, and the angle between the magnetic field and spin axis $\theta_\mathrm{PB}$. In the CSI model we include the parameter $s$ (where the CSM density as a function of distance behaves as $\rho \propto r^{-s}$), which we fix to $s=0$, indicating a CSM shell, and $s=2$, indicating a wind. In addition, this model includes the minimum inner radius of the CSM, $R_0$, the CSM mass $M_\mathrm{CSM}$, and the CSM density at $R_0$, $\rho$. We fix the density power-law parameters in the inner ($\rho_{ej} \propto r^{-\delta}$) and outer ($\rho_{ej} \propto r^{-n}$) ejecta, $\delta=1$ and $n=7$, respectively \citep[corresponding to normal values in H-poor ejecta:][]{chatz13,chevalier82}. Finally, the fallback model includes $\kappa$, the linear-to-power-law accretion transition time $t_{tr}$, and the accretion luminosity at the transition time $L_1$. In total, the \texttt{default}, \texttt{fallback}, and \texttt{csm} models have 9 free parameters and the \texttt{slsn} model has 12. These numbers include a nuisance parameter $\sigma$, which describes the added variance required to match the model being fitted. 

We also used \texttt{MOSFiT} to fit the light curve before rebrightening, ignoring the flattening after MJD\,=\,60000, in order to obtain an estimate of the parameters of the first peak, again with the assumption that it was a genuine SN. In this case, we used only the \texttt{default} model, with the same priors as for the second peak, except the ejecta velocity prior was set as [8000:15\,000]~km~s$^{-1}$, as informed by the $-$101~d spectrum. TESS data points were approximated as $i$-band photometry in this fit.

\begin{table}
\centering
\caption{Parameters and priors used in our \texttt{MOSFiT} fits.}
\label{tab:priors}
\begin{tabular}{ccc}
\hline
Parameter & Range & Distribution \\
\hline
\multicolumn{3}{c}{Common parameters} \\
\hline
$n_{H,\mathrm{host}}$ & [$10^{16}$ : $2\times10^{21}$] cm$^{-2}$ & log-uniform \\
$t_{\mathrm{expl}}$ & [$-$500 : 0] d & uniform \\
$T_\mathrm{min}$ & [1000 : 50\,000] K & log-uniform \\
$M_\mathrm{ej}$ & [0.1 : 100]~$\mathrm{M}_\odot$ & log-uniform \\
$v_\mathrm{ej}$ & [6000 : 12\,000]~km~s$^{-1}$ & uniform \\
\hline
\multicolumn{3}{c}{$^{56}$Ni model} \\
\hline
$f_{\mathrm{Ni}}$ & [$10^{-3}$ : 1.0] & log-uniform \\
$\kappa$ & [0.05 : 0.2]~cm$^2$~g$^{-1}$ & uniform \\
$\kappa_\gamma$ & [0.1 : $10^4$]~cm$^2$~g$^{-1}$ & log-uniform \\
\hline
\multicolumn{3}{c}{Magnetar model} \\
\hline
$\kappa$ & [0.05 : 0.2]~cm$^2$~g$^{-1}$ & uniform \\
$\kappa_\gamma$ & [0.1 : $10^4$]~cm$^2$~g$^{-1}$ & log-uniform \\
$P_{\mathrm{spin}}$ & [1 : 50] ms & uniform \\
$B_\perp$ & [0.1 : 50]~$\times10^{14}$ G & log-uniform \\
$M_\mathrm{NS}$ & [1.0 : 2.5]~$\mathrm{M}_\odot$ & uniform \\
$\theta_\mathrm{PB}$ & [0 : $\pi/2$] rad & uniform \\
\hline
\multicolumn{3}{c}{CSI model} \\
\hline
$n$ & 7 & fixed \\
$\delta$ & 1 & fixed \\
$s$ & 0 or 2 & fixed \\
$R_0$ & [0.1 : 1000] AU & log-uniform \\
$M_\mathrm{CSM}$ & [0.1 : 100]~$\mathrm{M}_\odot$ & log-uniform \\
$\rho$ & [$10^{-15}$ : $10^{-6}$]~cm$^{-3}$ & log-uniform \\
\hline
\multicolumn{3}{c}{Fallback model} \\
\hline
$\kappa$ & [0.05 : 0.2]~cm$^2$~g$^{-1}$ & uniform \\
$t_{tr}$ & [$10^{-4}$:100] d & log-uniform \\
$L_1$ & [$10^{50}$:$10^{57}$] erg s$^{-1}$ & log-uniform \\
\hline
\end{tabular}
\end{table}

\begin{figure*}
\centering
\includegraphics[width=0.47\linewidth]{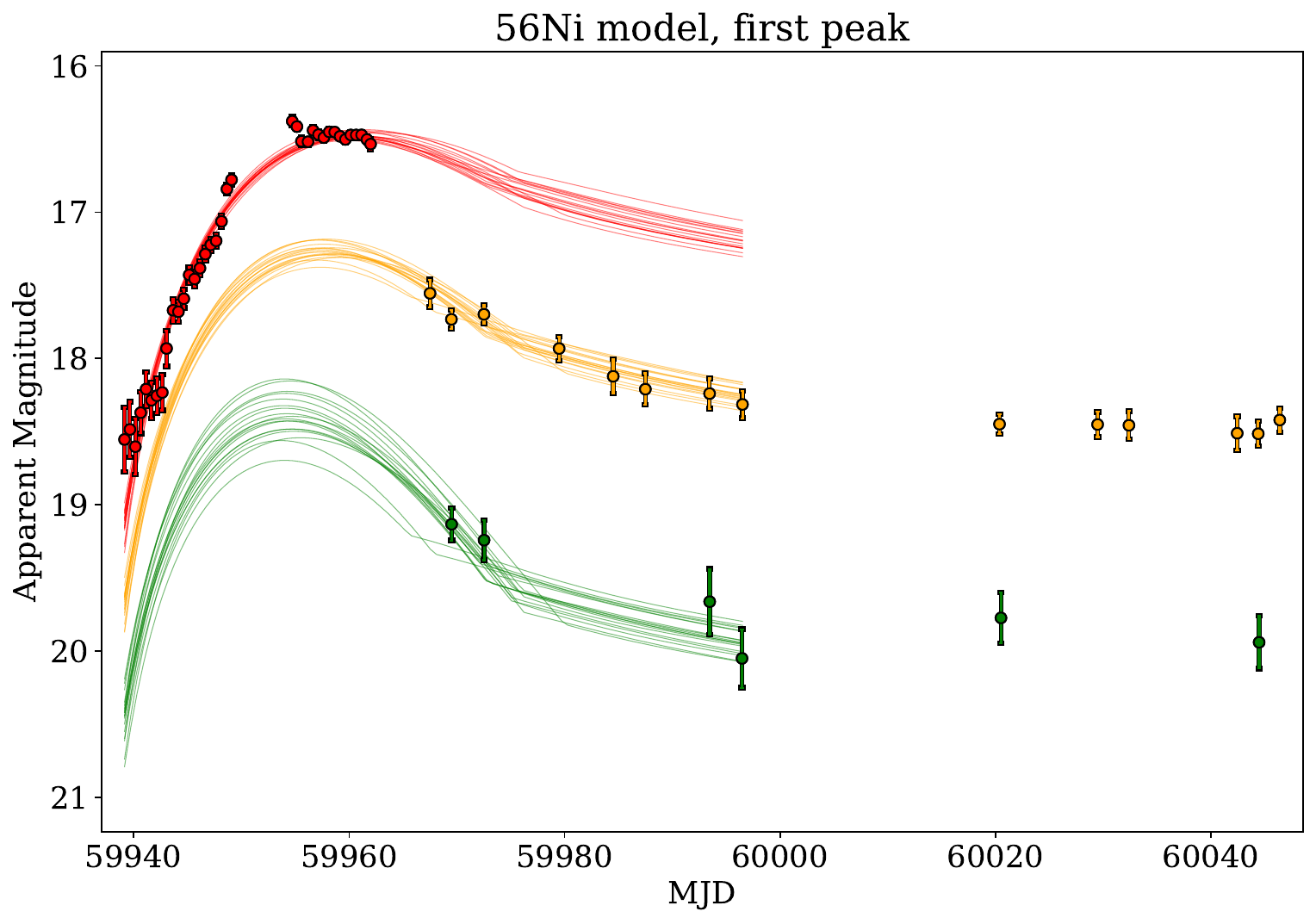}
\includegraphics[width=0.47\linewidth]{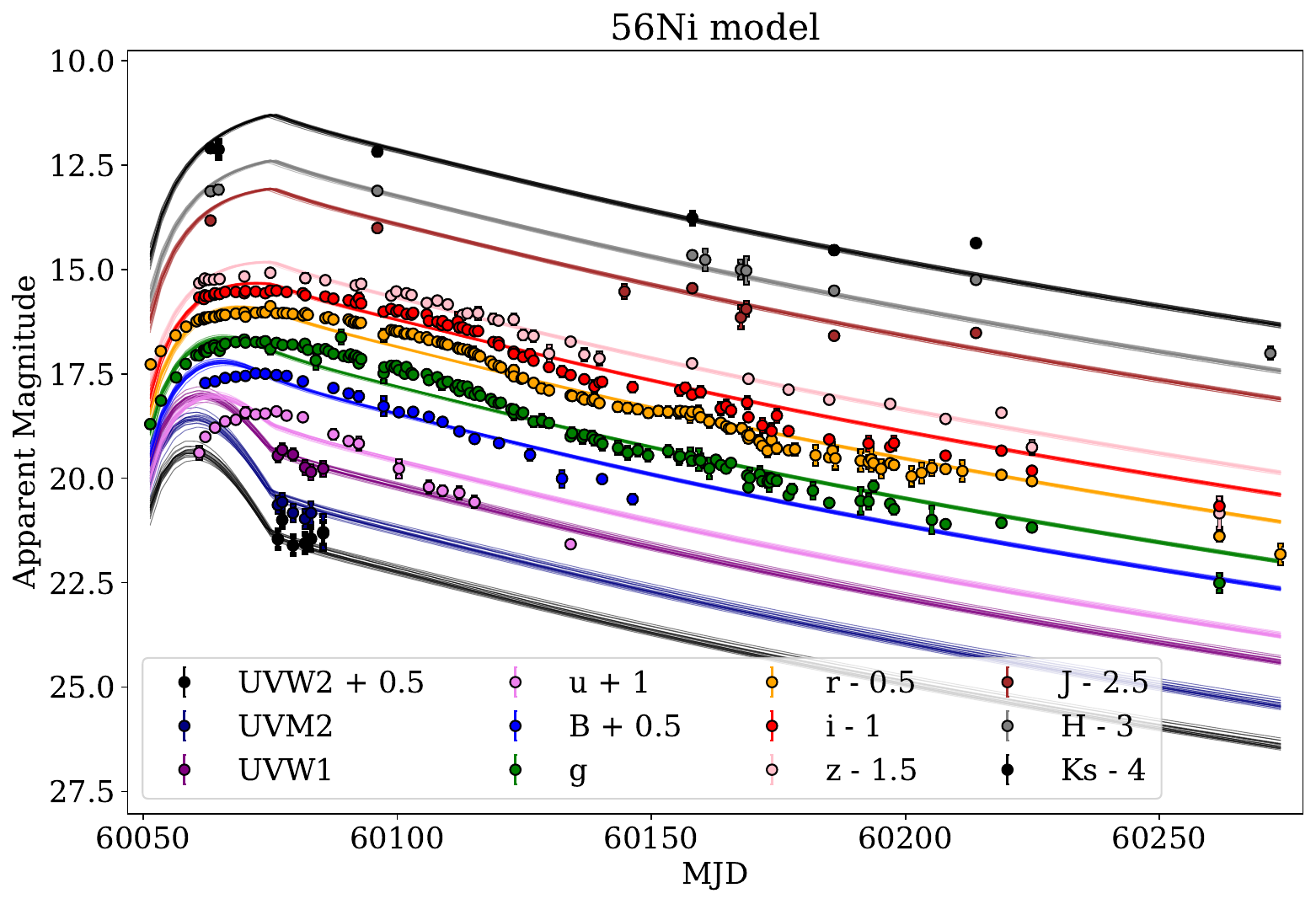}
\includegraphics[width=0.47\linewidth]{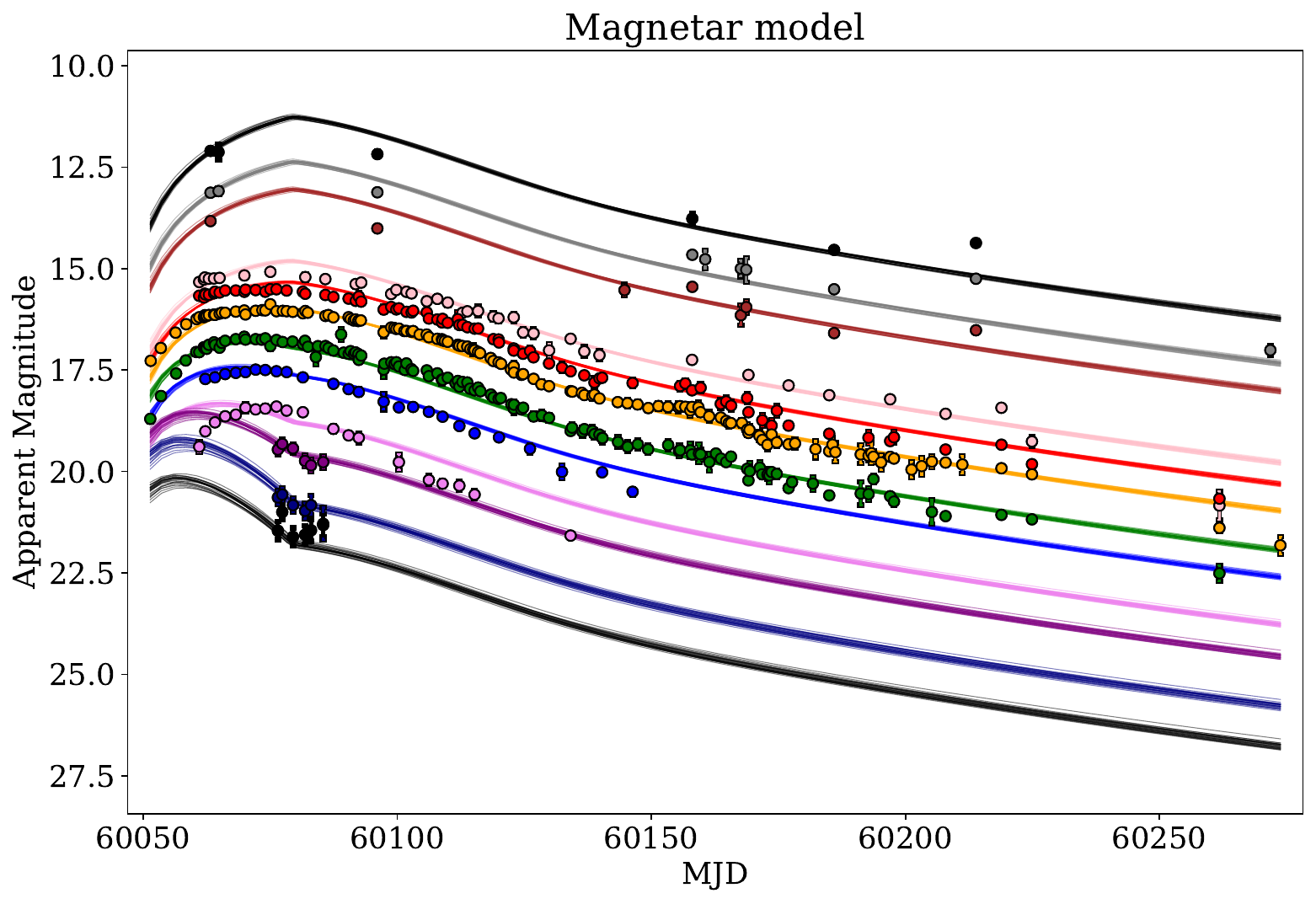}
\includegraphics[width=0.47\linewidth]{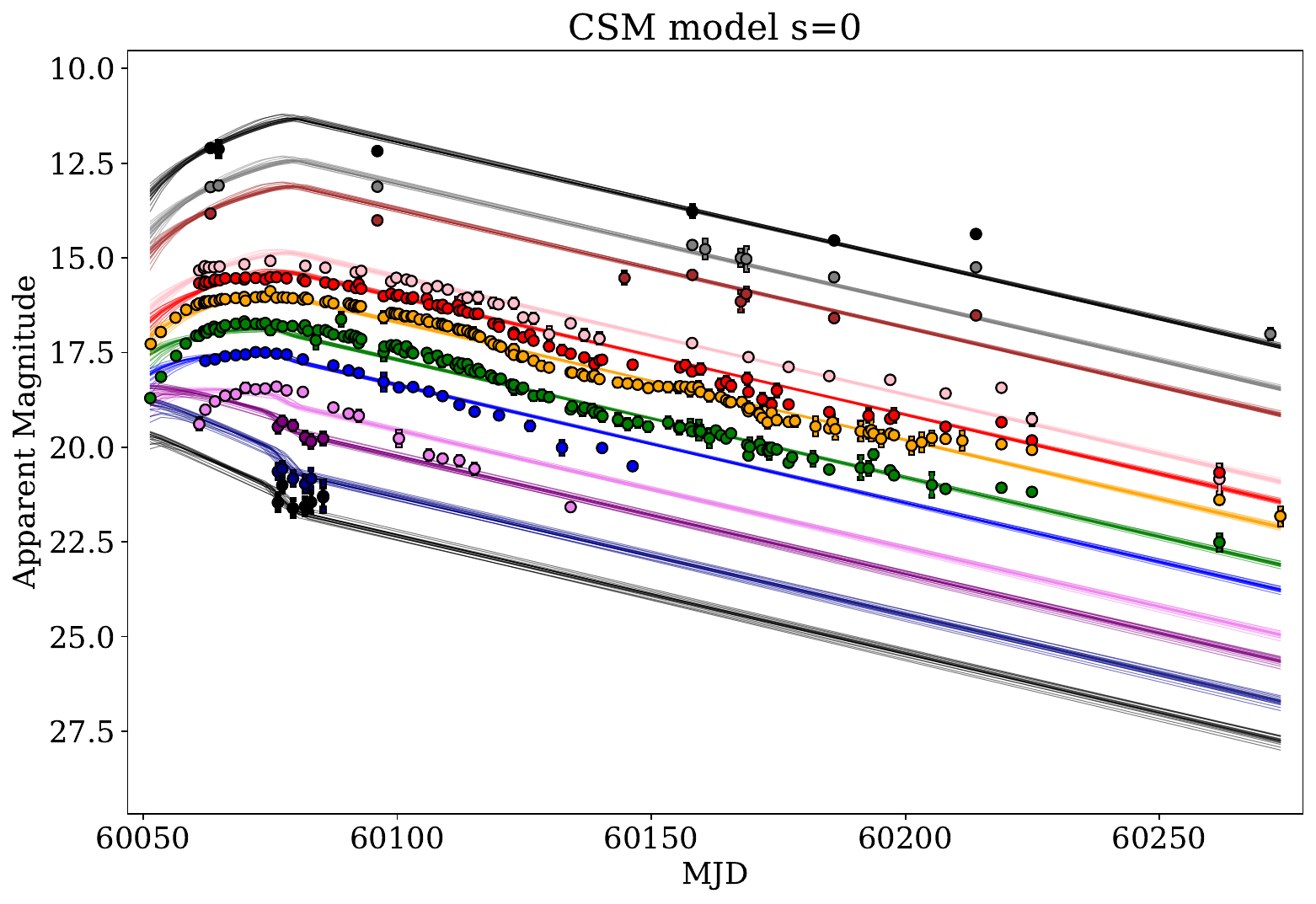}
\includegraphics[width=0.47\linewidth]{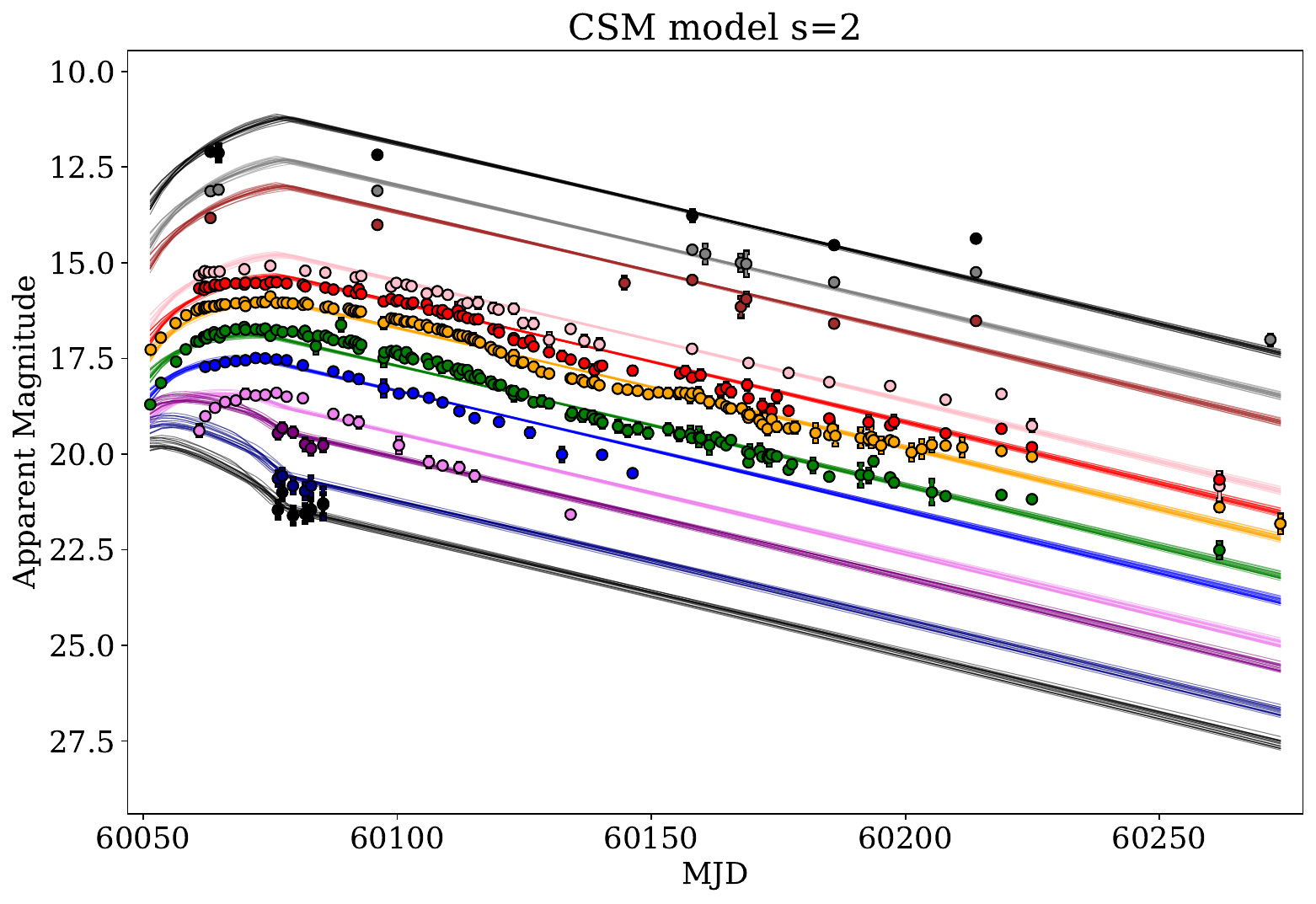}
\includegraphics[width=0.47\linewidth]{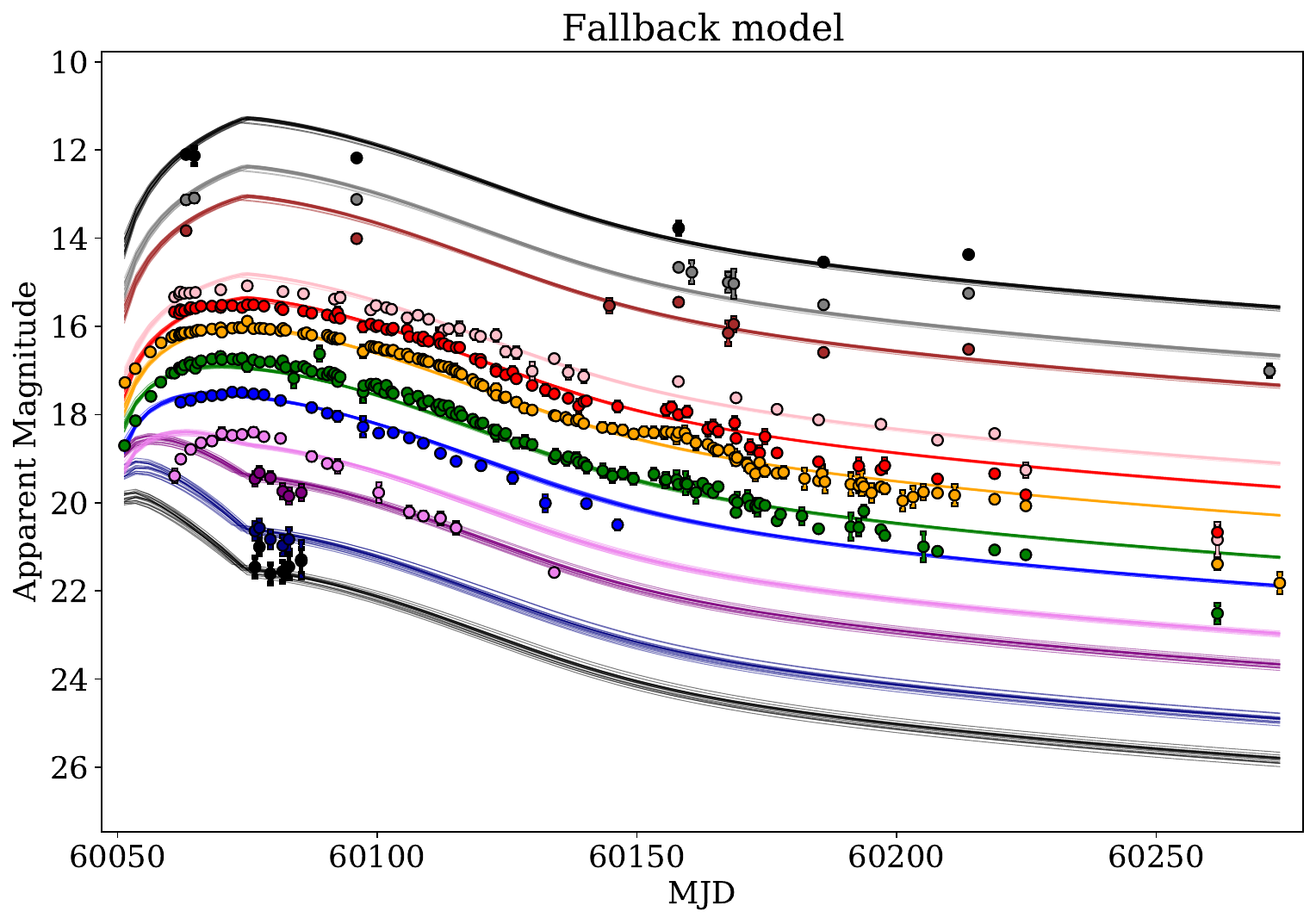}
\caption{\texttt{MOSFiT} light curve fits to the first and second peak of SN~2023aew using the different models described in the text. The flat part of the light curve after MJD=60000 was ignored in the first-peak fit, as the $^{56}$Ni model does not include such features. All models have trouble reproducing the simultaneity of the second peak in all optical bands, specifically failing to reproduce the $u$-band peak, but by eye and by \texttt{MOSFiT} score, the magnetar model provides the best fit.}
\label{fig:mosfit}
\end{figure*}

The results (median parameters and 1$\sigma$ uncertainties) of our fits are listed in Table \ref{tab:mosfitres}, and the model light curves are shown in Fig. \ref{fig:mosfit}. The corner plots of the fits are shown in Figs. \ref{fig:cornerni56_peak1} to \ref{fig:cornerFB}. All five models have trouble reproducing the second peak. Specifically, the main problem is that all optical bands show a simultaneous light-curve peak (Sect. \ref{sec:LC}), while in the \texttt{MOSFiT} models the blue bands, especially $u$, peak first. The $u$-band light curve shape is also difficult to reproduce. In the NUV, all our best fits include a much brighter peak prior to the observed epochs, which probably also influences the optical peaks. The ejecta nickel fraction $f_\mathrm{Ni}$ in the $^{56}$Ni decay model is practically 1, which further disfavors this model for the second peak. In addition, the fallback model results in a $t^{-5/3}$ decline, which overpredicts late-time fluxes, disfavoring this model as well. The magnetar model provides the best fit by eye and has the highest likelihood score returned by \texttt{MOSFiT}, corresponding to the logarithm of the Bayesian evidence $Z$, favoring it over the other models, but we note that the $\kappa$ and $\kappa_\gamma$ parameters of this model cluster at the edges of their priors. 

The $^{56}$Ni model fit to the first peak is fairly good; unlike the second peak, the features within the fitted time range are reproduced by the model. The median ejecta and Ni masses are $\sim$3.3 and $\sim$0.2~M$_\odot$, respectively -- fairly close to those found for SN~2003bg \citep[$\sim$4 and $\sim$0.2~M$_\odot$, respectively, according to][]{mazzali09}, which also exhibits a similar spectrum and peak luminosity. The caveat here is, of course, that the following flattening is not typical for a SN~IIb.

\begin{table*}
\centering
\caption{Scores and median posterior values of free parameters in our \texttt{MOSFiT} fits. } 
\label{tab:mosfitres}
\begin{tabular}{lcccccc}
\hline
Parameter & $^{56}$Ni {decay}, first peak & $^{56}$Ni decay & Magnetar & CSI ($s=0$) & CSI ($s=2$) & Fallback \\
\hline
log $n_{H,\mathrm{host}}$ (cm$^{-2}$) & $19.1^{+1.4}_{-1.8}$ & $20.78^{+0.06}_{-0.07}$ & $20.5^{+0.2}_{-0.4}$ & $19.5^{+0.9}_{-1.7}$ & $21.17\pm0.04$ & $21.01^{+0.07}_{-0.10}$ \\
$t_{\mathrm{expl}}$ (d) & $-3.9\pm0.3$ & $-3.7\pm0.3$ & $-6.0\pm0.4$ & $-10.5^{+1.5}_{-1.2}$ & $-6.9^{+1.0}_{-1.3}$ & $-3.2\pm0.4$ \\
log $T_\mathrm{min}$ (K) & $3.60\pm0.02$ & $3.85\pm0.01$ & $3.82\pm0.01$ & $3.80\pm0.01$ & $3.91\pm0.01$ & $3.89\pm0.02$ \\
log $M_\mathrm{ej}$ ($\mathrm{M}_\odot$) & $0.53^{+0.21}_{-0.17}$ & $-0.06\pm0.02$ & $0.77\pm0.02$ & $-0.96^{+0.06}_{-0.03}$ & $-0.88^{+0.11}_{-0.08}$ & $1.35^{+0.11}_{-0.17}$ \\
$v_\mathrm{ej}$ (km~s$^{-1}$) & $11900^{+1200}_{-1000}$ & $11910^{+70}_{-130}$ & $10200\pm400$ & $9000\pm400$ & $9700^{+500}_{-800}$ & $11700^{+200}_{-400}$\\
log $f_{\mathrm{Ni}}$ & $-1.2\pm0.2$ & $-0.01^{+0.01}_{-0.02}$ & - & - & - & -\\
$\kappa$ (cm$^2$~g$^{-1}$) & $0.14^{+0.04}_{-0.05}$ & $0.197^{+0.002}_{-0.004}$ & $0.197^{+0.002}_{-0.003}$ & - & - & $0.08^{+0.04}_{-0.02}$ \\
log $\kappa_\gamma$ (cm$^2$~g$^{-1}$) & $1.4\pm1.6$ & $-0.996^{+0.005}_{-0.003}$ & $-0.986^{+0.016}_{-0.010}$ & - & - & -\\
$P_{\mathrm{spin}}$ (ms) & - & - & $14.6\pm1.7$ & - & - & -\\
$B_\perp$ ($10^{14}$ G) & - & - & $20^{+18}_{-7}$ & - & - & -\\
$M_\mathrm{NS}$ ($\mathrm{M}_\odot$) & - & - & $2.2^{+0.2}_{-0.3}$ & - & - & -\\
$\theta_\mathrm{PB}$ (rad) & - & - & $0.46^{+0.20}_{-0.22}$ & - & - & -\\
log $R_0$ (AU) & - & - & - & $0.87^{+0.55}_{-0.85}$ & $1.35^{+0.42}_{-0.28}$ & - \\
log $M_\mathrm{CSM}$ ($\mathrm{M}_\odot$) & - & - & - & $0.32\pm0.02$ & $0.33^{+0.12}_{-0.05}$ & -\\
log $\rho$ (cm$^{-3}$) & - & - & - & $-12.18\pm0.04$ & $-11.50^{+0.50}_{-0.66}$ & -\\
log $t_{tr}$ & - & - & - & - & - & $-3.3^{+0.6}_{-0.5}$ \\
log $L_1$ & - & - & - & - & - & $53.65\pm0.05$ \\
\hline
Score (log $Z$) & 57.1 & 305.9 & 438.7 & 370.0 & 355.1 & 396.5\\
\hline 
\end{tabular}
\tablefoot{Model fits correspond to the second peak unless specified. Note that the score is affected by the number of data points, and the fit to the first peak should not be compared to other scores.}
\end{table*}

%--------------------------------------------------------------------
\section{Discussion}
\label{sec:disco}

We now discuss the features of SN~2023aew in the context of other double-peaked SE-SNe, and several possible scenarios that might conceivably explain these features. These scenarios include two separate SNe -- whether related to each other or not -- that are each responsible for one of the peaks and happen on the same line of sight; one SN preceded by an eruption; one SN followed by delayed CSI; one SN followed by delayed activity of a magnetar central engine; and one SN followed by accretion onto a newborn compact object. 

\subsection{SN 2023aew vs. other double-peaked SE-SNe}

Among the double-peaked SNe we have compared SN~2023aew against, it stands unique \citep[though we note that at least one similar object, SN~2023plg, exists according to][but no spectra of its first peak exist]{sharma24}. Other double-peaked SE-SNe, such as SNe~2019cad \citep{gutierrez21}, 2020bvc \citep{ho20,izzo20}, 2022jli \citep{moore23,chen23}, and 2022xxf \citep{kuncarayakti23}, and the ZTF sample of double-peaked SNe~Ibc from \citet{das23}, all show a shorter separation between the light-curve peaks and/or a much less pronounced second peak. Apart from SN~2022jli, they also show a faster spectroscopic and photometric evolution than SN~2023aew, as do normal single-peaked SNe~Ic. Out of these, the SN whose light curve most resembles SN~2023aew is SN~2020bvc -- the peak is less broad, but the peak luminosity and late-time bumpy light curve are similar -- but it shows much higher velocities and is classified as a SN~Ic-BL. 

Spectroscopically the best match to the late-time spectra of SN~2023aew is found in the Type Ic SN~2019cad: all features including the possible [N~{\sc ii}] feature\footnote{We note that \citet{gutierrez21} do not identify this feature.} at $\sim$6550~\AA~are replicated. However, this is only true in one epoch \citep[46~d after the second peak or 88~d after explosion;][]{gutierrez21}, before which SN~2019cad evolves faster than SN~2023aew. Later spectra of SN~2019cad are, unfortunately, not available. The $\sim$6550~\AA\, feature is not seen in the normal SNe~Ic we have examined \citep{filippenko95,hunter09,modjaz14}, nor in SN~2003bg at late times \citep{hamuy09}, nor in the double-peaked SN~2022xxf. SN~2022jli does show a feature at $\sim$6550~\AA; in this case, the line is attributed to H$\alpha$, and its profile is quite different from SN~2023aew, with a periodic change between a redshift and a blueshift compatible with the periodicity in the light curve of SN~2022jli \citep{chen23}. A strong [N~{\sc ii}] is more common in SNe~IIb from low-mass progenitor stars \citep{jerkstrand15,fang19}, and its absence in SNe~Ic is a sign that the progenitor star has lost its helium layer. Therefore, the presence of some helium in our spectra is qualitatively compatible with the [N~{\sc ii}] identification. 

Diverse mechanisms have been suggested for these SNe. \citet{das23} consider their sample to be the result of late-time eruptive mass loss: the first, short-lived peak in these objects was argued to be the shock breakout from a dense confined CSM or an extended envelope, followed by the second peak powered by $^{56}$Ni decay. Similarly, a shock breakout is likely responsible for the first peak of SN~2020bvc \citep{ho20,izzo20}. In SN~2022xxf, on the other hand, CSI involving a H- and He-free CSM, seen through late-time narrow peaks on top of broad lines, powers the second peak according to \citet{kuncarayakti23}. In SN~2019cad, \citet{gutierrez21} argue  that the double-peaked structure is the result of a two-component distribution of $^{56}$Ni (where the outer component could be created by jets), with possibly an additional contribution from a magnetar central engine. The possible magnetar would dominate the second peak. \citet{gomez21}, conversely, suggest that the first peak of the hybrid SLSN/Ic event SN~2019stc is powered by a magnetar and $^{56}$Ni and the second by CSI. Finally, \citet{chen23} suggest the second, slowly evolving peak of SN~2022jl is powered by the accretion of material from a companion star onto the newly formed compact object. The compact object would be orbiting the companion star on an eccentric orbit that it was kicked into by the explosion, explaining the periodicity of the light curve and the H$\alpha$ profile. 

While the presence of two peaks in it cannot be confirmed, some similarities to the second peak of SN~2023aew can be seen in the SLSN~I SN~2019szu \citep{aamer24}. An increasing temperature following a plateau is seen (albeit much higher than in SN~2023aew and not conforming to a single blackbody), and nebular lines appear in its spectra very early after the main peak. The SN is, however, much more luminous than SN~2023aew, and its inferred ejecta mass ($\sim$30~M$_\odot$) much higher than what we obtain from the \texttt{MOSFiT} models (Sect. \ref{sec:mosfit}). This SN also shows no hint of hydrogen or helium lines. \citet{aamer24} explain the plateau as shell-shell collisions from pulsational pair instability \citep[PPI;][]{woosley17}, while the main peak is inferred to be dominated by a magnetar.

As none of the diverse objects examined here provides a good analog to SN~2023aew, it is not clear which, if any, of these mechanisms are involved; however, the long separation between the peaks clearly argues against a shock breakout followed by a $^{56}$Ni peak and against a two-component $^{56}$Ni distribution, as does the fact that a $^{56}$Ni model fails to account for the shape of the second peak (see Sect. \ref{sec:mosfit}). We now discuss other possible physical scenarios for SN~2023aew.

\subsection{Physical scenarios}
\subsubsection{Two unrelated SNe}

At first glance, the simplest explanation for SN~2023aew is that each peak is a separate SN: first, a SE-SN spectroscopically similar to a SN~IIb, followed by another peculiar SN of Type~Ic from a different progenitor system along roughly the same line of sight. There is a $0\farcs090\pm0\farcs053$ ($\sim1.7\sigma$) difference between the centroids of the SN during the first and second peak (see Appendix \ref{sec:coords} for details), which, taken at face value, would tentatively support this scenario. This discrepancy is not statistically significant, but this is a conceivable, though very unlikely, scenario. The two SNe would have to be separated by only $\sim50$~pc (projected) in space and four months in time, not to mention their location at the outskirts of the host galaxy, where the star-formation rate and numbers of stars are low (Sect. \ref{sec:host}). 

The H$\alpha$ line flux caught on the $1\farcs0$ ($\sim$500-pc) slit in our GTC spectra is on the order of $2\times10^{-17}$~erg~s$^{-1}$~cm$^{-2}$. From this, even assuming this H$\alpha$ emission is concentrated within the 50-pc-radius region where the two SNe would be located, we get an H$\alpha$ luminosity of $L_{H\alpha} \sim10^{38}$~erg~s$^{-1}$ in this region, which corresponds to a star formation rate on the order of $10^{-3}\, \mathrm{M}_\odot$~yr$^{-1}$ \citep{kennicutt98}, and is low enough to be powered by only several O stars \citep{crowther13}. This, in turn, would translate into a CCSN rate on the order of $10^{-5}$~yr$^{-1}$ assuming the ratio between the CCSN rate and star formation rate in the nearby universe estimated empirically by \citet{botticella12}, $\sim0.01\, \mathrm{M}_\odot^{-1}$. In a four-month period, this means a chance probability on the order of only $10^{-10}$ for two CCSNe at this location, assuming a Poissonian process. This assumes a constant star formation (and CCSN) rate, however, which can result in an underestimate if the star formation, in reality, occurs in distinct episodes -- but at the same time, the star formation may occur more evenly over the 500-pc region covered by the slit, offsetting this effect.

This scenario presents some additional problems besides the low odds of the line-of-sight and timing coincidence: it would not only be a coincidence of two SNe, but of two peculiar SNe. The SN~Ic needs to have a low velocity, a slow spectral evolution with a post-peak appearance from the beginning, and a somewhat bumpy light curve that does not fit the $^{56}$Ni model (Sect. \ref{sec:mosfit}). Even the SN~IIb exhibits a flattening light curve inconsistent with $^{56}$Co decay, and SN~2003bg, which matches the properties of SN~2023aew at this epoch, shows an abnormally large kinetic energy and ejecta mass \citep{hamuy09,mazzali09}. For these reasons, we consider it even more unlikely that SN~2023aew consists of two unrelated SNe. 

\subsubsection{Two SNe from the same system}

A similar scenario that does not require an extremely coincidental chance alignment of unrelated SNe is that of two related SNe. One could conceive a scenario where the two stars in a binary system have very precisely the same mass, resulting in almost simultaneous explosions. Such a scenario could involve either two stars too distant to interact or a common-envelope evolution with a double core \citep[e.g., Fig. 6 of][]{vignagomez2018}. While roughly equal mass ratios ($q$) can, in principle, exist in binaries, they are somewhat rare \citep[even $q>0.7$ was rarely seen among massive stars by][]{mds17}, and the difference in masses would have to be extremely fine-tuned to produce SNe with a delay of months, as opposed to millennia or more. 

If we imagine an equal-mass binary system where the two stars never interact, we would expect the two SNe to be the same spectral type, which seems not to be the case. Although the spectra associated with the second peak seem to retain some relatively weak helium features, they most closely resemble SNe~Ic (Sect. \ref{sec:specs}) while the initial spectrum shows strong H$\alpha$ absorption. It is conceivable that the two progenitor stars never interact with each other, but that one or both has its own closer companion in a hierarchical triple or quadruple star system, resulting in different spectral types despite the same lifetime, but this again requires extreme fine-tuning.

Since most SE-SNe are the result of binary interactions (e.g., \citealt{smith2011}), we must consider the effect of Roche-lobe overflow on our hypothetical system. In a typical case where $q < 1$ and one star evolves before its companion, the mass transfer would increase the mass of the envelope of the companion, leading to rejuvenation and delayed death. In this case, we can obtain one stripped-envelope SN, but not two in rapid succession, and the first event would be expected to be more stripped, not less. On the other hand, in an equal-mass binary where both stars evolve simultaneously, one could recover two He stars. This scenario is invoked by \citet{vignagomez2018} as a step in one of their channels explaining the observed population of neutron star binaries. The challenge is then to explain the H$\alpha$ line at $-$101~d. Although \citet{vignagomez2018} do not expect to retain hydrogen in the envelope, this could be the result of approximations made in rapid population synthesis codes. As highlighted by \citet{laplace2020}, in detailed models that take into account stellar structure, some hydrogen can remain after a common envelope, which could explain the presence of hydrogen features. The hydrogen could then be blown away by the first SN. It is important to note that these considerations are qualitative and that detailed modeling is required to assess the feasibility of this scenario, which is beyond the scope of this paper. Even if the two-SN scenario is, in principle, feasible, the problem of simultaneous explosions and mass fine-tuning remains.

The problem of fine-tuning could be alleviated if one SN can trigger another in the same system. However, we are not aware of models that can achieve this. \citet{hirai18} studied the interaction between ejecta and a close companion star, which only resulted in the bloating and temporary brightening of the companion. In this study, the companions are hydrogen-rich, likely having gained mass from the primary on top of their own remaining hydrogen. One could, in principle, imagine another scenario where the ejecta of the first SN interacts with a stripped companion star instead (possibly after a common envelope phase), but fine-tuning of the mass ratio would still be required. Additionally, since only a small fraction of the kinetic energy of the first SN  -- itself a small fraction of the binding energy of the core -- can affect the secondary star, it is very unlikely that the nuclear burning could be appreciably sped up.

As a side note, whether the two progenitor stars interact or not, and whether the two SNe are separate or one is triggered by the other, we also note that the ejecta of the second SN should interact with that of the first. While the velocities are higher in the $-$101~d spectrum than later, and thus much of the ejecta from the first SN would be moving too fast to be caught up to by the second SN, some of it would be moving slower due to homologous expansion and could be encountered. Some could possibly even still be located around the progenitor of the second SN, and interaction with this material could conceivably slow down the ejecta of the second SN enough to explain the observed low velocities. While, to our knowledge, there are no published models for this, such an interaction between two massive matter "shells" could produce an extremely strong interaction, likely resulting in a SLSN and an interaction photosphere that would block the line of sight to the inner ejecta of the second SN. 

\subsubsection{Eruption followed by SN}

Numerous SNe have been observed with outbursts and eruptions immediately preceding a SN. Prominent examples of this behavior are the Type IIn SN~2009ip \citep[e.g.,][]{mauerhan13,pastorello13} and other similar events \citep[e.g.,][]{pastorello18,fransson22}, where the precursor eruption typically precedes the main peak by roughly a month. Although some debate remains over whether such transients are true SNe or merely caused by collisions of ejected CSM shells from separate eruptions \citep{fraser13,tpessi23}, the progenitor stars of SN~2009ip and the similar SN~2016jbu have declined below their pre-SN levels \citep{brennan22,smith22}, suggesting a terminal explosion. The observed broad, luminous, and bumpy light curve (Sect. \ref{sec:LC}) could be powered in part by interaction with this ejected material. The final eruption could also have removed (most of) the remaining hydrogen and helium of the progenitor star before its explosion, resulting in the SN~Ic spectrum of the second event. 

However, 2009ip-like SNe mostly exhibit precursors around $r$-band absolute magnitudes of $-15$~mag \citep{pastorello18} with at least one exception, SN~2019zrk, reaching $-16.5$~mag \citep{fransson22}, whereas the peak absolute magnitude of any possible eruption in SN~2023aew must reach a peak of $\sim-17.9$~mag in the $i$ band, which is even somewhat bright for a real SN~IIb \citep{richardson14} although similar to that of SN~2003bg \citep{hamuy09}. The delay between the eruption and the main peak is also longer by a factor of a few. %SN~1961V \citep[e.g.,][]{vandyk12}, a transient spectroscopically similar to SNe~IIn, reached a similar peak magnitude as the first peak. It may have been an eruption from a luminous blue variable star, but later studies have called this origin into question \citep{patton19,woosley22}.

SN~2019szu \citep{aamer24} does show an early plateau attributed to a pre-explosion shell-shell collision, reaching an absolute magnitude of $-18.7$~mag, and this plateau may have lasted a long time and/or been preceded by a brighter peak, similarly to SN~2023aew. In this case, however, the CSM velocity is estimated as 1500~km~s$^{-1}$. We observe absorption minimum (i.e., bulk) velocities of $>10\,000$~km~s$^{-1}$ in the H$\alpha$ and Ca~{\sc ii} lines in the first spectrum at $-101$~d. Such velocities are typically only seen in true SNe -- \citet{smith18} do find velocities in this ballpark in light echoes of the Great Eruption of $\eta$~Carinae, but most of the mass ejected in this eruption moved at a much lower velocity of $\sim600$~km~s$^{-1}$. A similar high-velocity component is also seen in SN~2009ip \citep{pastorello13}, but again, it is only formed by a small fraction of the total mass. Any putative eruption has to replicate the line profiles of a true SN without the narrower core that should contain the majority of the equivalent width of the lines similar to SN~2009ip and its Great Eruption. Even considering the low resolution of SEDM, the core should be unresolved, on the order of $3000$~km~s$^{-1}$ wide.

The first spectrum at $-101$~d resembles that of a post-peak SN~IIb, consistently with the light curve. This and the strength of the absorption lines suggest an ejecta mass and velocity, and hence kinetic energy, indicative of a true SN. This is also compatible with the \texttt{MOSFiT} fit to the first event (Sect. \ref{sec:mosfit}): the model parameters found from the fit are close to those of SN~2003bg, and the kinetic energy ($\sim\frac{3}{10}M_{ej}v_{ej}^2$) with the median parameter values is $\sim2.9\times10^{51}$~erg. Such an energy is theoretically achievable in an eruption with PPI \citep[][]{woosley17}. However, PPI eruptions energetic enough to resemble CCSNe are expected to occur in very massive stars with helium cores of $\gtrsim50$~M$_\odot$, with low shell velocities (a few thousand km~s$^{-1}$) more similar to SN~2019szu \citep{aamer24} or SNe~IIn, Ibn or Icn. 

We also note that the spectra of the second peak seem to exhibit multiple nebular lines starting around +70~d, at the same time as the light curve reaches the luminosity of the plateau that precedes the rebrightening, where lines from that component might re-emerge. These lines include [O~{\sc i}], Mg~{\sc i}], [Ca~{\sc ii}]+[O~{\sc ii}], and possibly [N~{\sc ii}]. [Ca~{\sc ii}]+[O~{\sc ii}], especially, seems to emerge very early, even before the second peak. This is reasonable if the first peak is in fact a true SN~IIb and the lines actually emerge at a later time relative to explosion. The velocities of the lines during the second peak are low compared to normal SNe~Ic, but more similar to what one would expect at late times from a SN~IIb \citep[e.g., SN~2011dh showed an H$\alpha$ velocity of $\sim-12\,000$~km~s$^{-1}$ at +20~d and absorption line velocities of 4000--7000~km~s$^{-1}$ at +100~d;][]{ergon14}. The first peak is red, but the $^{56}$Ni model in \texttt{MOSFiT} has no problem reproducing the color evolution with a low extinction. Considering all of this, we disfavor the eruption+SN scenario and consider the first peak a real SN. 

\subsubsection{SN with delayed CSI}

If the first peak is powered by a real SN~IIb, and the second peak cannot be a real SN~Ic from another star, we are left with scenarios where the second peak is instead caused by the delayed input from some mechanism that can inject extra energy into the SN well past the initial explosion. One such mechanism could be CSI, the preferred scenario for SN~2022xxf \citep{kuncarayakti23}. CSI is also seen at late times in SNe~IIb such as SN~1993J \citep[e.g.][]{patat95}, albeit not to the degree that SN~2023aew would require. The CSM could be produced in late-time eruptions. The ejecta mass we obtain from the \texttt{MOSFiT} fit to the first peak is much higher than in either of the CSI models, but since those models assume the second peak is a SN, this may not be a problem. The CSM could, in principle, be created through PPI, but in such a case the progenitor must be very massive compared to typical SN~IIb progenitors \citep{woosley17}, and most of the ejecta should fall back to create a black hole. The plateau following the SN~IIb and the bumps after the second peak suggest the presence of some interaction; a similar flattening has been seen in some interacting SE-SNe \citep[e.g.,][]{kuncarayakti22,ferrari24}. However, it is unclear if CSI is the dominant power source in SN~2023aew.

Interaction with a dense shell of CSM at a relatively large distance from the progenitor star, to account for the delayed second peak, presents some problems. The onset of the CSI powering the second peak ($\sim-$25~d) would be $\sim$110~d after the explosion. Assuming a velocity of $\sim12\, 500$~km~s$^{-1}$, observed in the first spectrum, for the outer ejecta, the distance to the main CSM shell would be $\gtrsim10^{16}$~cm. One problem in this scenario is the blackbody evolution (Sect. \ref{sec:BBparams}). While the temperature at the peak epochs is likely an overestimate because of the spotty NIR light curve interpolation, the dependence of the luminosity on temperature is much stronger than on the radius. Therefore the peak-epoch radius is unlikely to be much larger than $2\times10^{15}$~cm, and the radius is observed to stay fairly constant for about 50~d before continuing to decline. The photosphere never reaches distances even close to $10^{16}$~cm, where the dense shell should be located; even with a diluted photosphere, an increase in radius should be observed from the onset of CSI, but during the rebrightening phase, only the temperature increases. Instead, this behavior implies an internal mechanism heating the ejecta. 

We also see a decrease in absorption velocities from $>10\,000$~km~s$^{-1}$ at $-101$~d to $\sim6500$~km~s$^{-1}$ at $-$19~d, after which the velocity slowly declines to $\sim5500$~km~s$^{-1}$ over roughly 160 days. This implies that the matter above the photosphere is located further into the ejecta than at $-101$~d, while the interaction should be triggered by the ejecta further out and obscure some of the slower material from view. The interaction with a distant shell would also result in boxy emission line profiles \citep[e.g.,][]{patat95} that are not observed at this time. Since the SN in this scenario is of Type IIb and contains hydrogen and helium in its spectrum, the obvious implication is that both the outer ejecta of the progenitor star and its CSM must also primarily contain hydrogen and helium; therefore, the boxy emission should be seen in H$\alpha$, specifically. No narrow emission lines similar to those in SN~2022xxf \citep{kuncarayakti23} are seen at late times, either, except a faint H$\alpha$ attributable to the host galaxy, accompanied by [O~{\sc iii}]~$\lambda\lambda4959,5007$ of similar strength.

CSI can provide an internal heating mechanism if, say, a disk-shaped CSM is engulfed by the ejecta. A disk-interaction model has been proposed for various transients \citet{smith15,andrewssmith18,pursiainen22,pursiainen23}. A hypothetical disk located at $\sim10^{16}$~cm to account for the delay would have the same problems as a shell at this distance. However, it is possible that the interaction begins somewhat earlier, but most emission from it takes until $\sim$110~d from the explosion to break out. If the outer ejecta encounters the inner edge of the disk $\sim$25~d after the explosion, this corresponds to $\sim2\times10^{15}$~cm, the blackbody radius we see at the end of the plateau. The main breakout of emission would, in this scenario, not happen until the outer ejecta becomes optically thin enough. Before this, we would only see the plateau, once the SN itself stops dominating the total emission. The breakout could cause the photosphere to stall as it recedes close to the interaction site, and would cause the increase in observed temperature. The reprocessing site of the CSI energy would move inward in the ejecta over time, creating the absorption lines we see. \citet{nagao20} model the light curves of SNe~II with disk CSI and find that the observed temperature and luminosity rise very rapidly once the interaction site is uncovered, while being affected only slightly before this happens -- but we should in this case also see lines from the interaction region, unless only the ejecta around this region is still optically thick. A possible H$\alpha$ line (albeit broad and without H$\beta$) is seen in the spectra, and at late times, it looks somewhat boxy or double-peaked, as might be expected from CSI -- but it only appears at +40~d or so, and it is also quite possible that the feature is dominated by [N~{\sc ii}] instead. Furthermore, the ejecta of a SN~IIb is less massive ($\sim$3~M$_\odot$ according to \texttt{MOSFiT}) and less opaque (fewer free electrons) than in a SN~II, making it more difficult to hide the interaction until $\sim$110~d.

\subsubsection{SN with delayed magnetar engine}

Another possible internal heating mechanism for replicating the blackbody evolution -- a shrinking or stagnant photosphere but a rising temperature during the rebrightening -- could conceivably be a central engine, such as the spin-down of a magnetar \citep{kasenbildsten10}. The observed evolution might be the result of a combination of a delayed onset of the engine and the shrinking of the photosphere with the decreasing optical depth, which is also needed in the CSI scenario. 

The \texttt{MOSFiT}-implemented magnetar model \citep{nicholl17} is unable to fit the temperature evolution of the second peak well (see Sect. \ref{sec:mosfit}), even though it is the best of the models used here. This manifests as a different rise time and peak epoch in each optical band because of cooling, while in SN~2023aew, the peak is observed to be simultaneous in all bands because of a rising blackbody temperature. The \texttt{slsn} model in \texttt{MOSFiT}, however, assumes the ejecta structure of an infant SN, without 100~d of expansion. The photosphere at the epoch of rebrightening was located at $\sim2\times10^{15}$~cm and shrinking, possibly allowing hotter inner layers to emerge over time and changing the typical temperature evolution of the magnetar-powered SN model. After the second peak, the magnetar model is better equipped to replicate the light curve, although at very late times the light-curve features would somewhat deviate from it (due to, for example, a decreasing trapping fraction). In this scenario, relatively weak CSI could produce the bumps in the light curve.

More importantly, however, to our knowledge, no magnetar-powered model includes a delay between the SN and the onset of the central engine -- caused by the spindown of the newborn magnetar through magnetic dipole radiation -- let alone a delay of over 100~d (or possibly $\sim$50, similarly to the disk CSI scenario). Magnetar scenarios, even those that have been invoked for double-peaked SNe such as SN~2005bf \citep{maeda07} and SN~2019cad \citep{gutierrez21}, assume an immediate onset of the magnetic braking. It is not clear if a long delay can be produced in reality, and in the case of SN~2023aew, it is needed. We thus disfavor this scenario.

\subsubsection{SN with delayed accretion}

A third way to heat the ejecta from within could be fallback accretion onto a newly formed black hole. Such a scenario was considered for SN~2005bf by \citet{maeda07}, but ultimately disfavored because of the long delay required for the onset of the accretion. \citet{moriya18} also suggest fallback accretion to power the highly energetic SN~II OGLE-2014-SN-073: in this case, the accretion would begin immediately, but the peak of the light curve would not be reached until $\sim$100~d after the explosion because of a large ejecta mass. 

In the case of SN~2023aew, the delay before the onset of the main accretion episode (possibly preceded by weaker accretion that causes the flattening light curve before the rebrightening) must be even longer. The velocity of the slow inner ejecta should be fine-tuned to be just below the escape velocity in such a way that falling back onto the black hole and/or creating the accretion disk would take months. A part of the inner ejecta would receive additional energy deposited by the accretion disk wind and be re-accelerated. As a result, the density of the inner ejecta would stay higher than expected from $\sim$100~d of homologous expansion, and it would be possible to have a combination of nebular lines from the outer ejecta and a photospheric spectrum from the inner ejecta. We also note that the line profiles of forbidden oxygen and calcium lines are usually not both double-peaked because of their different distribution in the ejecta \citep{fang23}, but in SN~2023aew neither feature is single-peaked at late times (Fig.~\ref{fig:naid}). This may suggest that the forbidden lines are only emitted by the part of the ejecta that did not fall back onto the compact object, perhaps supporting this scenario. If the accretion does not proceed smoothly, bumps in the light curve (and the fast decline compared to the \texttt{MOSFiT} \texttt{fallback} model; see Sect. \ref{sec:mosfit}) could be explained as well, but an additional contribution from CSI is possible. We encourage detailed modeling of this scenario to determine its plausibility.

The total radiated energy during the second peak, within the observed timescale and wavelength range, is $\sim4\times10^{49}$~erg, and the peak luminosity is $\sim8\times10^{42}$~erg~s$^{-1}$ (Sect. \ref{sec:bolom}). The outflow energy that can be extracted from accretion in a disk wind is $E_{acc} = \epsilon  M c^2$ (and the corresponding energy injection rate is $\frac{dE_{acc}}{dt} = \epsilon \frac{dM}{dt} c^2$), where $\epsilon \sim 10^{-3}$ is the radiative efficiency of the accretion \citep{dexterkasen13}. This corresponds to the luminosity of a directly observed accretion disk, which in our scenario is hidden by the SN ejecta. Assuming that this energy is wholly deposited into the ejecta and roughly 1\% can escape as radiation, we can estimate that in order to power the observed light curve, the accreted mass needs to be on the order of a few~M$_\odot$ (note that this is only the part of the fallback that forms the accretion disk and does not fall directly into the black hole). The accretion rate $\frac{dM}{dt}$ at the peak would be a few $\times\,10^{-2}$~M$_\odot$~d$^{-1}$ with the same assumptions. As the exact processes are unclear, these numbers are simply order-of-magnitude estimates. In this scenario, the fallback of the inner ejecta would lessen the amount of $^{56}$Ni available for powering the first event, but since the inner ejecta is probably aspherically distributed \citep{fang23}, this may not present a problem. The progenitor star would likely be much more massive than those of normal SNe~IIb -- it would need to have at least several M$_\odot$ of ejecta divided between the accreted inner ejecta, the accelerated inner ejecta and the outer ejecta responsible for the first peak.

The Eddington luminosity of an accreting compact object is $L_{Edd} \approx 1.2\times 10^{38} \frac{M_{CO}}{\mathrm{M_\odot}}$~erg~s$^{-1}$, where $M_{CO}$ is the mass of the object. For a 10-M$_\odot$ black hole, one thus obtains $L_{Edd} \approx 1.2\times10^{39}$~erg~s$^{-1}$, roughly a factor of 6700 below even the observed bolometric luminosity at the peak, let alone the total required outflow energy. A similarly super-Eddington luminosity was noted for the Type~Ic SN~2022jli by \citet{chen23} when using a neutron star as the compact object; SN~2023aew was somewhat more luminous than SN~2022jli, requiring a larger compact-object mass or an even more dramatically super-Eddington accretion. The mechanisms of such accretion are unclear \citep{brightman19}, but examples are known, from slightly super-Eddington supermassive black holes through some X-ray binaries \citep[one of which exceeds its $L_{Edd}$ by a factor of $\sim$500;][]{israel17} to accretion-powered long gamma-ray bursts (LGRBs). In LGRBs, accretion rates onto stellar-mass black holes can momentarily range from 0.01 to several M$_\odot$~s$^{-1}$ \citep[e.g.,][]{kumarzhang15}, orders of magnitude above our estimate for SN~2023aew. The SNe accompanying such events tend to be of Type~Ic-BL \citep[e.g., SN~1998bw;][]{galama98}. We see no very broad lines indicating an outflow similar to the SNe accompanying LGRBs (indeed, the velocity during the second peak is lower than in most SNe~Ic); instead, in SN~2023aew any Ic-BL-like outflow or jet would either be directed outside our line of sight or be too weak to accelerate the (inner) ejecta to such velocities. 
 
A compact object kicked in the direction of a secondary star, similarly to the onset of the second peak in SN~2022jli \citep{chen23}, could also trigger accretion, possibly falling inside the envelope of the secondary and causing an outflow of hydrogen-poor material without the periodicity of SN~2022jli. %In the latter case, though, one might expect the compact object to accrete more strongly than required by SN~2023aew. 
We note that the bumps in the SN~2023aew light curve could have a $\sim$75-day periodicity, but with only two bumps this is difficult to say for certain, and such a period would imply a considerably larger orbital separation from the companion than in SN~2022jli, where the period was $\sim$12~d, resulting in weaker accretion except near the pericenter of an eccentric orbit.

%--------------------------------------------------------------------
\section{Conclusions}
\label{sec:concl}

We spectroscopically and photometrically followed up the peculiar, double-peaked SE-SN, SN~2023aew, in the optical and NIR. This object first shows behavior similar to a SN~IIb -- albeit somewhat luminous ($M_{TESS,\mathrm{peak}} = -17.88\pm0.12$~mag), red and with weak helium lines -- followed by a plateau $\sim$50~d after the first peak and a rebrightening at $\sim$100~d after the first peak. The second peak resembles a slowly evolving SN~Ic with a bumpy light curve and a peak absolute magnitude ($M_r = -18.75\pm0.04$~mag) similar to SNe~Ic-BL. We analyzed the data, performed fits to the spectral energy distributions and light curves and compared the behavior of SN~2023aew to other SNe~Ic and double-peaked SE-SNe. We then explored various physical scenarios to attempt to explain the anomalous behavior of the object. Our conclusions are as follows:

\begin{itemize}
    \item SN~2023aew, among the double-peaked SE-SNe examined here, presents a unique combination of properties, with very widely separated peaks and a slow spectroscopic evolution. The first peak before the flattening is in many ways (spectrum, ejecta mass, luminosity) similar to the Type IIb SN~2003bg. Compared to normal SNe~Ic, the second peak exhibits a low expansion velocity in addition to the slow spectroscopic and photometric evolution.
    \item The second peak is difficult to fit with a $^{56}$Ni decay or magnetar central engine model. A part of this difficulty is the simultaneous peak in all optical bands, caused by an increasing rather than decreasing temperature -- which, furthermore, only reaches $\sim$8000~K at peak -- during the rebrightening. Meanwhile, the blackbody radius slowly shrinks during almost the entire observed light curve.
    \item The simplest explanation for the two widely separated peaks, two SE-SNe coincident in both time ($\sim$120~d) and space ($\lesssim$50~pc projected), is extremely unlikely, considering not only the low star-formation rate in the vicinity of SN~2023aew, but also its peculiarity. Meanwhile, a scenario of two SE-SNe from the same system avoids this problem but requires extremely precise fine-tuning and has difficulty explaining the properties of the two SNe. Therefore, we do not consider a two-SN scenario plausible.
    \item The line velocities, luminosity and kinetic energy of the first peak are difficult to explain if it is a pre-SN eruption instead of a true SN, whereas these properties and the time scale are compatible with a SN~IIb. We thus consider the first peak to be associated with a SN, followed by delayed energy input by some mechanism.
    \item This mechanism is unlikely to be interaction with a distant ($\gtrsim10^{16}$~cm) CSM shell, for several reasons. No spectroscopic signatures of interaction are seen; hiding the interaction within a photosphere at such a distance would be difficult; and absorption lines would be formed in the outer ejecta, which is not the case. 
    \item The blackbody parameter evolution instead indicates a source of heating from within the ejecta. This could be emission from embedded CSI with a disk of CSM engulfed by the outer ejecta. The main emission could stay hidden until the outer ejecta becomes optically thin enough. However, the difficulty of hiding the interaction until the second peak and a lack of CSM lines at this time argue against this scenario.
    \item If the internal source is a magnetar central engine instead, the onset of the magnetar spindown has to be delayed by months through an as-yet unknown and unprecedented mechanism.
    \item Fallback accretion onto a newly formed compact object could launch a disk wind that re-accelerates the inner ejecta. The long delay before the onset of the accretion requires a fine-tuned, low inner ejecta velocity. 
\end{itemize}

While we can therefore exclude some of the scenarios we have considered, such as two-SN scenarios and pre-SN eruptions, we cannot definitively establish the power source of the second, brighter peak. CSI with a distant shell and some internal heating mechanisms present problems with the time scale and/or the spectroscopic evolution, but a scenario where the second peak is powered by accretion onto a black hole may be the least problematic. We encourage detailed modeling studies of SN~2023aew, outside the scope of this paper, in order to shed more light on this unprecedented SN.

\begin{acknowledgements}

We thank the anonymous referee for their comments on the manuscript, and Luc Dessart for helpful discussions. \\
HK was funded by the Research Council of Finland projects 324504, 328898, and 353019. TN acknowledges support from the Research Council of Finland projects 324504 and 328898. RK acknowledges support via Research Council of Finland (grant 340613).
MF is supported by a Royal Society - Science Foundation Ireland University Research Fellowship.
HFS is supported by the Eric and Wendy Schmidt A.I. in Science Fellowship.
KM acknowledges support from JSPS KAKENHI grant (JP20H04737) and the JSPS Open Partnership Bilateral Joint Research Project between Japan and Finland (JPJSBP120229923). 
MDS is funded by the Independent Research Fund Denmark (IRFD, grant number  10.46540/2032-00022B.
CPG acknowledges financial support from the Secretary of Universities and Research (Government of Catalonia) and by the Horizon 2020 Research and Innovation Programme of the European Union under the Marie Sk\l{}odowska-Curie and the Beatriu de Pin\'os 2021 BP 00168 programme,
from the Spanish Ministerio de Ciencia e Innovaci\'on (MCIN) and the Agencia Estatal de Investigaci\'on (AEI) 10.13039/501100011033 under the PID2020-115253GA-I00 HOSTFLOWS project, and the program Unidad de Excelencia Mar\'ia de Maeztu CEX2020-001058-M. Y-ZC is supported by the National Natural Science Foundation of China (NSFC, Grant No. 12303054), the Yunnan Fundamental Research Projects (Grant No. 202401AU070063) and the International Centre of Supernovae, Yunnan Key Laboratory (No. 202302AN360001). MP acknowledges support from a UK Research and Innovation Fellowship (MR/T020784/1). PC acknowledges support via Research Council of Finland grant 340613.
SM acknowledges support from the Research Council of Finland project 350458.
NER, AP, SB, AR and GV acknowledge support from the PRIN-INAF 2022, ‘Shedding light on the nature of gap transients: from the observations to the models’. NER also acknowledges partial support from MIUR, PRIN 2017 (grant 20179ZF5KS). AR also acknowledges financial support from the INAF GRAWITA Large Program Grant. \\

We acknowledge ESA Gaia, DPAC and the Photometric Science Alerts Team (http://gsaweb.ast.cam.ac.uk/alerts). The Starlink software \citep{currie14} is currently supported by the East Asian Observatory. \\

Partially based on observations made with the Nordic Optical Telescope, owned in collaboration by the University of Turku and Aarhus University, and operated jointly by Aarhus University, the University of Turku and the University of Oslo, representing Denmark, Finland and Norway, the University of Iceland and Stockholm University at the Observatorio del Roque de los Muchachos, La Palma, Spain, of the Instituto de Astrofisica de Canarias. The data presented here were obtained in part with ALFOSC, which is provided by the Instituto de Astrofisica de Andalucia (IAA) under a joint agreement with the University of Copenhagen and NOT. \\

This study is based on observations made with the Gran Telescopio Canarias (GTC), installed in the Spanish Observatorio del Roque de los Muchachos of the Instituto de Astrofísica de Canarias, in the island of La Palma. \\

Access to the Las Cumbres Observatory was made possible via an allocation by OPTICON (program 22A/012, PI Stritzinger). This project has received funding from the European Union's Horizon 2020 research and innovation programme under grant agreement No 101004719. \\

The GROWTH India Telescope (GIT) is a 70-cm telescope with a 0.7-degree field of view, set up by the Indian Institute of Astrophysics (IIA) and the Indian Institute of Technology Bombay (IITB) with funding from  Indo-US Science and Technology Forum and the Science and Engineering Research Board, Department of Science and Technology, Government of India. It is located at the Indian Astronomical Observatory (IAO, Hanle). We acknowledge funding by the IITB alumni batch of 1994, which partially supports the operation of the telescope. Telescope technical details are available at https://sites.google.com/view/growthindia/. \\

Operation of Nayuta telescope and NIC is partially supported by the Optical and Infrared Synergetic Telescopes for Education and Research (OISTER) program funded by the MEXT of Japan. \\

The spectra taken by the Seimei telescope are obtained under the KASTOR (Kanata And Seimei Transient Observation Regime) campaign. The Seimei telescope at the Okayama Observatory is jointly operated by Kyoto University and the Astronomical Observatory of Japan (NAOJ), with assistance provided by the OISTER program. \\
      
\end{acknowledgements}

\bibliographystyle{aa}
\bibliography{kangas_23aew_revised2}
\clearpage

\begin{appendix}

\section{Supplementary tables}

\FloatBarrier

\begin{table}[h!]
\centering
\caption{Log of our GIT imaging observations of SN~2023aew.}
\label{tab:growth}
\begin{tabular}{cccc}
\hline
MJD & Filter & Magnitude & Error \\
 & & (mag) & (mag) \\
\hline
60061.9	&	$g$	&	16.94	&	0.03	\\
60061.9	&	$i$	&	16.71	&	0.04	\\
60061.9	&	$r$	&	16.70	&	0.03	\\
60061.9	&	$z$	&	16.77	&	0.06	\\
60062.9	&	$g$	&	16.88	&	0.03	\\
60062.9	&	$i$	&	16.64	&	0.03	\\
60062.9	&	$r$	&	16.64	&	0.02	\\
60062.9	&	$z$	&	16.75	&	0.05	\\
60063.9	&	$g$	&	16.82	&	0.03	\\
60063.9	&	$i$	&	16.61	&	0.04	\\
60063.9	&	$r$	&	16.63	&	0.03	\\
60063.9	&	$z$	&	16.74	&	0.05	\\
60069.9	&	$g$	&	16.68	&	0.07	\\
60069.9	&	$i$	&	16.57	&	0.03	\\
60069.9	&	$r$	&	16.53	&	0.04	\\
60069.9	&	$z$	&	16.67	&	0.07	\\
60081.8	&	$g$	&	16.78	&	0.03	\\
60081.9	&	$i$	&	16.62	&	0.04	\\
60081.9	&	$r$	&	16.55	&	0.04	\\
60081.9	&	$z$	&	16.71	&	0.08	\\
%60083.9	&	$g$	&	17.18	&	0.19	\\
60085.8	&	$g$	&	16.91	&	0.03	\\
60085.8	&	$i$	&	16.65	&	0.04	\\
60085.8	&	$r$	&	16.67	&	0.03	\\
60085.8	&	$z$	&	16.76	&	0.06	\\
60088.9	&	$g$	&	16.62	&	0.15	\\
%60088.9	&	$i$	&	16.74	&	0.29	\\
%60088.9	&	$r$	&	17.97*	&	0.90	\\
60090.8	&	$g$	&	17.04	&	0.04	\\
%60090.8	&	$i$	&	16.19*	&	0.57	\\
60090.8	&	$r$	&	16.74	&	0.04	\\
%60090.8	&	$z$	&	17.19*	&	0.45	\\
60091.8	&	$g$	&	17.07	&	0.03	\\
60091.8	&	$i$	&	16.80	&	0.03	\\
60091.8	&	$r$	&	16.75	&	0.04	\\
60091.8	&	$z$	&	16.88	&	0.07	\\
60092.8	&	$r$	&	16.78	&	0.04	\\
60092.9	&	$g$	&	17.15	&	0.06	\\
60092.9	&	$i$	&	16.81	&	0.04	\\
60092.9	&	$z$	&	16.85	&	0.08	\\
60098.8	&	$g$	&	17.31	&	0.03	\\
60098.8	&	$i$	&	16.95	&	0.03	\\
60098.8	&	$r$	&	16.95	&	0.03	\\
60098.8	&	$z$	&	17.13	&	0.07	\\
60099.8	&	$g$	&	17.31	&	0.03	\\
60099.8	&	$i$	&	17.01	&	0.03	\\
60099.8	&	$r$	&	16.99	&	0.04	\\
60099.8	&	$z$	&	17.03	&	0.07	\\
60101.8	&	$g$	&	17.34	&	0.04	\\
60101.8	&	$i$	&	17.07	&	0.03	\\
60101.8	&	$r$	&	17.04	&	0.03	\\
60101.8	&	$z$	&	17.07	&	0.05	\\
60102.8	&	$g$	&	17.48	&	0.06	\\
60102.8	&	$i$	&	17.08	&	0.03	\\
60102.8	&	$r$	&	17.06	&	0.03	\\
60102.8	&	$z$	&	17.11	&	0.06	\\
\hline
\end{tabular}
\tablefoot{All magnitudes are in the AB system.}
\end{table}

\setcounter{table}{0}

\begin{table}
\centering
\caption{Log of our GIT imaging observations of SN~2023aew (continued).}
\label{tab:growth2}
\begin{tabular}{cccc}
\hline
MJD & Filter & Magnitude & Error \\
 & & (mag) & (mag) \\
\hline
60105.8	&	$g$	&	17.51	&	0.03	\\
60105.8	&	$i$	&	17.08	&	0.03	\\
60105.8	&	$r$	&	17.13	&	0.03	\\
60105.8	&	$z$	&	17.31	&	0.08	\\
60107.8	&	$r$	&	17.23	&	0.04	\\
60107.9	&	$g$	&	17.58	&	0.03	\\
60107.9	&	$i$	&	17.25	&	0.04	\\
60107.9	&	$z$	&	17.25	&	0.07	\\
60108.8	&	$g$	&	17.73	&	0.12	\\
60108.8	&	$i$	&	17.33	&	0.07	\\
60108.8	&	$r$	&	17.24	&	0.04	\\
%60108.8	&	$z$	&	16.04*	&	0.25	\\
60109.9	&	$g$	&	17.69	&	0.03	\\
60109.9	&	$i$	&	17.34	&	0.03	\\
60109.9	&	$r$	&	17.30	&	0.04	\\
60109.9	&	$z$	&	17.34	&	0.07	\\
60111.9	&	$g$	&	17.77	&	0.07	\\
60111.9	&	$i$	&	17.25	&	0.19	\\
60111.9	&	$r$	&	17.38	&	0.05	\\
60112.8	&	$g$	&	17.79	&	0.03	\\
60112.8	&	$i$	&	17.39	&	0.03	\\
60112.8	&	$r$	&	17.41	&	0.03	\\
60112.8	&	$z$	&	17.59	&	0.08	\\
60113.8	&	$g$	&	17.81	&	0.03	\\
60113.8	&	$i$	&	17.44	&	0.03	\\
60113.8	&	$r$	&	17.43	&	0.03	\\
60113.8	&	$r$	&	17.43	&	0.03	\\
60113.8	&	$z$	&	17.55	&	0.08	\\
%60114.9	&	$i$	&	17.57	&	0.35	\\
60114.9	&	$r$	&	17.49	&	0.12	\\
60115.8	&	$g$	&	17.93	&	0.05	\\
60115.9	&	$i$	&	17.48	&	0.06	\\
60115.9	&	$r$	&	17.55	&	0.07	\\
60115.9	&	$z$	&	17.55	&	0.12	\\
60118.9	&	$g$	&	18.17	&	0.04	\\
60118.9	&	$i$	&	17.74	&	0.06	\\
60118.9	&	$r$	&	17.77	&	0.05	\\
60118.9	&	$z$	&	17.69	&	0.09	\\
60119.9	&	$g$	&	18.21	&	0.04	\\
60119.9	&	$i$	&	17.75	&	0.04	\\
60119.9	&	$r$	&	17.81	&	0.04	\\
60119.9	&	$z$	&	17.73	&	0.08	\\
60122.9	&	$g$	&	18.36	&	0.09	\\
60122.9	&	$i$	&	18.02	&	0.06	\\
60122.9	&	$r$	&	18.06	&	0.06	\\
%60122.9	&	$z$	&	18.05	&	0.67	\\
60124.7	&	$i$	&	18.09	&	0.06	\\
60124.7	&	$z$	&	18.07	&	0.10	\\
60124.8	&	$g$	&	18.43	&	0.05	\\
60124.8	&	$r$	&	18.10	&	0.04	\\
60126.8	&	$g$	&	18.65	&	0.09	\\
60126.8	&	$i$	&	18.19	&	0.06	\\
60126.8	&	$r$	&	18.22	&	0.05	\\
60126.8	&	$z$	&	18.10	&	0.11	\\
\hline
\end{tabular}
\end{table}

\setcounter{table}{0}

\begin{table}
\centering
\caption{Log of our GIT imaging observations of SN~2023aew (continued).}
\label{tab:growth3}
\begin{tabular}{cccc}
\hline
MJD & Filter & Magnitude & Error \\
 & & (mag) & (mag) \\
\hline
60129.8	&	$g$	&	18.68	&	0.09	\\
60129.9	&	$i$	&	18.34	&	0.08	\\
60129.9	&	$r$	&	18.40	&	0.06	\\
60129.9	&	$z$	&	18.52	&	0.18	\\
60136.8	&	$g$	&	18.96	&	0.04	\\
60136.8	&	$i$	&	18.63	&	0.05	\\
60136.8	&	$r$	&	18.62	&	0.04	\\
60136.8	&	$z$	&	18.54	&	0.12	\\
60138.8	&	$g$	&	19.08	&	0.10	\\
60138.8	&	$i$	&	18.81	&	0.13	\\
60138.8	&	$r$	&	18.61	&	0.13	\\
%60138.8	&	$z$	&	18.45	&	0.26	\\
60139.8	&	$g$	&	19.09	&	0.05	\\
60139.8	&	$i$	&	18.71	&	0.06	\\
60139.8	&	$r$	&	18.70	&	0.05	\\
60139.8	&	$z$	&	18.63	&	0.12	\\
60155.6	&	$g$	&	19.48	&	0.18	\\
60155.6	&	$r$	&	18.89	&	0.07	\\
60155.6	&	$z$	&	18.64	&	0.15	\\
60155.7	&	$i$	&	18.89	&	0.09	\\
60156.7	&	$g$	&	19.51	&	0.21	\\
60156.7	&	$i$	&	18.83	&	0.09	\\
60156.7	&	$r$	&	18.91	&	0.08	\\
60156.7	&	$z$	&	18.70	&	0.17	\\
%60157.6	&	$i$	&	18.91	&	0.22	\\
%60157.6	&	$z$	&	18.65	&	0.25	\\
60157.7	&	$g$	&	19.48	&	0.18	\\
60157.7	&	$r$	&	19.00	&	0.14	\\
60159.7	&	$g$	&	19.58	&	0.14	\\
60159.7	&	$i$	&	18.94	&	0.11	\\
60159.7	&	$r$	&	19.04	&	0.08	\\
60159.7	&	$z$	&	18.64	&	0.17	\\
60161.8	&	$i$	&	18.63	&	0.25	\\
60161.8	&	$r$	&	18.76	&	0.22	\\
60162.7	&	$g$	&	19.56	&	0.07	\\
%60162.7	&	$r$	&	17.32*	&	0.37	\\
60163.6	&	$i$	&	19.34	&	0.12	\\
60163.6	&	$z$	&	18.74 &	0.20	\\
60163.7	&	$g$	&	19.68	&	0.06	\\
60163.7	&	$r$	&	19.17	&	0.06	\\
60164.7	&	$g$	&	19.77	&	0.08	\\
60164.7	&	$i$	&	19.27	&	0.12	\\
60164.7	&	$r$	&	19.26	&	0.06	\\
60164.7	&	$z$	&	18.70	&	0.19	\\
60165.7	&	$g$	&	19.63	&	0.05	\\
60165.7	&	$i$	&	19.38	&	0.11	\\
60165.7	&	$r$	&	19.31	&	0.06	\\
%60165.7	&	$z$	&	18.98*	&	0.24	\\
%60167.8	&	$i$	&	19.27	&	0.50 \\
60167.8	&	$r$	&	19.31	&	0.09	\\
%60167.8	&	$z$	&	19.15*	&	0.30	\\
60168.8	&	$g$	&	19.94	&	0.06	\\
60168.8	&	$i$	&	19.19	&	0.12	\\
60168.8	&	$r$	&	19.47	&	0.15	\\
%60168.8	&	$z$	&	19.20*	&	0.32	\\
%60170.8	&	$g$	&	19.33*	&	0.24	\\
%60170.8	&	$i$	&	18.84*	&	0.34	\\
%60170.8	&	$r$	&	19.50	&	0.35	\\
60171.8	&	$g$	&	20.07	&	0.08	\\
60171.8	&	$i$	&	19.74	&	0.14	\\
60171.8	&	$r$	&	19.72	&	0.08	\\
%60171.8	&	$z$	&	19.78*	&	0.38	\\
60172.8	&	$g$	&	20.08	&	0.12	\\
%60172.8	&	$i$	&	19.78	&	0.23	\\
60172.8	&	$r$	&	19.84	&	0.13	\\
%60172.8	&	$z$	&	20.12*	&	0.72	\\
60173.6	&	$i$	&	19.87	&	0.15	\\
%60173.6	&	$z$	&	19.10*	&	0.26	\\
60173.7	&	$g$	&	20.02	&	0.06	\\
60173.7	&	$r$	&	19.59	&	0.07	\\
\hline
\end{tabular}
\end{table}

\setcounter{table}{0}

\begin{table}
\centering
\caption{Log of our GIT imaging observations of SN~2023aew (continued). }
\label{tab:growth4}
\begin{tabular}{cccc}
\hline
MJD & Filter & Magnitude & Error \\
 & & (mag) & (mag) \\
\hline
60174.7	&	$g$	&	20.06	&	0.09	\\
60174.7	&	$i$	&	19.50	&	0.14	\\
60174.7	&	$r$	&	19.79	&	0.10	\\
60177.7	&	$g$	&	20.26	&	0.09	\\
%60179.8	&	$g$	&	20.11	&	0.27	\\
%60179.8	&	$z$	&	17.80*	&	0.73	\\
%60181.8	&	$g$	&	20.30	&	0.17	\\
%60181.8	&	$r$	&	18.98*	&	0.30	\\
%60182.8	&	$g$	&	20.64	&	0.31	\\
%60182.8	&	$i$	&	19.88	&	0.26	\\
%60182.8	&	$r$	&	20.37	&	0.28	\\
%60182.8 &	$z$	&	19.37*	&	0.38	\\
%60185.6	&	$i$	&	18.32*	&	1.21	\\
60185.6	&	$r$	&	19.84	&	0.18	\\
%60185.6	&	$z$	&	19.36*	&	1.06	\\
60185.7	&	$g$	&	20.53	&	0.36	\\
%60188.6	&	$i$	&	18.21*	&	1.06	\\
%60188.6	&	$r$	&	19.58	&	0.59	\\
%60188.6	&	$z$	&	18.12*	&	0.32	\\
%60188.8	&	$g$	&	20.13	&	0.30	\\
%60190.8	&	$i$	&	20.30	&	0.30	\\
%60190.8	&	$r$	&	20.44	&	0.25	\\
%60191.7	&	$g$	&	20.29	&	0.22	\\
60191.7	&	$r$	&	20.21	&	0.22	\\
%60191.8	&	$i$	&	20.53	&	0.30	\\
60191.8	&	$z$	&	18.80*	&	0.19	\\
60192.7	&	$g$	&	20.56	&	0.16	\\
60192.7	&	$i$	&	20.17	&	0.16	\\
60192.7	&	$r$	&	20.11	&	0.13	\\
%60192.7	&	$z$	&	19.96*	&	0.31	\\
%60193.7	&	$g$	&	20.19	&	0.10	\\
%60193.7	&	$i$	&	19.95	&	0.27	\\
%60193.7	&	$r$	&	20.14	&	0.14	\\
%60193.7	&	$z$	&	18.73*	&	0.28	\\
%60194.8	&	$z$	&	19.35*	&	0.25	\\
%60196.8	&	$i$	&	20.72*	&	0.36	\\
%60196.8	&	$z$	&	19.27*	&	0.29	\\
60197.7	&	$g$	&	20.74	&	0.09	\\
60197.7	&	$i$	&	20.16	&	0.15	\\
60197.7	&	$r$	&	20.18	&	0.09	\\
%60197.7	&	$z$	&	19.52*	&	0.25	\\
%60214.7	&	$z$	&	18.99*	&	0.21	\\
%60216.7	&	$i$	&	19.95	&	0.24	\\
%60244.7	&	$i$	&	20.56	&	0.40	\\

\hline
\end{tabular}
\tablefoot{The point marked with an asterisk (*) is an outlier and not used in our analysis.}
\end{table}

\begin{table}
\centering
\caption{UV magnitudes of SN~2023aew.}
\label{tab:uvot}
\begin{tabular}{cccc}
\hline
MJD & Filter & Magnitude & Error \\
 & & (mag) & (mag) \\
\hline
 60076.5 & UVW1 & 19.46 & 0.13 \\
 60076.5 & UVM2 & 20.64 & 0.18 \\
 60076.5 & UVW2 & 20.96 & 0.22 \\
 60077.3 & UVW1 & 19.32 & 0.11 \\
 60077.3 & UVM2 & 20.57 & 0.17 \\
 60077.3 & UVW2 & 20.50 & 0.18 \\
 60079.5 & UVW1 & 19.43 & 0.11 \\
 60079.5 & UVM2 & 20.83 & 0.18 \\
 60079.5 & UVW2 & 21.11 & 0.22 \\
 60081.8 & UVW1 & 19.74 & 0.14 \\
 60081.8 & UVM2 & 20.97 & 0.20 \\
 60081.8 & UVW2 & 21.06 & 0.23 \\
 60083.0 & UVW1 & 19.85 & 0.15 \\
 60083.0 & UVM2 & 20.83 & 0.24 \\
 60083.0 & UVW2 & 20.95 & 0.27 \\
 60085.4 & UVW1 & 19.77 & 0.15 \\
 60085.4 & UVM2 & 21.29 & 0.40 \\
 60085.4 & UVW2 & 20.82 & 0.28 \\
\hline
\end{tabular}
\tablefoot{All magnitudes are based on the public \emph{Swift}/UVOT data, in the AB system, and a host-galaxy measurement from December 2023 has been subtracted.}
\end{table}

\begin{table}
\centering
\caption{Log of our LCOGT $uB$-band imaging observations of SN~2023aew.}
\label{tab:lcogtB}
\begin{tabular}{cccc}
\hline
MJD & Filter & Magnitude & Error \\
 & & (mag) & (mag) \\
\hline
60062.2 & $u$ & 18.01 & 0.03 \\
60062.2 & $B$ & 17.13 & 0.04 \\
60064.1 & $u$ & 17.79 & 0.05 \\
60064.1 & $B$ & 17.09 & 0.02 \\
60066.1 & $u$ & 17.64 & 0.07 \\
60066.1 & $B$ & 17.01 & 0.02 \\
60068.2 & $u$ & 17.60 & 0.07 \\
60068.2 & $B$ & 16.98 & 0.03 \\
60070.1 & $u$ & 17.43 & 0.10 \\
60070.1 & $B$ & 16.96 & 0.04 \\
60072.1 & $u$ & 17.47 & 0.07 \\
60072.1 & $B$ & 16.90 & 0.02 \\
60074.0 & $u$ & 17.45 & 0.07 \\
60074.0 & $B$ & 16.91 & 0.02 \\
60076.2 & $u$ & 17.40 & 0.07 \\
60076.2 & $B$ & 16.94 & 0.02 \\
60078.2 & $u$ & 17.50 & 0.06 \\
60078.2 & $B$ & 16.96 & 0.02 \\
60081.4 & $u$ & 17.54 & 0.05 \\
60081.4 & $B$ & 17.09 & 0.02 \\
60087.4 & $u$ & 17.95 & 0.08 \\
60087.4 & $B$ & 17.25 & 0.03 \\
60090.4 & $u$ & 18.11 & 0.07 \\
60090.4 & $B$ & 17.38 & 0.03 \\
60092.4 & $u$ & 18.17 & 0.12 \\
60092.4 & $B$ & 17.45 & 0.08 \\
60097.3 & $B$ & 17.69 & 0.20 \\
60100.3 & $u$ & 18.77 & 0.19 \\
60100.3 & $B$ & 17.83 & 0.06 \\
60103.1 & $B$ & 17.82 & 0.04 \\
60106.2 & $u$ & 19.21 & 0.11 \\
60106.2 & $B$ & 17.94 & 0.03 \\
60108.9 & $u$ & 19.30 & 0.08 \\
60108.9 & $B$ & 18.06 & 0.03 \\
60112.2 & $u$ & 19.35 & 0.11 \\
60112.2 & $B$ & 18.29 & 0.05 \\
60115.2 & $u$ & 19.57 & 0.11 \\
60115.2 & $B$ & 18.65 & 0.04 \\
60120.0 & $B$ & 18.57 & 0.04 \\
60126.1 & $B$ & 18.85 & 0.10 \\
60132.4 & $B$ & 19.42 & 0.17 \\
60140.3 & $B$ & 19.43 & 0.05 \\
60146.3 & $B$ & 19.91 & 0.08 \\
\hline
\end{tabular}
\tablefoot{All magnitudes are in the AB system.}
\end{table}

\begin{table}
\centering
\caption{Log of our LCOGT $gri$-band imaging observations of SN~2023aew.}
\label{tab:lcogtR}
\begin{tabular}{cccc}
\hline
MJD & Filter & Magnitude & Error \\
 & & (mag) & (mag) \\
\hline
60062.2 & $g$ & 16.97 & 0.02 \\
60062.2 & $r$ & 16.67 & 0.02 \\
60062.2 & $i$ & 16.64 & 0.03 \\
60064.1 & $g$ & 16.88 & 0.02 \\
60064.1 & $r$ & 16.64 & 0.02 \\
60064.1 & $i$ & 16.57 & 0.04 \\
60066.1 & $g$ & 16.78 & 0.02 \\
60066.1 & $r$ & 16.59 & 0.02 \\
60066.1 & $i$ & 16.54 & 0.03 \\
60068.2 & $g$ & 16.74 & 0.03 \\
60068.2 & $r$ & 16.56 & 0.02 \\
60068.2 & $i$ & 16.54 & 0.02 \\
60070.1 & $g$ & 16.75 & 0.04 \\
60070.1 & $r$ & 16.63 & 0.03 \\
60070.1 & $i$ & 16.52 & 0.04 \\
60072.1 & $g$ & 16.74 & 0.02 \\
60072.1 & $r$ & 16.54 & 0.02 \\
60072.1 & $i$ & 16.53 & 0.03 \\
60074.0 & $g$ & 16.72 & 0.01 \\
60074.0 & $r$ & 16.53 & 0.02 \\
60074.0 & $i$ & 16.57 & 0.03 \\
60076.2 & $g$ & 16.76 & 0.02 \\
60076.2 & $r$ & 16.55 & 0.02 \\
60076.2 & $i$ & 16.51 & 0.02 \\
60078.2 & $r$ & 16.55 & 0.01 \\
60078.2 & $i$ & 16.74 & 0.02 \\
60081.4 & $g$ & 16.86 & 0.02 \\
60081.4 & $r$ & 16.60 & 0.03 \\
60081.4 & $i$ & 16.57 & 0.02 \\
60087.4 & $g$ & 17.02 & 0.03 \\
60087.4 & $r$ & 16.70 & 0.02 \\
60087.4 & $i$ & 16.70 & 0.02 \\
60090.4 & $g$ & 17.10 & 0.03 \\
60090.4 & $r$ & 16.70 & 0.02 \\
60090.4 & $i$ & 16.74 & 0.03 \\
60092.4 & $g$ & 17.25 & 0.06 \\
60092.4 & $r$ & 16.78 & 0.06 \\
60092.4 & $i$ & 16.69 & 0.10 \\
60097.3 & $g$ & 17.49 & 0.20 \\
60097.3 & $r$ & 17.07 & 0.12 \\
60097.3 & $i$ & 17.01 & 0.08 \\
60100.3 & $g$ & 17.41 & 0.04 \\
60100.3 & $r$ & 16.98 & 0.03 \\
60100.3 & $i$ & 16.98 & 0.03 \\
60103.1 & $g$ & 17.52 & 0.03 \\
60103.1 & $r$ & 17.04 & 0.03 \\
60103.1 & $i$ & 17.04 & 0.03 \\
60106.2 & $g$ & 17.65 & 0.02 \\
60106.2 & $r$ & 17.19 & 0.02 \\
60106.2 & $i$ & 17.23 & 0.04 \\
60108.9 & $g$ & 17.74 & 0.03 \\
60108.9 & $r$ & 17.27 & 0.02 \\
60108.9 & $i$ & 17.24 & 0.03 \\
\hline
\end{tabular}
\tablefoot{All magnitudes are in the AB system.}
\end{table}

\setcounter{table}{3}

\begin{table}
\centering
\caption{Log of our LCOGT $gri$-band imaging observations of SN~2023aew (continued).}
\label{tab:lcogtR}
\begin{tabular}{cccc}
\hline
MJD & Filter & Magnitude & Error \\
 & & (mag) & (mag) \\
\hline
60112.2 & $g$ & 17.91 & 0.04 \\
60112.2 & $r$ & 17.40 & 0.03 \\
60112.2 & $i$ & 17.39 & 0.05 \\
60115.2 & $g$ & 18.00 & 0.03 \\
60115.2 & $r$ & 17.51 & 0.02 \\
60115.2 & $i$ & 17.48 & 0.04 \\
60120.0 & $i$ & 17.82 & 0.04 \\
60126.1 & $i$ & 18.03 & 0.09 \\
60132.4 & $i$ & 18.44 & 0.10 \\
60140.3 & $i$ & 18.69 & 0.07 \\
60146.3 & $i$ & 18.82 & 0.09 \\
\hline
\end{tabular}
\end{table}

\begin{table}
\centering
\caption{Log of our LT/IO:O imaging observations of SN~2023aew.}
\label{tab:ltioo}
\begin{tabular}{cccc}
\hline
MJD & Filter & Magnitude & Error \\
 & & (mag) & (mag) \\
\hline
60062.1 & $g$ & 16.97 & 0.03 \\
60062.1 & $r$ & 16.65 & 0.03 \\
60062.1 & $i$ & 16.62 & 0.03 \\
60062.1 & $z$ & 16.72 & 0.04 \\
60065.0 & $g$ & 16.95 & 0.05 \\
60065.0 & $r$ & 16.61 & 0.04 \\
60065.0 & $i$ & 16.59 & 0.03 \\
60065.0 & $z$ & 16.73 & 0.04 \\
60075.0 & $g$ & 16.91 & 0.08 \\
60075.0 & $r$ & 16.38 & 0.03 \\
60075.0 & $i$ & 16.51 & 0.03 \\
60075.0 & $z$ & 16.58 & 0.04 \\
60122.9 & $g$ & 18.46 & 0.14 \\
60122.9 & $r$ & 17.91 & 0.09 \\
60122.9 & $i$ & 17.97 & 0.07 \\
60122.9 & $z$ & 17.70 & 0.10 \\
 \hline
\end{tabular}
\tablefoot{All magnitudes are in the AB system.}
\end{table}

\begin{table}
\centering
\caption{Log of our NOT/ALFOSC imaging observations of SN~2023aew.}
\label{tab:alfosc}
\begin{tabular}{cccc}
\hline
MJD & Filter & Magnitude & Error \\
 & & (mag) & (mag) \\
\hline
60061.0 & $u$ & 18.39 & 0.13 \\
60061.0 & $g$ & 17.06 & 0.02 \\
60061.0 & $r$ & 16.70 & 0.02 \\
60061.0 & $i$ & 16.67 & 0.02 \\
60061.0 & $z$ & 16.83 & 0.02 \\
60134.1 & $u$ & 20.58 & 0.05 \\
60134.1 & $g$ & 19.00 & 0.01 \\
60134.1 & $r$ & 18.52 & 0.01 \\
60134.1 & $i$ & 18.53 & 0.01 \\
60134.1 & $z$ & 18.23 & 0.02 \\
60158.0 & $g$ & 19.58 & 0.03 \\
60158.0 & $r$ & 18.91 & 0.03 \\
60158.0 & $i$ & 19.00 & 0.03 \\
60158.0 & $z$ & 18.75 & 0.04 \\
60169.1 & $g$ & 20.22 & 0.05 \\
60169.1 & $r$ & 19.55 & 0.03 \\
60169.1 & $i$ & 19.54 & 0.03 \\
60169.1 & $z$ & 19.12 & 0.04 \\
60177.0 & $g$ & 20.41 & 0.02 \\
60177.0 & $r$ & 19.83 & 0.02 \\
60177.0 & $i$ & 19.87 & 0.02 \\
60177.0 & $z$ & 19.38 & 0.04 \\
60185.0 & $g$ & 20.59 & 0.04 \\
60185.0 & $r$ & 20.00 & 0.03 \\
60185.0 & $i$ & 20.07 & 0.03 \\
60185.0 & $z$ & 19.62 & 0.04 \\
60197.0 & $g$ & 20.61 & 0.03 \\
60197.0 & $r$ & 20.14 & 0.02 \\
60197.0 & $i$ & 20.25 & 0.03 \\
60197.0 & $z$ & 19.72 & 0.04 \\
60207.9 & $g$ & 21.10 & 0.02 \\
60207.9 & $r$ & 20.28 & 0.02 \\
60207.9 & $i$ & 20.46 & 0.03 \\
60207.9 & $z$ & 20.08 & 0.04 \\
60218.9 & $g$ & 21.07 & 0.03 \\
60218.9 & $r$ & 20.42 & 0.03 \\
60218.9 & $i$ & 20.34 & 0.03 \\
60218.9 & $z$ & 19.93 & 0.04 \\
60224.9 & $g$ & 21.18 & 0.03 \\
60224.9 & $r$ & 20.57 & 0.03 \\
60224.9 & $i$ & 20.82 & 0.05 \\
60224.9 & $z$ & 20.76 & 0.13 \\
60261.8 & $g$ & 22.51 & 0.20 \\
60261.8 & $r$ & 21.89 & 0.09 \\
60261.8 & $i$ & 21.67 & 0.07 \\
60261.8 & $z$ & 22.34 & 0.36 \\
60273.8 & $r$ & 22.32 & 0.22 \\
60338.3 & $r$ & 23.53 & 0.42 \\
\hline
\end{tabular}
\tablefoot{All magnitudes are in the AB system.}
\end{table}

\begin{table}
\centering
\caption{TESS-Red magnitudes of SN~2023aew.}
\label{tab:tess}
\begin{tabular}{ccc}
\hline
MJD & Magnitude & Error \\
 & (mag) & (mag) \\
\hline
59936.7 & 19.31 & 0.38 \\
59937.2 & 18.45 & 0.24 \\
59937.7 & 18.66 & 0.18 \\
59938.2 & 19.13 & 0.23 \\
59938.7 & 19.35 & 0.24 \\
59939.2 & 19.55 & 0.23 \\
59939.7 & 19.49 & 0.19 \\
59940.2 & 19.60 & 0.20 \\
59940.7 & 19.37 & 0.15 \\
59941.2 & 19.21 & 0.12 \\
59941.7 & 19.29 & 0.13 \\
59942.2 & 19.26 & 0.13 \\
59942.7 & 19.23 & 0.13 \\
59943.2 & 18.93 & 0.13 \\
59943.7 & 18.67 & 0.08 \\
59944.2 & 18.68 & 0.08 \\
59944.7 & 18.59 & 0.07 \\
59945.2 & 18.43 & 0.06 \\
59945.7 & 18.46 & 0.06 \\
59946.2 & 18.38 & 0.06 \\
59946.7 & 18.29 & 0.05 \\
59947.2 & 18.22 & 0.05 \\
59947.7 & 18.20 & 0.05 \\
59948.2 & 18.06 & 0.04 \\
59948.7 & 17.84 & 0.03 \\
59949.2 & 17.78 & 0.04 \\
59954.7 & 17.38 & 0.04 \\
59955.2 & 17.41 & 0.03 \\
59955.7 & 17.52 & 0.03 \\
59956.2 & 17.52 & 0.03 \\
59956.7 & 17.44 & 0.03 \\
59957.2 & 17.47 & 0.03 \\
59957.7 & 17.49 & 0.03 \\
59958.2 & 17.45 & 0.03 \\
59958.7 & 17.45 & 0.03 \\
59959.2 & 17.48 & 0.02 \\
59959.7 & 17.50 & 0.03 \\
59960.2 & 17.47 & 0.03 \\
59960.7 & 17.47 & 0.03 \\
59961.2 & 17.47 & 0.03 \\
59961.7 & 17.50 & 0.03 \\
59962.0 & 17.53 & 0.04 \\
\hline
\end{tabular}
\tablefoot{All magnitudes are in the AB system and based on public Sector 60 TESS data. The photometry has been combined into 12-hour bins.}
\end{table}

\begin{table}
\centering
\caption{Log of our NOT/NOTCam imaging observations of SN~2023aew.}
\label{tab:notcam}
\begin{tabular}{cccc}
\hline
MJD & Filter & Magnitude & Error \\
 & & (mag) & (mag) \\
\hline
60096.1 & $J$ & 16.51 & 0.05 \\
60096.1 & $H$ & 16.12 & 0.06 \\
60096.1 & $Ks$ & 16.18 & 0.07 \\
60158.1 & $J$ & 17.95 & 0.05 \\
60158.1 & $H$ & 17.66 & 0.06 \\
60158.1 & $Ks$ & 17.77 & 0.14 \\
60186.0 & $J$ & 19.09 & 0.05 \\
60186.0 & $H$ & 18.51 & 0.06 \\
60186.0 & $Ks$ & 18.54 & 0.07 \\
60213.9 & $J$ & 19.02 & 0.05 \\
60213.9 & $H$ & 18.25 & 0.04 \\
60213.9 & $Ks$ & 18.37 & 0.06 \\
60271.8 & $H$ & 20.01 & 0.12 \\
\hline
\end{tabular}
\tablefoot{All magnitudes are in the Vega system.}
\end{table}

\begin{table}
\centering
\caption{Log of our Nayuta/NIC imaging observations of SN~2023aew.}
\label{tab:hyogo}
\begin{tabular}{cccc}
\hline
MJD & Filter & Magnitude & Error \\
 & & (mag) & (mag) \\
\hline
60064.8 & $H$ & 16.09 & 0.08 \\
60064.8 & $Ks$ & 16.13 & 0.20 \\
60144.7 & $J$ & 18.03 & 0.14 \\
60160.6 & $H$ & 17.77 & 0.23 \\
60167.6 & $J$ & 18.65 & 0.25 \\
60167.6 & $H$ & 18.00 & 0.20 \\
60168.7 & $J$ & 18.45 & 0.15 \\
60168.7 & $H$ & 18.03 & 0.30 \\
\hline
\end{tabular}
\tablefoot{All magnitudes are in the Vega system.}
\end{table}

\begin{table}
\centering
\caption{Log of our TNG/NICS imaging observations of SN~2023aew.}
\label{tab:nics}
\begin{tabular}{cccc}
\hline
MJD & Filter & Magnitude & Error \\
 & & (mag) & (mag) \\
\hline
60063.2 & $J$ & 16.33 & 0.04 \\
60063.2 & $H$ & 16.13 & 0.07 \\
60063.2 & $Ks$ & 16.10 & 0.07 \\
\hline
\end{tabular}
\tablefoot{All magnitudes are in the Vega system.}
\end{table}

\begin{table}
\centering
\caption{Log of spectroscopic observations used in this paper.}
\label{tab:specs}
\begin{tabular}{ccccc}
\hline
MJD & Epoch & Grism/grating & $R$ & Exposure \\
 & (d) &  &  & (s) \\
\hline
LT/SPRAT & & & & \\
\hline
60055.0 & $-$19.4 & Wasatch600 & 350 & 600 \\
60062.2 & $-$12.4 & Wasatch600 & 350 & 600 \\
60065.0 & $-$9.7 & Wasatch600 & 350 & 600 \\
60075.1 & 0.2 & Wasatch600 & 350 & 600 \\
\hline
NOT/ALFOSC & & & & \\
\hline
60061.0 & $-$13.6 & \#4 & 360 & 900 \\
60068.1 & $-$6.6 & \#4 & 360 & 900 \\
60077.2 & 2.2 & \#4 & 360 & 900 \\
60084.1 & 9.0 & \#4 & 360 & 900 \\
60098.2 & 22.7 & \#4 & 360 & 1800 \\
60117.2 & 41.3 & \#4 & 360 & 2100 \\
60133.0 & 56.7 & \#4 & 360 & 2100 \\
60169.1 & 91.9 & \#4 & 360 & 1800 \\
\hline
INT/IDS & & & & \\
\hline
60095.1 & 19.7 & R300V & 1100 & 900 \\
60096.2 & 20.8 & R300V & 1100 & 900 \\
60124.1 & 48.0 & R300V & 1100 & 900 \\
60126.1 & 50.0 & R300V & 1100 & 900 \\
\hline
TNG/NICS & & & & \\
\hline
60064.1 & $-$10.5 & $JH$ & 500 & 3600 \\
60064.2 & $-$10.4 & $HK$ & 500 & 3600 \\
\hline
Seimei/KOOLS & & & & \\
\hline
60060.7 & $-$13.9 & VPH-Blue & 500 & 1200 \\
60061.6 & $-$13.0 & VPH-Blue & 500 & 1800 \\
60074.7 & $-$0.2 & VPH-Blue & 500 & 1800 \\
60081.7 & 6.6 & VPH-Blue & 500 & 2700 \\
60085.7 & 10.5 & VPH-Blue & 500 & 1800 \\
60112.8 & 37.0 & VPH-Blue & 500 & 1647 \\
\hline
Gemini/GMOS & & & & \\
\hline
 60147.3 & 70.6 & R400 & 640 & 6000 \\
\hline
GTC/OSIRIS & & & & \\
\hline
 60184.0 & 106.4 & R1000B & 610 & 1800 \\
 60184.0 & 106.4 & R1000R & 670 & 1800 \\
 60208.9 & 130.7 & R1000B & 610 & 1800 \\
 60208.9 & 130.7 & R1000R & 670 & 1800 \\
 60226.9 & 148.3 & R1000B & 610 & 1800 \\
 60226.9 & 148.3 & R1000R & 670 & 1800 \\
\hline
\end{tabular}
\tablefoot{The resolution is reported at the blaze wavelength of each grism/grating.}
\end{table}

\FloatBarrier

\section{MOSFiT corner plots}

\begin{figure}[h!]
\centering
\includegraphics[width=1.8\linewidth]{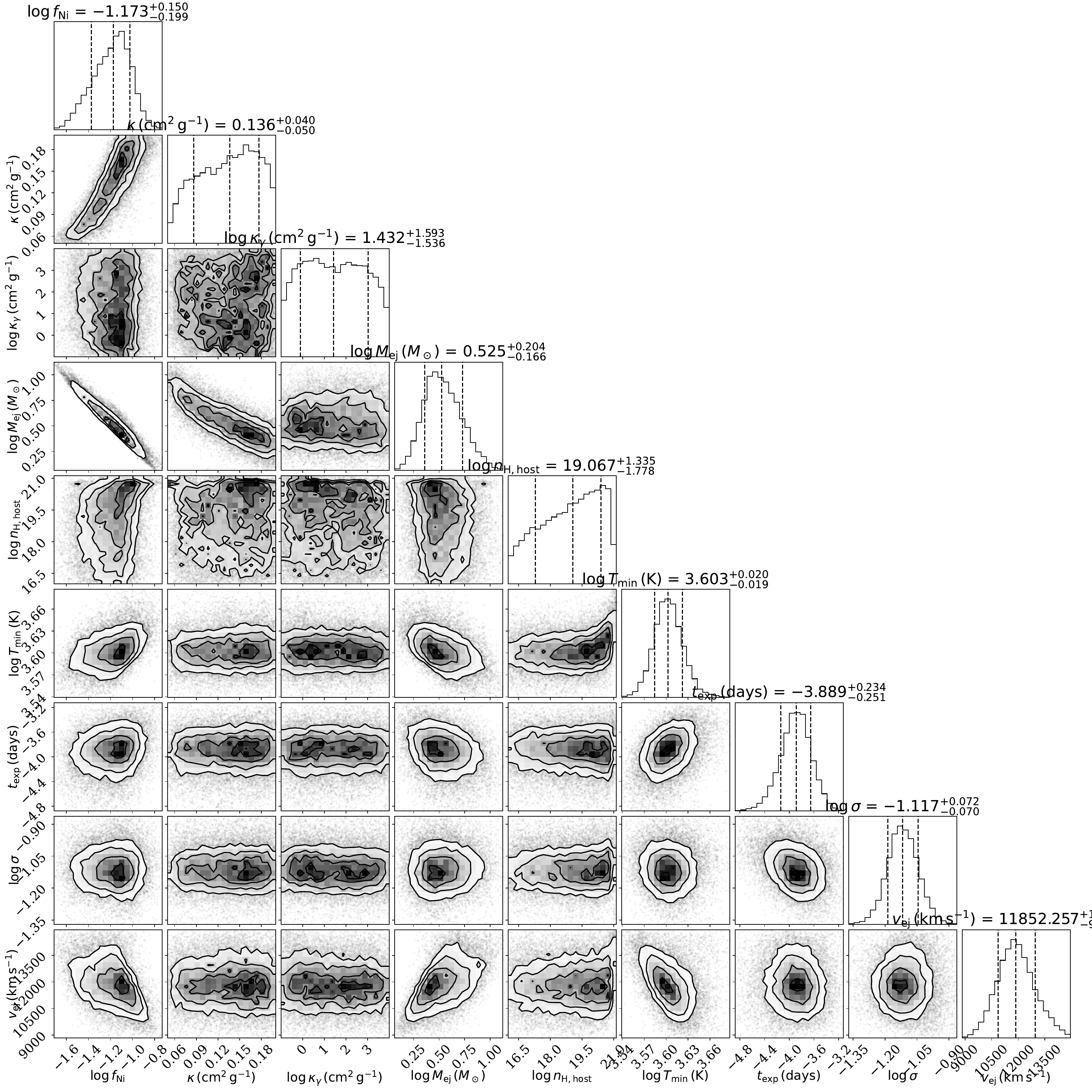}
\caption{\texttt{MOSFiT} corner plot for the $^{56}$Ni decay fit of the first peak.}
\label{fig:cornerni56_peak1}
\end{figure}

\onecolumn

\begin{figure}
\centering
\includegraphics[width=0.9\linewidth]{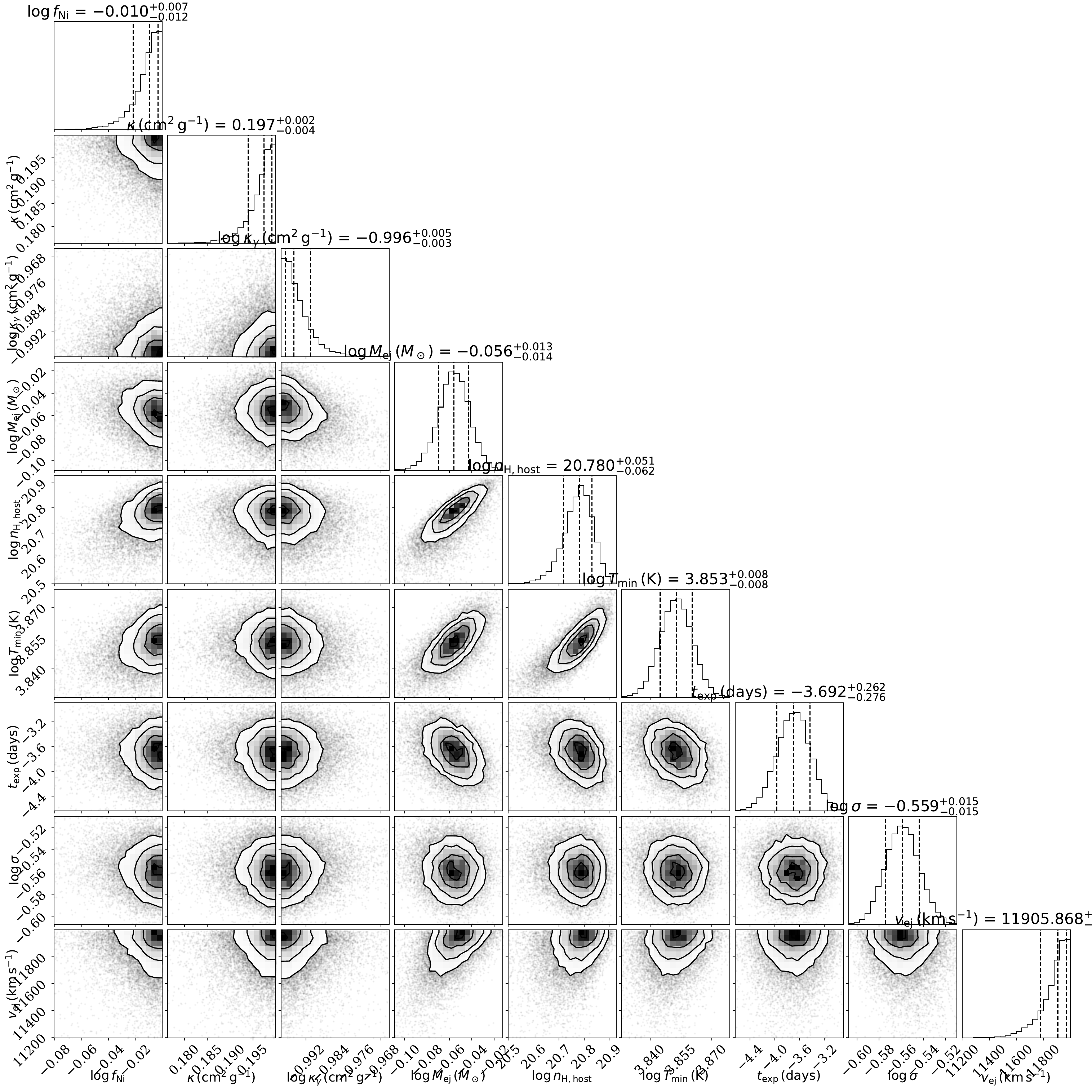}
\caption{\texttt{MOSFiT} corner plot for the $^{56}$Ni decay fit of the second peak.}
\label{fig:cornerni56}
\end{figure}

\begin{figure}
\centering
\includegraphics[width=0.9\linewidth]{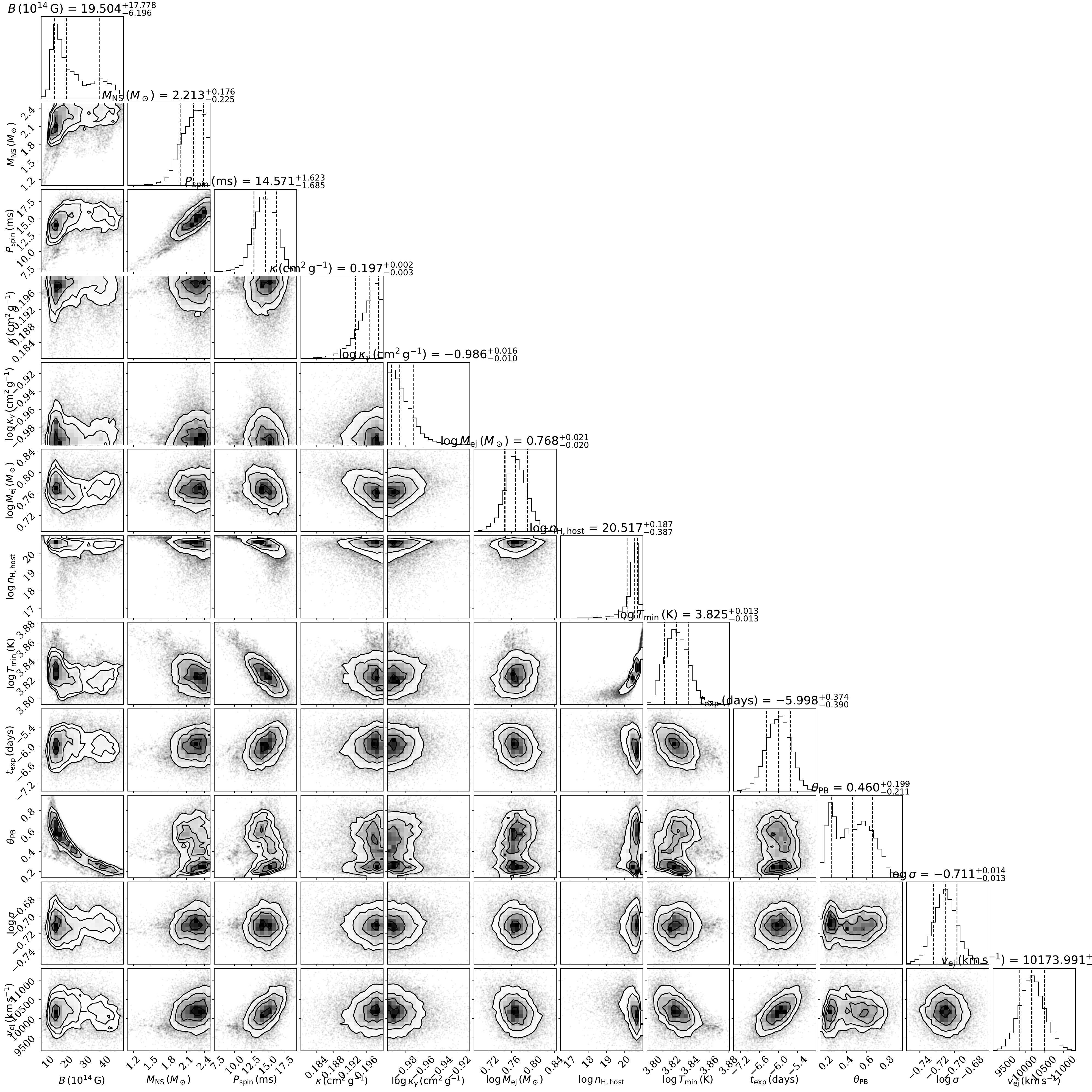}
\caption{\texttt{MOSFiT} corner plot for the magnetar spindown fit of the second peak.}
\label{fig:cornermag}
\end{figure}

\begin{figure}
\centering
\includegraphics[width=0.9\linewidth]{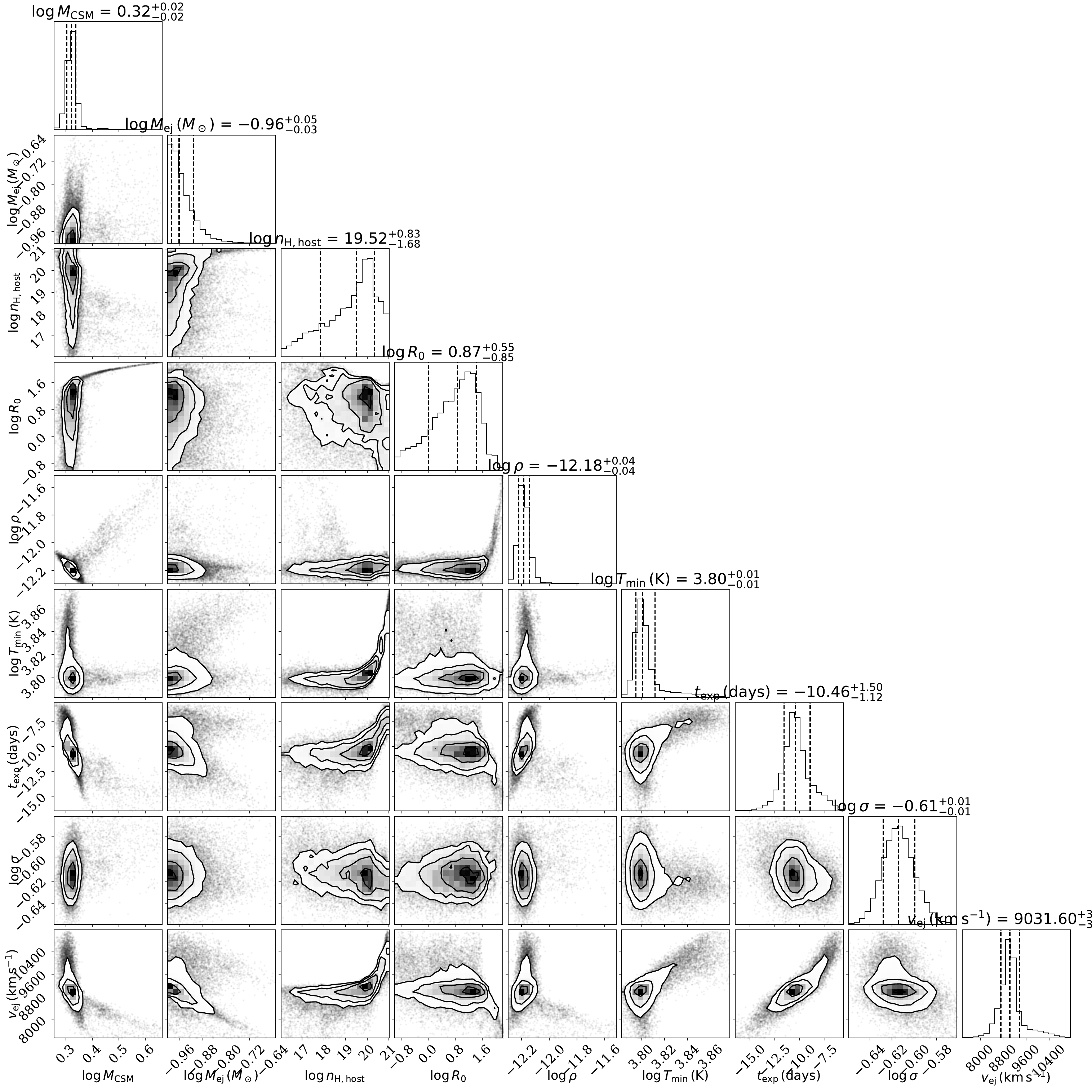}
\caption{\texttt{MOSFiT} corner plot for the CSI fit of the second peak when $s=0$.}
\label{fig:cornercms0}
\end{figure}

\begin{figure}
\centering
\includegraphics[width=0.9\linewidth]{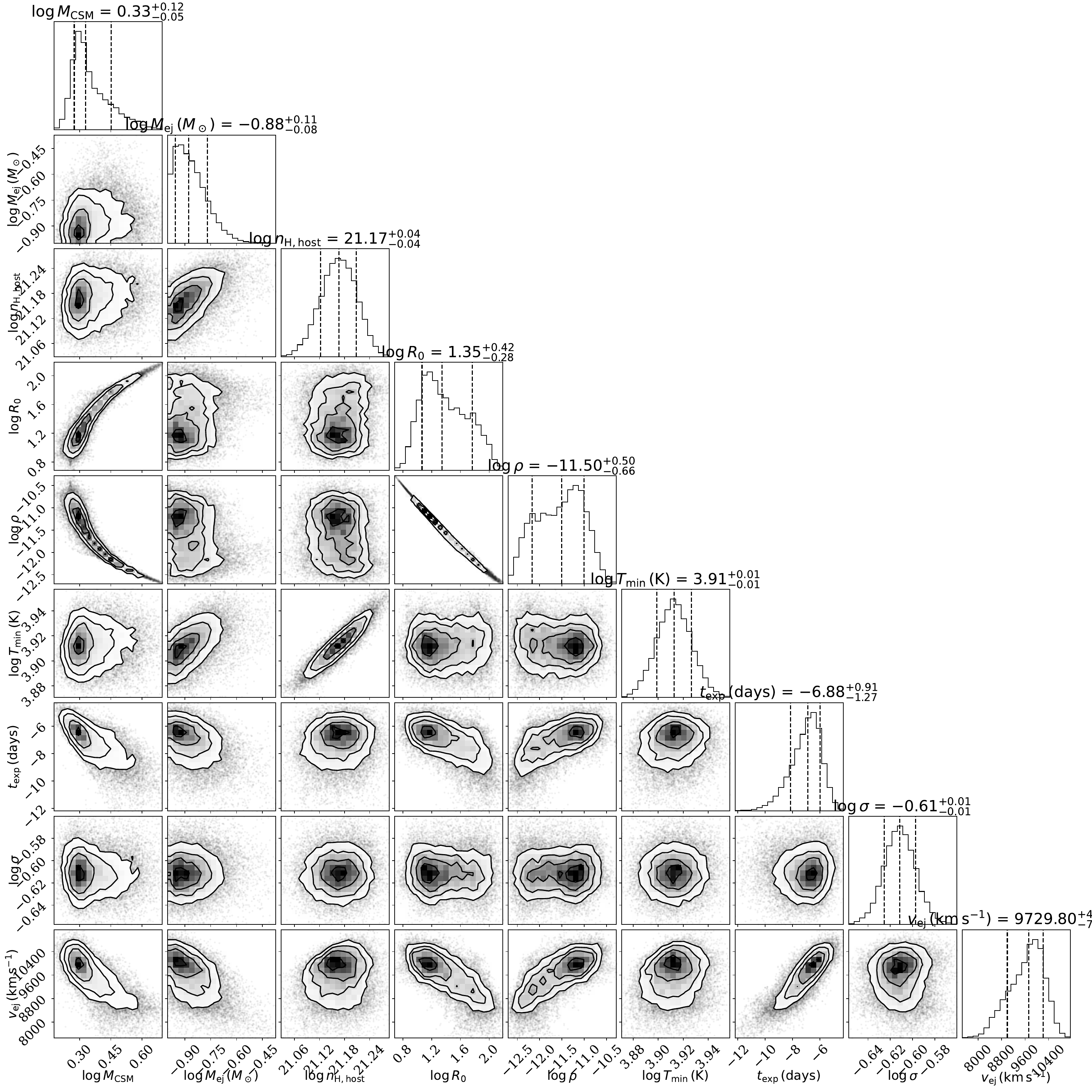}
\caption{\texttt{MOSFiT} corner plot for the CSI fit of the second peak when $s=2$.}
\label{fig:cornercsms2}
\end{figure}

\begin{figure}
\centering
\includegraphics[width=0.9\linewidth]{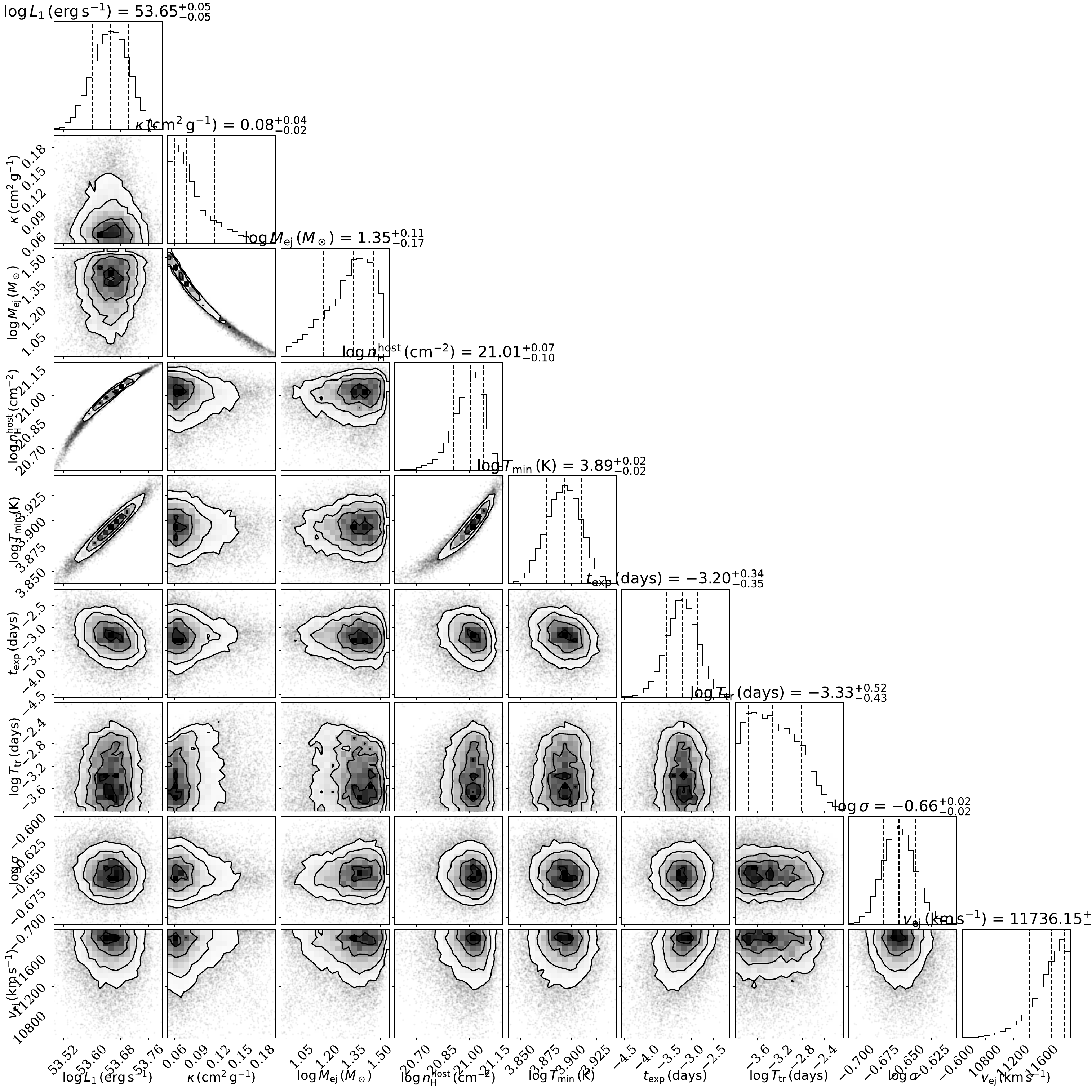}
\caption{{\texttt{MOSFiT} corner plot for the fallback fit of the second peak.}}
\label{fig:cornerFB}
\end{figure}

\FloatBarrier

\section{Coordinate discrepancy}
\label{sec:coords}

Considering the differences between the spectra of the first and second peak and the rapidity of the rise to the second peak, the two peaks of SN 2023aew might, in fact, conceivably be two SNe on slightly different lines of sight, although such a spatial and temporal coincidence would be extremely unlikely, especially taking into account the faintness and low star formation rate of the host galaxy (see Sect. \ref{sec:disco}). In order to check whether this is the case, we used the coordinates of Gaia23ate in the \emph{Gaia} Alerts Index ($\alpha_1$=17:40:51.37; $\delta_1$=66:12:22.75). These correspond to the first scan of the first peak. The uncertainty of \emph{Gaia} coordinates is 0\farcs055 per scan \citep{hodgkin21}. We also searched for highly precise \emph{Gaia} coordinates of stars within the field, using 26 stars within 2\arcmin~of the SN with astrometric uncertainties between 0.02 and 3 mas as reported in the \emph{Gaia} Archive.\footnote{\url{https://gea.esac.esa.int/archive/}} This uncertainty is negligible compared to that of the position of SN~2023aew. We then used the \texttt{IRAF} tasks \texttt{ccmap}, \texttt{cctran} and \texttt{ccsetwcs} to calibrate the astrometry in our ALFOSC and NOTCam images from 2023 Apr 26, 2023 Jul 08 and 2023 May 30; the root-mean-square (RMS) errors from these fits were between 0\farcs03~and 0\farcs06. After rejecting outliers, between 15 and 20 stars visible in the field were used in each fit.

From our NOT images, we measured the average coordinates of the second peak as $\bar{\alpha_2}$=17:40:51.356; $\bar{\delta_2}$=66:12:22.820. The uncertainty of any discrepancy is $\sqrt{\sigma_{1}^2 + \sigma_{2}^2}$, where $\sigma_{1}$ is the \emph{Gaia} uncertainty of the first peak coordinates (0\farcs055) and $\sigma_{2}$ the uncertainty of the average NOT centroid measurement, $\sigma_2 = \sqrt{\Sigma \sigma_i^2}/N$, where $N$ is the number of measurements (9). Each individual measurement error $\sigma_i$ is calculated as
\begin{equation}
    \sigma_i = d_i \left( \frac{\sigma_{i,\mathrm{RA}}}{\Delta \mathrm{RA}_i} + \frac{\sigma_{i,\mathrm{Dec}}}{\Delta \mathrm{Dec}_i} \right),
\end{equation}
where $d$ is the angular distance between the peaks in measurement $i$, $\Delta \mathrm{RA}$ is the angular distance in the right ascension direction between the peaks (i.e., not the difference in $\alpha$ directly) and $\Delta \mathrm{Dec}$ the declination difference, $\sigma_\mathrm{RA}$ the RMS error of the astrometry in $\alpha$ and $\sigma_\mathrm{Dec}$ in $\delta$. This is because $d = \sqrt{\Delta \mathrm{RA}^2 + \Delta \mathrm{Dec}^2}$. We then obtain an offset of $0\farcs104\pm0\farcs080$, with a significance of 1.3$\sigma$.

Additionally, we used the $gr$-band centroid coordinates reported for the ZTF photometry by ALeRCE, assuming that the astrometry is consistent between each epoch (the accuracy of the absolute astrometry is irrelevant here). We took the averages of the $\alpha$ and $\delta$ coordinates of all epochs before MJD\,=\,60043 (17 measurements) to represent the first peak and those of all epochs between MJD\,=\,60044 and MJD\,=\,60156 (76 measurements) to represent the second peak. The uncertainties in $\alpha$ and $\delta$ were taken to be the standard errors of the mean. The resulting coordinate discrepancy between the two peaks is $0\farcs078\pm0\farcs068$, also barely over $1\sigma$, but in the same direction (negative $\alpha$, positive $\delta$) as the \emph{Gaia}-NOT discrepancy. 

Finally, combining the $\alpha$ and $\delta$ offsets from the \emph{Gaia}-NOT comparison and the ZTF measurements, we obtain a total offset of $0\farcs090\pm0\farcs053$ if the ZTF and NOT offsets are given equal weight. The significance of this discrepancy is only $\sim$1.7$\sigma$ -- as such, even if there is any coordinate offset between the two objects, it is too small to significantly detect in the data available to us. There is also a possibility of underestimated errors because of a systematic error in the \emph{Gaia} coordinates of the first peak, as the offset from the NOT images relies on this measurement. Such an issue could further decrease the significance.

\end{appendix}

\end{document}